\documentclass[twocolumn]{autart}    
\usepackage[T1]{fontenc}
\usepackage[utf8]{inputenc}
\usepackage{cite}
\usepackage{amsmath,amssymb,amsfonts}
\usepackage{graphicx}
\usepackage{caption}
\usepackage{subcaption}
\usepackage{algpseudocode}
\usepackage{textcomp}
\usepackage{xcolor}
\usepackage{stfloats}
\usepackage{booktabs}

\usepackage{textcomp}
\usepackage{multirow}
\usepackage{mathtools}
\usepackage[thinc]{esdiff}
\usepackage{blkarray}
\usepackage{fixltx2e}
\usepackage{tabularx}
\usepackage{arydshln,leftidx,mathtools}
\usepackage{enumerate}
\usepackage{times}
\usepackage{bm}
\usepackage{algorithm}
\usepackage{algpseudocode}
\usepackage{hyperref}
\makeatletter
\@ifundefined{c@part}{\newcounter{part}}{}
\providecommand\thepart{\arabic{part}}

\makeatother

\newcommand{\eos}{\ensuremath{\hfill\square}}

\newtheorem{assumption}{Assumption}[section]

\newtheorem{theorem}{Theorem}[section]

\newtheorem{lemma}{Lemma}
\newtheorem{corollary}{Corollary}
\newtheorem{remark}{Remark}
\newenvironment{proof}{%
  \par\noindent\textit{Proof.}\ %
}{%
  \eos\par
}

\newcommand{\norm}[1]{\left\lVert#1\right\rVert}
\newcommand{\innerproduct}[2]{\langle #1, #2 \rangle}
\newcommand{\AsN}[2]{\operatorname{As}\!\mathcal N\!\left(#1,#2\right)}
\newcommand{\Vc}[1]{\text{Vec}\left(#1\right)}

\begin{document}

\begin{frontmatter}

\title{Weighted null space fitting (WNSF): A link between the prediction error method and subspace identification\thanksref{footnoteinfo}} 
\thanks[footnoteinfo]{This work was supported by VINNOVA Competence Center AdBIOPRO, contract [2016-05181] and by the Swedish Research Council through the research environment NewLEADS (New Directions in Learning Dynamical Systems), contract [2016-06079], and contract 2019-04956.}

\author[Sweden]{Jiabao He}\ead{jiabaoh@kth.se},    
\author[HK]{S. Joe Qin}\ead{joeqin@ln.edu.hk},               
\author[Sweden]{H\r{a}kan Hjalmarsson}\ead{hjalmars@kth.se}  
\address[Sweden]{Division of Decision and Control Systems, KTH Royal Institute of Technology, Sweden}  
\address[HK]{School of Data Science, Lingnan University, Hong Kong}


\begin{abstract}                          
Subspace identification methods (SIMs) have proven to be very useful and numerically robust for building state-space models. While most SIMs are consistent, few if any can achieve the
efficiency of the maximum likelihood estimate (MLE). Conversely, the prediction error method (PEM) with a quadratic criteria is equivalent to MLE, but it comes with non-convex optimization problems and requires good initialization points. This contribution proposes a weighted null space fitting (WNSF) approach for estimating state-space models, combining some key advantages of the two aforementioned mainstream approaches. It starts with a least-squares estimate of a high-order ARX model, and then a multi-step least-squares procedure reduces the model to a state-space model on canoncial form. It is demonstrated through statistical analysis that when a canonical parameterization is admissible, the proposed method is consistent and asymptotically efficient, thereby making progress on the long-standing open problem about the existence of an asymptotically efficient SIM. Numerical and practical examples are provided to illustrate that the proposed method performs favorable in comparison with SIMs.
\end{abstract}

\end{frontmatter}

\section{Introduction} \label{Sct1}

The prediction error method (PEM) and subspace identification methods (SIMs) are two of the mainstream approaches in system identification. Originating from the maximum likelihood estimator (MLE) \cite{Astrom1965numerical}, PEM minimizes a cost function based on prediction errors, the differences between observed outputs and their predictions based on the model and past data. When the noise is Gaussian, PEM with a quadratic cost function is equivalent to MLE. Importantly, its asymptotic covariance reaches the Cram\'er-Rao lower bound (CRLB), making PEM an asymptotically efficient estimator \cite{Ljung1976consistency,Caines1976prediction}. A comprehensive overview of PEM, including both numerical and theoretical perspectives, is available in \cite{Ljung1999system}. PEM is widely used as a benchmark in system identification, with implementations in software like MATLAB \cite{Ljung1995system}. However, there is one key issue that may hinder successful application of PEM, namely the risk of converging to a local minimum rather than a global minimum of the cost function, which is generally non-convex. Addressing this requires local nonlinear optimization algorithms and good initial estimates. This problem is excacerbated for multi-input multi-output (MIMO) models, which typically require extensive parametrizations, leading to many false local minima.

On the other hand, originating from the celebrated Ho-Kalman algorithm \cite{Ho1966effective}, SIMs are known for its numerical robustness and convenient parameterization for MIMO models. Although there exist many variants, including but not limited to \cite{Larimore1990canonical,Van1994n4sid,Verahegen1992subspace,Qin2005novel,Jansson2003subspace,Chiuso2007role,Yu2019constrained}, most SIMs can be unified into a common framework which typically involves least-squares and singular value decomposition (SVD) \cite{Van1995unifying}. While SIMs are appealing due to their state-space representation, which is highly convenient for estimation, filtering, prediction and control, as well as their numerical robustness, certain open problems remain unsolved. For instance, the question of whether there are subspace methods that are asymptotically efficient in the presence of exogenous inputs is still unresolved, even some 60 years after this family of methods was introduced.

The primary motivation of this work is to introduce a new method for identifying linear time-invariant (LTI) systems in state-space form. This method serves as a bridge between PEM and SIMs: It offers statistical properties (consistency and asymptotic efficiency)  matching PEM and numerical robustness comparable to SIMs. Our method builds upon the foundation of existing approaches that aim to address the aforementioned drawbacks of PEM and SIMs. We will not attempt to fully review this vast field, but we highlight some of the milestones.

\subsection{Related Work}  \label{Sct1.1}

Instrumental variable methods (IVMs) \cite{Soderstrom2002instrumental} can ensure consistency in a large variety of settings without encountering non-convexity issues. Moreover, asymptotic efficiency can be achieved for certain settings via iterative algorithms \cite{Stoica1983optimal,Young2008refined}, but not for closed-loop data. 

Some methods involve fixing certain parameters within the cost function to transform it into a quadratic optimization problem, allowing the estimate to be obtained using (weighted) least-squares. In subsequent iterations, the fixed coefficients are replaced with estimates from the previous step, either in the weighting process or during a filtering step. This approach gives rise to iterative least-squares methods, which date back to \cite{Sanathanan1963transfer}. Some representative methods are the iterative quadratic maximum likelihood (IQML) method \cite{Evans1973optimal,Shaw1994optimal,Lemmerling2001iqml}, the Steiglitz-McBride method \cite{Steiglitz1965technique}, and the Box-Jenkins Steiglitz-McBride (BJSM) algorithm \cite{Zhu2016box}. 
Although this class of iterative methods bypasses non-convex optimization problems, asymptotic efficiency is only guaranteed in specific scenarios, such as using open-loop data. Additionally, to be efficient, the number of iterations is required to be infinite.

Besides iterative least-squares methods, there are some multi-step least-squares methods which require a finite number of least-squares to obtain an estimate with certain statistical properties. The rationale behind this procedure is that, in certain cases, each step corresponds to a convex optimization problem or a numerically reliable procedure. An important feature of these methods is that a more flexible model is often estimated in an intermediate step, followed by a model reduction step to obtain a model of interest. To ensure asymptotic efficiency, it is crucial that the intermediate model serves as a sufficient statistic, at least as the sample size grows and the model reduction step is conducted in a statistically sound manner. Some of the representative methods are indirect PEM \cite{Soderstrom1991indirect}, Durbin's first and second methods \cite{Durbin1960fitting,Durbin1959efficient}, and the weighted null space fitting (WNSF) method \cite{Galrinho2018parametric}. For a comprehensive overview of these methods, we refer to \cite{Galrinho2018system}. These methods have been applied to several structured models, such as output-error (OE), auto-regressive moving-average with exogenous inputs (ARMAX) models \cite{Hannan1984multivariate,Reinsel1992maximum,Poskitt1995relationship,Dufour2014asymptotic}, and Box-Jenkins (BJ) models in the left matrix fraction description (MFD) form \cite{Poskitt1989a,Poskitt1990estimation}, but not to state-space models, which is the gap this work aims to address.

During the half century since the publication of the Ho-Kalman algorithm \cite{Ho1966effective}, numerous efforts have been made to develop improved SIMs. Some significant contributions include estimating a Hankel matrix of Markov parameters directly in a unstructured manner \cite{Larimore1990canonical,Van1994n4sid,Verahegen1992subspace}, estimating multiple high-order ARX (HOARX) models in parallel \cite{Qin2005novel,Chiuso2007role}, and addressing the bias issue in closed-loop settings \cite{Verhaegen1993application,Jansson2003subspace,Ljung1996subspace,Qin2003closed,Chiuso2005consistency}. For a thorough expos\'es of SIMs, we refer to \cite{Qin2006overview,Veen2013closed}. When reducing a high order model to a state-space model, most SIMs focus on estimating the range space of the Hankel matrix via SVD. Meanwhile, a few exceptions exist, such as the null space fitting method in \cite{Viberg1997analysis,Swindlehust1995subspace,Jansson1996linear}, where an optimal estimate of the null space of the observability matrix is obtained by a two-step weighted least-squares (WLS). The null space fitting method enables the possibility to derive an optimal weighting compared to classical SIMs, which is an important heuristic for our method. However, since the optimal weighting matrix depends on the true observability matrix which is unknown, this method still requires a SVD step to explicitly obtain the observability matrix. Given the close relationship between SVD and the total least-squares (TLS) problem, the approximate realization problem was treated as a special global TLS problem in \cite{Markovsky2005application}, where a kernel representation of the system is used. Related studies can be found in \cite{De1993structured,Markovsky2007overview}. While the TLS solution has
the potential of improving the accuracy in small
samples, it can be shown as in \cite{Stoica1995weighted,He2025range} that the TLS and least-squares estimates have the same asymptotic properties. Recently, it was highlighted in \cite{De2019least, De2020least} that the least-squares optimal realization of autonomous LTI systems can be reformulated as a multi-parameter eigenvalue problem. This problem can be solved by applying forward shift recursions to a given set of multivariate polynomial equations, generating so-called block Macaulay matrices. A key concept therein is the elimination of the state vector by leveraging the Cayley-Hamilton theorem \cite[Th. 2.4.3.2]{Horn2012matrix}, with similar ideas also discussed in \cite{Nicolai2023realizing}. This perspective sheds some new light in understanding the identification of a state-space model. However, the solution of the proposed eigenvalue problem demands large-scale numerical linear algebra algorithms, and these methods are not yet applicable to larger sample sizes. Regarding the statistical properties of SIMs, asymptotic results on their consistency and asymptotic normality have been established in the literature \cite{Deistler1995consistency,Peternell1996statistical,Jansson1998consistency,Bauer1999consistency,Knudsen2001consistency,Bauer2000analysis,Gustafsson2002subspace,Bauer2005asymptotic,Chiuso2004asymptotic,Chiuso2005consistency,Chiuso2007relation,Chiuso2007role}. More recently, their statistical properties have been further investigated in the non-asymptotic regime \cite{Tsiamis2019finite,Oymak2021revisiting,He2025finite,Bakshi2023new}. In particular, the canonical variate analysis (CVA) \cite{Larimore1990canonical} method achieves the optimal accuracy in the absence of exogenous inputs \cite{Larimore1996statistical}, however, there is no formal proof to show that it is not asymptotically efficient when exogenous inputs are involved \cite{Chiuso2007role}. Currently, the quest for an asymptotically efficient SIM is still open \cite{Qin2006overview,Chiuso2007relation}. 

To identify factors hindering asymptotic efficiency in SIMs, our recent work \cite{He2025range} examines some prototype realization algorithms within a least-squares framework. It reveals that the SVD-based method corresponds to a TLS solution. Under mild assumptions, this estimator is consistent but not the best linear unbiased estimator (BLUE). Due to the low-rank property of the true Hankel matrix, it is crucial to utilize appropriate weighting matrices to enhance the statistical performance of realization algorithms. As recognized in the literature of SIMs \cite{Van2012subspace}, determining optimal weighting matrices for SVD-based methods remains a challenging task. A more recent contribution in this direction is presented in \cite{Mo2025probabilistic}, which introduces a MLE framework with an instrumental variables interpretation, aiming to minimize the covariance of latent prediction errors. However, their analysis focuses on vector autoregressive models rather than state-space models. Notably, the problem of designing an optimal weighting matrix, in the asymptotic MLE sense \cite{Wahlberg1989model}, can be solved in the null space. In \cite{He2025range}, we introduce an optimal realization algorithm for matrix $A$ of SISO systems, which bypasses the SVD step by directly estimating the null space of the Hankel matrix through a two-step least-squares procedure. This algorithm serves as a prototype for the method developed in the this work.

\subsection{Contributions} \label{Sct1.2}

This work has its origin in \cite{Galrinho2018parametric}, where the WNSF method for SISO BJ models was proposed. The proposed method, hereafter referred to as WNSF\textsubscript{SS} (with "SS" denoting state-space models), uses two features of the aforementioned methods. The first feature is starting with an estimate of a HOARX model which contains Markov parameters. This HOARX model captures the behavior of the true system with sufficient accuracy and serves as an approximate sufficient statistic, at least as the sample size grows. Subsequently, model reduction is performed via a multi-step least-squares procedure to obtain a state-space model. The WNSF\textsubscript{SS} method offers favorable computational properties compared to methods like PEM. Moreover, we conduct a rigorous statistical analysis of WNSF\textsubscript{SS} for single-output systems, focusing on the consistency and asymptotic efficiency. Another interesting feature of WNSF\textsubscript{SS} is that it estimates the null space of the Hankel matrix, parameterized by the coefficients of the system's characteristic polynomial, rather than the range space typically estimated by most SIMs using SVD. By working with the null space, WNSF\textsubscript{SS} enables an explicit derivation of the optimal weighting, a key factor in achieving asymptotic efficiency.

In summary, WNSF\textsubscript{SS} is a novel realization-based estimation method for state-space models, combining key statistical and numerical features of PEM and SIMs. Specifically, WNSF\textsubscript{SS} is consistent and asymptotically efficient both for open and closed loop data and we demonstrate in numerical simulations that WNSF\textsubscript{SS} is competitive in comparison with state-of-the-art methods for finite sample sizes.

\subsection{Structure} \label{Sct1.3}

The disposition of this paper is as follows: We present preliminaries, including models and assumptions in  Section~\ref{Sct2}. In Section~\ref{Sct3}, we introduce the WNSF\textsubscript{SS} method with SISO systems. In Section~\ref{Sct4}, we generalize WNSF\textsubscript{SS} to MIMO systems. In Section~\ref{Sct5}, we provide asymptotic properties of the methods. In Section~\ref{Sct6}, we compare the performance of WNSF\textsubscript{SS} on numerical examples and the benchmark data sets DaiSy~\cite{De1997daisy}. In Section~\ref{Sct7}, we discuss the relations between WNSF\textsubscript{SS} and PEM, SIMs and existing variants of WNSF methods. Finally, the paper is concluded in Section~\ref{Sct8}. All proofs and technical lemmas are provided in the Appendix.

\subsection{Notations} \label{Sct1.4}

(1) For a matrix $X$ with appropriate dimensions, $X^\top$, $X^{*}$, $X^{-1}$, $X^\dagger$, $\lVert X \rVert$, $\lVert X \rVert_F$, $\rho(X)$,  ${\text{rank}}(X)$, ${\text{trace}}(X)$,  ${\text{Null}}(X)$ and $\text{dim}\left(\text{Null}(X)\right)$ denote its transpose, complex conjugate transpose, inverse, Moore$\mbox{-}$Penrose pseudo-inverse, spectral norm, Frobenius norm, spectral radius, rank, trace, null space and dimension of the null space, respectively. The notation $X_1\otimes X_2$ is the Kronecker product of matrices $X_1$ and $X_2$, and $\text{diag}\left\{X_1,X_2\right\}$ is a diagonal matrix having $X_1$ and $X_2$ on its diagonal. The notation $\text{Vec}(X)$ denotes the vectorization of $X$ by row. Moreover, $I_k \in \mathbb{R}^{k\times k}$ and $0$ are the identity and zero matrices of appropriate dimensions.

(2) ${\mathbb{E}\left\{x_k\right\}}$ is the expectation of a random vector $x_k$, and $\bar {\mathbb{E}}\left\{x\right\}$ is defined by $\bar {\mathbb{E}}\left\{x\right\} := \mathop {\lim }\limits_{N \to \infty } \frac{1}{N}\sum\limits_{t = 1}^N {\mathbb{E}\left\{x_k\right\}}$. The notation $x \sim \mathcal{N}(\mu,\Sigma)$ means that a random vector $x$ is normally distributed with mean $\mu$ and covariance $\Sigma$, and $x_N \sim \AsN{\mu}{\Sigma}$ means that $x_N$ converges in distribution to $\mathcal{N}(\mu,\Sigma)$ as $N \to \infty$ $\text{w.p.1}$, where $N \to \infty$ $\text{w.p.1}$ means $N$ tends to infinity with probability one. The notation $x_N \simeq y_N$ means that $x_N$ asymptotically equal to $y_N$. Moreover, $x_N = \mathcal{O}(f_N)$ means that $\exists M$ such that $\limsup\limits_{N \to \infty }\frac{x_N}{f_N} \leq M$.

(3) $q^{-1}$ is the backward time-shift operator, and $\mathcal{V}_n(q)$ is defined by $\mathcal{V}_n(q) := \begin{bmatrix}q^{-1}&q^{-2}&\cdots&q^{-n}\end{bmatrix}^\top$. $\mathcal{T}_{n,m}(G(q))$ is the Toeplitz matrix of size $n \times m (m\leq n)$ with the first column $\begin{bmatrix}g_0&g_1&\cdots&g_{n-1}
\end{bmatrix}^\top$ and the first row $\begin{bmatrix}g_0&0&\cdots&0\end{bmatrix}$, and $\innerproduct{G(q)}{H(q)} := \frac{1}{2\pi}\int_{-\pi}^{\pi} G(e^{iw})H^{*}(e^{-iw})\,dw$, where $G(q) = \sum_{k=0}^{\infty}g_kq^k$ and $H(q) = \sum_{k=0}^{\infty}h_kq^k$ are transfer functions of appropriate sizes. Moreover, $\norm{G(q)}_{\mathcal{H}_\infty}:=\sup_w\norm{G(e^{iw})}$, and $\norm{G(q)}_{\mathcal{H}_2}:=\sqrt{\frac{1}{2\pi}\int_{-\infty}^{\infty}\norm{G(e^{iw})}_F^2dw}$.

(4) For $\theta$, a quantity of interest, $\hat \theta$ denotes an estimate of $\theta$, and $\tilde{\theta}$ denotes the estimation error, i.e., $\tilde{\theta} = \hat \theta -  \theta$.

(5) The notations $c_1$, $c_2$, $\cdots$ stand for universal constants. 

\section{Preliminaries} \label{Sct2}

Consider the following discrete-time LTI system on the innovations form:
\begin{subequations} \label{E1}
	\begin{align}
		x_{k + 1} &= Ax_{k}  + Bu_{k} + Ke_{k}, \label{E1a}\\
		y_{k} &= Cx_{k} + e_{k}, \label{E1b}		
	\end{align}
\end{subequations}
where $x_{t}\in \mathbb{R}^{n_x}$, $u_{t}\in \mathbb{R}^{n_u}$, $y_{t}\in \mathbb{R}^{n_y}$ and $e_{t}\in \mathbb{R}^{n_y}$ are the system state, input, output and innovation, respectively. By replacing $e_{k}$ in \eqref{E1a} with $y_{k} - Cx_{k}$, the system \eqref{E1} can be expressed in its predictor form:
\begin{subequations} \label{E2}
	\begin{align}
		x_{k + 1} &= A_Kx_{k}  + B_Kz_{k}, \label{E2a}\\
		y_{k} &= Cx_{k} + e_{k}, \label{E2b}		
	\end{align}
\end{subequations}
where $A_K = A-KC$, $B_K = \begin{bmatrix}B&K\end{bmatrix}$ and $z_{k} = \begin{bmatrix}u_{k}^\top&y_{k}^\top\end{bmatrix}^\top$. As pointed out in \cite{Qin2006overview}, the innovations form and the predictor form are equivalent, and both can represent the input and output data $\left\{u_k, y_k\right\}$ exactly. Same as SSARX \cite{Jansson2003subspace}, for the convenience of the closed-loop identification and ARX modeling, we use the predictor form \eqref{E2} to illustrate our method.

The main focus of this work is to estimate system matrices $A$, $C$, $B$ and $K$, using input and output data $\left\{u_k, y_k\right\}_{k=1}^{\bar N}$ from a single trajectory, where $\bar N$ is the total number of samples. We have the following assumption about the true system.

\begin{assumption} [System] \label{Asp1}
	The system \eqref{E1} is stable and minimal, i.e., the spectral radius of $A$ and $A_K$ satisfy $\rho(A) \leq 1$ and $\rho(A_K) < 1$, and $(A,\begin{bmatrix}B&K\end{bmatrix})$ is controllable and $(A,C)$ is observable. Moreover, the system order ${n_x}$ is known to the user.
\end{assumption}

We allow for the closed-loop data where the input $\left\{u_k\right\}$ has a stochastic part. Defining $\mathcal{F}_{k-1}$ to be the $\sigma-$algebra generated by $\left\{e_j,u_j,j\leq k-1\right\}$, we then have the following assumptions about the noise and input.

\begin{assumption} [Noise] \label{Asp2}
	The innovations $\{e_k\}$ is a stochastic process that satisfies
	\begin{equation*}
		\mathbb{E}(e_k|\mathcal{F}_{k-1}) = 0,\  \mathbb{E}(e_k^2|\mathcal{F}_{k-1}) = \sigma_e^2I\footnote{While our method can be extended to heteroskedastic innovations, we confine our analysis to the homoskedastic case to streamline the proof of asymptotic efficiency.}, \ \mathbb{E}(|e_k|^{10})\leq c.
	\end{equation*}
\end{assumption}

%
%
%


\begin{assumption} [Input] \label{Asp3}
	The input $\{u_k\}$ is defined by $u_k = -F_y(q)y_k + r_k$ under the following conditions \footnote{If $F_y(q)=0$, it means that data comes from an open-loop operation.}:
	
	(1) The sequence $\{r_k\}$ is independent of $\{e_k\}$, $f_N$-quasi-stationary with $f_N = \sqrt{\frac{{\text{log}} N}{N}}$, and uniformly bounded\footnote{For definitions of $f_N$-Quasi-Stationarity and $f_N$-Stability, see \cite{Ljung1992asymptotic}.}.
	
	(2) With $\Psi_r(z)=\psi_r(z)\psi_r(z^{-1})$ the spectral factorization of $\{r_k\}$ and $\psi_r(z)$ causal, $\psi_r(q)$ is bounded-input-bounded-output (BIBO) stable.
	
	(3) The closed-loop system is $f_N$-stable with $f_N = 1/\sqrt{N}$.
	
	(4) The transfer function $F_y(z)$ is bounded on the unite circle.
	
	(5) The spectral density of $\{\begin{bmatrix}r_k&e_k\end{bmatrix}^\top\}$ is coercive, i.e., bounded from below by the matrix $\delta I$ for some $\delta > 0$.
\end{assumption}

\section{Weighted Null-Space Fitting} \label{Sct3}

We now introduce the WNSF\textsubscript{SS} method. For simplicity, in this section we use SISO systems to illustrate major steps of our method. An extension to MIMO systems is later given in Section \ref{Sct4}. To begin with, we introduce the following observer canonical form \cite{Kailath1980linear} for a SISO system \eqref{E2}:
\begin{subequations} \label{E13}
	\begin{align}
		A_K &= \begin{bmatrix}
			-{a}_{1}&1&0&\cdots&0\\
			-{a}_{2}&0&1&\cdots&0\\
			\vdots&\vdots&\vdots&\ddots&\vdots\\
			-{a}_{n_x}&0&0&\cdots&0
		\end{bmatrix}, \\
		C &= \begin{bmatrix}
			1&0&0&\cdots&0
		\end{bmatrix}, \\
		B &= \begin{bmatrix}
			b_{1}&b_{2}&b_{3}&\cdots&b_{n_x}
		\end{bmatrix}^\top, \\
	    K &= \begin{bmatrix}
			k_{1}&k_{2}&k_{3}&\cdots&k_{n_x}
		\end{bmatrix}^\top,
	\end{align}
\end{subequations}
where $a_1,\dots,{a}_{n_x}$ are coefficients of the characteristic polynomial of matrix $A_K$. Our focus is to estimate unknown parameters $a_1,\dots,{a}_{n_x}$, $b_1,\dots,b_{n_x}$, and $k_1,\dots,{k}_{n_x}$ in a statistically optimal way. 
The WNSF\textsubscript{SS} algorithm achieves this through a multi-step least-squares procedure. First, a HOARX is identified via OLS, where the model order is allowed to grow with the number of samples. In the subsequent steps, the non-parametric HOARX estimate and its covariance are exploited to identify the state-space model in \eqref{E13}, where matrix $A_K$ is first obtained using a two-step least-squares procedure, after which matrices $B$ and $K$ are estimated in an analogous manner.

\begin{remark}
	Unlike most SIMs which build a black-box state-space model, WNSF builds a model on canonical form \eqref{E13}, where matrices $A_K$ and $C$ have certain structures. Since each SISO state-space model satisfying Assumption \ref{Asp1} has its unique observer canonical form \eqref{E13}, our result does not lose generality. It should, however, be noted that estimating polynomial coefficients is numerically difficult for high order systems \cite{Viberg1997analysis}.
\end{remark}

\subsection{Multi-Step Least-Squares} \label{Sct3.1}
 We now detail each step of WNSF\textsubscript{SS}. 
 
\textbf{Step 1 (HOARX Modeling):} Based on the predictor form \eqref{E2}, the output is given by
\begin{equation} \label{E15}
	{y_k} = C{\left({qI - A_K} \right)^{-1}}B_K{z_k} + {e_k} 
	= \sum\limits_{i = 1}^\infty {g_i}{z_{k - i}}  + {e_k},
\end{equation}
where predictor Markov parameters $g_i = C{A_K^{i - 1}}B_K$. After selecting a sufficiently large order $n$, the model \eqref{E15} is truncated to a HOARX model
\begin{equation} \label{E16}
	{y_k} \approx  \sum\limits_{i = 1}^n  {g_i{z_{k - i}}}  + {e_k} =  \bm{g}_n \bm{z}_n(k) + {e_k},
\end{equation}
where $\bm{g}_n = \begin{bmatrix}g_1 & \cdots &g_n\end{bmatrix}$, $\bm{z}_n(k) = \begin{bmatrix}z_{k-1}^\top & \cdots & z_{k-n}^\top\end{bmatrix}^{\top}$. Based on \eqref{E16}, an estimate of the first $n$ Markov parameters is
\begin{equation} \label{E17}
	\hat{\bm{g}}_n = r_n R_n^{-1},
\end{equation}
where
\begin{subequations} \label{E18}
	\begin{align}		
		r_n &:= \frac{1}{N}\sum\limits_{t = 1}^N {{y_k\bm{z}_n^\top(k)}}, \label{E16a}\\
		R_n &:= \frac{1}{N}\sum\limits_{k = 1}^N {{\bm{z}_n(k)}\bm{z}_n^\top(k)}, \label{E16b}
	\end{align}
\end{subequations}
where $ N =\bar N-n+1$. According to \cite{Ljung1992asymptotic}, we have
\begin{subequations} \label{E19}
	\begin{align}		
		r_n \to \bar r_n &:= \bar {\mathbb{E}}\left[ {y_k\bm{z}_n^\top(k)}\right], {\rm{as}} \ N \to \infty \ {\rm{w.p.1}}, \label{E17a} \\
		R_n \to \bar R_n &:= \bar {\mathbb{E}}\left[ {\bm{z}_n(k)}\bm{z}_n^\top(k) \right], {\rm{as}} \ N \to \infty \ {\rm{w.p.1}}, \label{E17b}
	\end{align}
\end{subequations}
which further imply
\begin{equation} \label{E20}
	\hat{\bm{g}}_n \to \bar{\bm{g}}_n :=  \bar r_n \bar R_n^{-1}, {\rm{as}} \ N \to \infty \ {\rm{w.p.1}}.
\end{equation}
When the order of the HOARX model is sufficiently large, the truncation bias of \eqref{E16} is negligible. Then, for the estimation error $\tilde{{\bm{g}}}_n := \hat{\bm{g}}_n - {\bm{g}}_n$, it can be shown that $\norm{\tilde{{\bm{g}}}_n} \to 0$,  ${\rm{as}} \ N \to \infty \ {\rm{w.p.1}}$. Moreover, the asymptotic distribution of $\tilde{{\bm{g}}}_n$ can be approximated as
\begin{equation} \label{E21}
	\sqrt N \tilde{{\bm{g}}}_n \sim \AsN{0}{\sigma _e^2{{\bar R_n}^{-1}}}.
\end{equation}

\textbf{Step 2 (OLS for $A_K$):} With the HOARX model in Step 1, we proceed to show how to get a parametric state-space model \eqref{E13}. Unlike most SIMs that concentrate on the range space of the extended observability matrix $\mathcal{O}_f$, we shift our focus to its null space, which is essentially parameterized by coefficients of the characteristic polynomial of matrix $A_K$. According to the Cayley-Hamilton theorem \cite[Th. 2.4.3.2]{Horn2012matrix}, we have
\begin{equation} \label{E22}
	A_K^{n_x} + a_1 A_K^{{n_x}-1} + \cdots + a_{{n_x}-1} A_K + a_{{n_x}}I = 0.
\end{equation}
Moreover, the extended observability matrix is given by 
\begin{equation} \label{E23}
	{\mathcal{O}}_{n_x} = \begin{bmatrix}
		{{C^{\top}}}&{{{\left({CA_K} \right)}^{\top}}}& \cdots &{{{\left({C{A_K^{{n_x}}}}\right)}^{\top}}}
	\end{bmatrix}^{\top} \in \mathbb{R}^{(n_x+1)\times n_x}.
\end{equation}
Under Assumption \ref{Asp1}, we have $\text{rank}\left(\mathcal{O}_{n_x}\right) = n_x$, and thus, $\text{dim}\left(\text{Null}(\mathcal{O}_{n_x}^\top)\right) = 1$. Using equation \eqref{E22}, we have 
\begin{equation} \label{E24}
	\begin{bmatrix}
		a_{n_x}&a_{{n_x}-1}& \cdots &a_1&1\end{bmatrix} \mathcal{O}_{n_x} = 0,
\end{equation}
i.e., the null space of $\mathcal{O}_{n_x}$ is completely parameterized by the coefficients $\left\{a_i\right\}_{i=1}^{n_x}$. For simplicity of illustration, we define
\begin{equation} \label{E25}
	\bm{a} := \begin{bmatrix}
		a_{n_x}&a_{{n_x}-1}& \cdots &a_1\end{bmatrix}.
\end{equation}
Similar to SIMs, we construct a Hankel matrix using the first $n$ Markov parameters:
\begin{equation} \label{E26}
	\mathcal{H}_{{n_x}n} =  
	\left[\begin{array}{cccc}
		{{g_1}} & {{g_2}} & \cdots & {{g_p}}\\[6pt]
		{{g_2}} & {{g_3}} & \cdots & {{g_{p + 1}}}\\[2pt]
		\vdots & \vdots & \ddots & \vdots \\[2pt]
		\hdashline[2pt/2pt]
		{g_{n_x + 1}} & {g_{n_x + 2}} & \cdots & {{g_{n}}}
	\end{array}\right] 
	:= 
	\left[\begin{array}{c}
		\mathcal{H}_{n_xn}^{+}\\[2pt] 
		\hdashline[2pt/2pt]
		\mathcal{H}_{n_xn}^{-}
	\end{array}\right],
\end{equation}
where the column number $p = n - {n_x}$. It is well known that the above Hankel matrix is the product of the extended observability matrix ${\mathcal{O}}_{n_x}$ and controllability matrix $\mathcal{C}_p$, i.e.,
\begin{equation} \label{E27}
	\mathcal{H}_{{n_x}n} =  {\mathcal{O}}_{n_x}\mathcal{C}_p,
\end{equation}
where $\mathcal{C}_p = \begin{bmatrix}B_K&{A_KB_K}& \cdots &{{A_K^{p}}B_K}\end{bmatrix}$. A key observation is that the left null space of the extended observability matrix $\mathcal{O}_{n_x}$ is also the left null space of the Hankel matrix $\mathcal{H}_{{n_x}n}$, i.e., $\begin{bmatrix}\bm{a}&1\end{bmatrix}\mathcal{H}_{{n_x}n} = 0$, which implies
\begin{equation} \label{E28}
	\bm{a} \mathcal{H}_{n_xn}^{+} + \mathcal{H}_{n_xn}^{-} = 0.
\end{equation}
After replacing true Markov parameters in ${\mathcal{H}}_{n_xn}$ with their estimates given in Step~1, we obtain an OLS estimate of $\bm{a}$
\begin{equation} \label{E29}
	\hat{\bm{a}}_{\text{ols}} = -{\hat {\mathcal{H}}}_{n_xn}^{-}({\hat {\mathcal{H}}}_{n_xn}^{+})^{\top}\left({\hat {\mathcal{H}}}_{n_xn}^{+}({\hat {\mathcal{H}}}_{n_xn}^{+})^{\top}\right)^{-1}.
\end{equation}

\textbf{Step 3 (WLS for $A_K$):} Now we refine the initial estimate $\hat{\bm{a}}_{\text{ols}}$ in Step 2 by using the asymptotic distribution of $\tilde{{\bm{g}}}_n$ in \eqref{E21}. The residual of $\bm{a}{\hat {\mathcal{H}}}_{n_xn}^{+} + \hat {\mathcal{H}}_{n_xn}^{-}$ is
\begin{equation} \label{E30}
	\bm{a}{\hat {\mathcal{H}}}_{n_xn}^{+} + \hat {\mathcal{H}}_{n_xn}^{-} - \left(\bm{a}{{\mathcal{H}}}_{n_xn}^{+} + {\mathcal{H}}_{n_xn}^{-}\right) = \begin{bmatrix}\bm{a}&1\end{bmatrix}\tilde {\mathcal{H}}_{n_xn},
\end{equation}
where $\tilde {\mathcal{H}}_{n_xn} := \hat {\mathcal{H}}_{n_xn} - {\mathcal{H}}_{n_xn}$. 
Since $\tilde {\mathcal{H}}_{n_xn}$ is a Hankel matrix, we rewrite \eqref{E30} as
\begin{equation} \label{E31}
	\begin{bmatrix}\bm{a}&1\end{bmatrix} \tilde {\mathcal{H}}_{n_xn} =  \tilde{\bm{g}}_n{\mathcal{K}_n}(\bm{a}),
\end{equation}
where ${\mathcal{K}_n}(\bm{a}) = {{\mathcal{T}_{n,p}}(\bm{a})} \otimes I$, and ${{\mathcal{T}_{n,p}}(\bm{a})}$ is a Toeplitz matrix with compatible dimension, having $\begin{bmatrix}\bm{a}&1&0&\cdots &0\end{bmatrix}^\top$ on its first column and $\begin{bmatrix}a_{n_x}&0&\cdots&0\end{bmatrix}$ on its first row. According to \eqref{E21}, we conclude that the distribution of the residual \eqref{E31} is
\begin{equation} \label{E32}
	\sqrt N \tilde{\bm{g}}_n{\mathcal{K}_n}(\bm{a})\sim 
	\AsN{0}{\sigma _e^2 \bar\Lambda_n(\bm{a})},
\end{equation}
where $\bar\Lambda_n(\bm{a})  = {\mathcal{K}_n^\top}(\bm{a}){{\bar R_n}^{-1}}{\mathcal{K}_n}(\bm{a})$. Taking ${\bar\Lambda_n^{-1}({\bm{a}})}$ as the optimal weighting, where in practice $\bm{a}$ and ${\bar R_n}$ are replaced with their consistent estimates $\hat{\bm{a}}_{\text{ols}}$ and ${R_n}$ from Steps 2 and 1, giving ${\hat\Lambda_n^{-1}(\hat{\bm{a}}_{\text{ols}})}$, we refine the estimate of $\bm{a$} with WLS
\begin{equation} \label{E34}
		\begin{split}
		\hat{\bm{a}}_{\text{wls}} =& -{\hat {\mathcal{H}}}_{n_xn}^{-}{\hat\Lambda_n^{-1}(\hat{\bm{a}}_{\text{ols}})}({\hat {\mathcal{H}}_{n_xn}^{+}})^{\top}\\
		&\times \left({\hat {\mathcal{H}}_{n_xn}^{+}} {\hat\Lambda_n^{-1}(\hat{\bm{a}}_{\text{ols}})}({\hat{\mathcal{H}}_{n_xn}^{+}})^{\top}\right)^{-1}.
	\end{split}	
\end{equation}
As demonstrated in \cite{Galrinho2018parametric}, replacing $\bm{a}$ with its consistent estimate $\hat{\bm{a}}_{\text{ols}}$ will not affect the asymptotic optimality of $\hat{\bm{a}}_{\text{wls}}$. However, it is possible to continue iterating, which may improve the estimate for finite samples. 

With the optimal estimate of coefficients $\left\{{a_i}\right\}_{i=1}^{n_x}$ in hand, we return to the observer canonical form \eqref{E13}. This yields an estimate of $A_K$ (with $C$ already known). We then apply a similar procedure to estimate $B$ and $K$.

\textbf{Step 4 (OLS for $B_K$):} In most literature of SIMs, with available estimates of $A_K$ and $C$, the following one-step ahead predictor is constructed:
\begin{equation} \label{E35}
	\hat y_k(B,K) = C (qI-\hat A_K)^{-1}(Bu_k+Ky_k),
\end{equation}
which is linear in $B$ and $K$. Then, estimates of $B$ and $K$ are given by OLS. This method is claimed to be optimal, but its statistical property is unclear yet. We now provide a new method which uses two-step least-squares to estimate matrices $B$ and $K$. First, we notice that
\begin{equation}
	\mathcal{O}_{n-1} B_K = \begin{bmatrix}g_0^\top&g_1^\top&\cdots &g_{n-1}^\top\end{bmatrix}^\top,
\end{equation}
where $\mathcal{O}_{n-1}$ is the extended observability matrix. After vectorization by row, we have that
\begin{equation}
	\Vc{B_K}\left(\mathcal{O}_{n-1}^\top\otimes I_2\right) = \bm{g}_n,
\end{equation}
which is further denoted by
\begin{equation} \label{estimation_B_K_prototype}
	\bm{\eta}\Phi_n = \bm{g}_n,
\end{equation}
where
\begin{equation*}
	\begin{split}
		\Phi_n &= \mathcal{O}_{n-1}^\top\otimes I_2 \in \mathbb{R}^{2n_x\times2n}, \\
		\bm{\eta}&= \Vc{B_K} = \begin{bmatrix}b_1&k_1&b_2&k_2&\cdots&b_{n_x}&k_{n_x}\end{bmatrix}.
	\end{split}
\end{equation*} 
With the estimate of $A_K$ in Step~3, an estimate of the extended observability matrix ${\mathcal{O}}_{n-1}$ is given by
\begin{equation} \label{Extended-Obs-wls}
	\hat{\mathcal{O}}_{n-1} = \begin{bmatrix}
		C^\top&(C\hat A_K)^\top& \cdots&(C{\hat A_K^{n-1}})^\top\end{bmatrix}^\top.
\end{equation}
After replacing $\mathcal{O}_{n-1}$ and $\bm{g}_n$  in \eqref{estimation_B_K_prototype} with their estimates in \eqref{Extended-Obs-wls} and \eqref{E17}, an OLS estimate of $\bm{\eta}$ is given by
\begin{equation} \label{estimation_B_K_ols}
	\hat{\bm{\eta}}_{\text{ols}} = \hat{\bm{g}}_n{\hat\Phi}_n^\top\left({\hat\Phi}_n{\hat\Phi}_n^\top\right)^{-1}.
\end{equation}

\textbf{Step 5 (WLS for $B_K$):} As in Step~3, we now refine the estimate of $\bm{\eta}$ with WLS. Since $\bm{\eta}{\Phi}_n$ can also be expressed as $ \bm{\eta}{\Phi}_n = \Vc{\mathcal{O}_{n-1}}\left(I_n\otimes B_K\right)$, the residual of $\hat{\bm{g}}_n - \bm{\eta}{\hat\Phi}_n$ can be rewritten as 
\begin{equation} \label{residual_B_K_wls}
	\hat{\bm{g}}_n - \bm{\eta}{\hat\Phi}_n - \left({\bm{g}}_n - \bm{\eta}{\Phi}_n\right) =\tilde{\bm{g}}_n - \Vc{\tilde{\mathcal{O}}_{n-1}}\left(I_n\otimes B_K\right),
\end{equation}
where $\tilde{\mathcal{O}}_{n-1} = \hat{\mathcal{O}}_{n-1} - {\mathcal{O}}_{n-1}$. We now show that the error $\Vc{\tilde{\mathcal{O}}_{n-1}}$ scales linearly with the error $\tilde{\bm{g}}_n$. We first study each error term in $\hat{\mathcal{O}}_{n-1}$, which is
\begin{equation} \label{error-obs-each-term}
	\begin{split}
		C(\hat A_K^{k} - A_K^{k}) &= C\left(\hat A_K - A_K + A_K\right)^{k} - CA_K^{k} \\
		&\simeq C \sum_{i=0}^{k-1}A_K^{i}(\hat A_K - A_K)A_K^{k-i-1} \\
		& = \text{Vec}(\tilde A_K)
		\left(\sum_{i=0}^{k-1}(CA_K^{i})^\top\otimes A_K^{k-i-1}\right) \\
		& = -\tilde{\bm{a}}_{\text{wls}} \bar P \bar I \left(\sum_{i=0}^{k-1}(CA_K^{i})^\top\otimes A_K^{k-i-1}\right) \\
		& = \tilde{\bm{a}}_{\text{wls}}S_k(\bm{a}),
	\end{split}
\end{equation}
where for $k = 1,2,\dots,n-1$,
\begin{equation*}
	\begin{split}
		\tilde A_K &= \hat A_K - A_K, \\
		\bar P &= \begin{bmatrix}
			0&0& \cdots &1\\
			0&0& \cdots &0\\
			\vdots & \vdots & \ddots & \vdots \\
			1&0& \cdots &0
		\end{bmatrix} \in \mathbb{R}^{n_x\times n_x}, \\
	    \bar I &= I_{n_x}\otimes \begin{bmatrix}1&0& \cdots &0\end{bmatrix} \in \mathbb{R}^{n_x\times n_x^2}, \\
	    S_k(\bm{a}) &= -\bar P \bar I \left(\sum_{i=0}^{k-1}(CA_K^{i})^\top\otimes A_K^{k-i-1}\right) \in \mathbb{R}^{n_x^2\times n_x}.
	\end{split}    
\end{equation*}
In \eqref{error-obs-each-term}, the asymptotic equivalence holds because the higher-order terms involving higher powers of $\tilde A_K$ decay mush faster than $\tilde A_K$, and can therefore be considered negligible. The result in \eqref{error-obs-each-term} shows that the error $C(\hat A_K^{k} - A_K^{k})$ scales linearly with the error $\tilde{\bm{a}}_{\text{wls}}$. After vectorizing ${\mathcal{O}}_{n-1}$ by row, we further conclude that the total error $\Vc{\tilde{\mathcal{O}}_{n-1}}$ also scales linearly with $\tilde{\bm{a}}_{\text{wls}}$, i.e., 
\begin{equation} \label{error_obs_wls}
	\Vc{\tilde{\mathcal{O}}_{n-1}} \simeq \tilde{\bm{a}}_{\text{wls}}\mathcal{S}_{n}(\bm{a}),
\end{equation}
where $\mathcal{S}_{n}(\bm{a}) = \begin{bmatrix} 0&S_1(\bm{a})&\cdots&S_{n-1}(\bm{a}) \end{bmatrix}$. Furthermore, for the estimation error  $\tilde{\bm{a}}_{\text{wls}}$ in Step~3, we have that
\begin{equation} \label{error_a_wls}
	\tilde{\bm{a}}_{\text{wls}}= -\tilde{{\bm{g}}}_n{\mathcal{K}_n}(\bm{a}){\hat\Lambda_n^{-1}(\hat{\bm{a}}_{\text{ols}})}({\hat {\mathcal{H}}_{n_xn}^{+}})^{\top} {{\hat{M}}^{-1}(\hat{\bm{g}}_n,\hat{\bm{a}}_{\text{ols}})},
\end{equation}
where ${\hat{M}}^{-1}(\hat{\bm{g}}_n,\hat{\bm{a}}_{\text{ols}}) := {\hat {\mathcal{H}}_{n_xn}^{+}} {\hat\Lambda_n^{-1}(\hat{\bm{a}}_{\text{ols}})}({\hat{\mathcal{H}}_{n_xn}^{+}})^{\top}$. Substituting \eqref{error_a_wls} into \eqref{error_obs_wls}, we conclude that the error $\Vc{\tilde{\mathcal{O}}_{n-1}}$ scales linearly with the error $\tilde{\bm{g}}_n$. As a result, the residual \eqref{residual_B_K_wls} can be rewritten as
\begin{equation} \label{residual_B_K_wls_final}
	\tilde{\bm{g}}_n - \Vc{\tilde{\mathcal{O}}_{n-1}}\left(I_n\otimes B_K\right) \simeq \tilde{\bm{g}}_n {\mathcal{K}_n}(\bm{a},\bm{\eta}),
\end{equation}
where 
\begin{equation*}
	\begin{split}
		{\mathcal{K}_n}(\bm{a},\bm{\eta}) = &I_n + {\mathcal{K}_n}(\bm{a}){\hat\Lambda_n^{-1}(\hat{\bm{a}}_{\text{ols}})}({\hat {\mathcal{H}}_{n_xn}^{+}})^{\top}{{\hat{M}}^{-1}(\hat{\bm{g}}_n,\hat{\bm{a}}_{\text{ols}})} \times \\ &\mathcal{S}_{n}(\bm{a})\left(I_n\otimes B_K\right).
	\end{split}	
\end{equation*}
According to \eqref{E21}, we conclude that the distribution of the residual \eqref{residual_B_K_wls_final} is
\begin{equation} \label{residual_B_K_wls_dist}
	\sqrt N \tilde{\bm{g}}_n{\mathcal{K}_n}(\bm{a},\bm{\eta}) \sim \AsN{0}{\sigma _e^2 \bar\Lambda_n(\bm{a},\bm{\eta})},
\end{equation}
where 
\begin{equation} \label{residual_B_K_wls_weigthing}
	\bar\Lambda_n(\bm{a},\bm{\eta})  = {\mathcal{K}_n^\top}(\bm{a},\bm{\eta}){{\bar R_n}^{-1}}{\mathcal{K}_n}(\bm{a},\bm{\eta}).
\end{equation}
Taking ${\bar\Lambda_n^{-1}({\bm{a}},\bm{\eta})}$ as the optimal weighting, where in practice $\bm{a}$, $\bm{\eta}$ and ${\bar R_n}$ are replaced with their consistent estimates $\hat{\bm{a}}_{\text{wls}}$, $\hat{\bm{\eta}}_{\text{ols}}$ and ${R_n}$ from Steps 3, 4 and 1, giving ${\hat\Lambda_n^{-1}(\hat{\bm{a}}_{\text{wls}},\hat{\bm{\eta}}_{\text{ols}})}$, we refine the estimate of $\bm{\eta}$ with WLS
\begin{equation} \label{estimation_B_K_wls}
	\hat{\bm{\eta}}_{\text{wls}} = \hat{\bm{g}}_n \hat\Lambda_n^{-1}(\hat{\bm{a}}_{\text{wls}},\hat{\bm{\eta}}_{\text{ols}}) {\hat\Phi}_n^\top\left({\hat\Phi}_n\hat\Lambda_n^{-1}(\hat{\bm{a}}_{\text{wls}},\hat{\bm{\eta}}_{\text{ols}}){\hat\Phi}_n^\top\right)^{-1}.
\end{equation}
In this way, optimal estimates of matrices $B$ and $K$ are obtained. Together with the optimal estimate for matrix $A_K$ in Step~3, an optimal estimate for matrix $A$ is given by $\hat A = \hat A_K + C\hat K$.

WNSF\textsubscript{SS} is summarized in Algorithm~1 below.
\begin{algorithm} \label{Alg1}
	\caption{WNSF\textsubscript{SS}: State-Space System Identification Using Weighted Null Space Fitting.}
	\begin{algorithmic}[1]
		\Procedure{Multi-Step Least-Squares}{}\\
		\textbf{inputs:} Dimension of state $n_x$, order of HOARX $n$, input and output data $\left\{u_k,y_k\right\}_{k=1}^{\bar N}$.\\
		\textbf{outputs:} System matrices $\hat A$, $\hat C$, $\hat B$ and $\hat K$.
		\State Step 1 (OLS for HOARX): Initial estimate of Markov parameters  $\hat{\bm{g}}_n$ from an HOARX model using OLS~\eqref{E17}.
		\State Step 2 (OLS for $A_K$): Construct the Hankel matrix $\hat {\mathcal{H}}_{{n_x}n}$, and estimate the coefficients ${\bm{a}}$ using OLS~\eqref{E29}.
		\State Step 3 (WLS for $A_K$): Construct the weighting matrix in \eqref{E32}, and re-estimate ${\bm{a}}$ using WLS~\eqref{E34}.
		\State Step 4 (OLS for $B$ and $K$): Construct extended observability matrix $\hat{\mathcal{O}}_{n-1}$ using matrices $\hat A_K$ and $C$, then estimate matrices $B$ and $K$ using OLS~\eqref{estimation_B_K_ols}. 
		\State Step 5 (WLS for $B$ and $K$): Construct the weighting matrix in \eqref{residual_B_K_wls_dist}, and re-estimate $B$ and $K$ using WLS~\eqref{estimation_B_K_wls}. \\
		\textbf{return} $\hat A = \hat A_K + C\hat K$, $C$, $\hat B$ and $\hat K$, where $\hat A_K$ is in Step~3, $C$ is trivial, and $\hat B$ and $\hat K$ are in Step~5.
		\EndProcedure
	\end{algorithmic}
\end{algorithm}

\begin{remark}
	Extension to multi-input-single-output (MISO) systems: The key requirement to apply WNSF is that there is a linear relation between the HOARX parameters and the parameters of system matrices. As shown in \eqref{E28}, such a relation is trivial for SISO systems. A further extension of Algorithm~1 to MISO systems is straightforward. To illustrate, we first introduce the following observer canonical form for MISO systems:
	\begin{subequations} \label{E13-MISO}
		\begin{align}
			A_K &= \begin{bmatrix}
				\times&1&0&\cdots&0\\
				\times&0&1&\cdots&0\\
				\vdots&\vdots&\vdots&\ddots&\vdots\\
				\times&0&0&\cdots&0
			\end{bmatrix}, \\
			C &= \begin{bmatrix}
				1&0&0&\cdots&0
			\end{bmatrix}, \\
			B &= \begin{bmatrix}
				\times&\times&\times&\cdots&\times \\
				\vdots&\vdots&\vdots&\ddots&\vdots \\
				\times&\times&\times&\cdots&\times\\
			\end{bmatrix}^\top \in \mathbb{R}^{n_x \times n_u}, \\
			K &= \begin{bmatrix}
				\times&\times&\times&\cdots&\times
			\end{bmatrix}^\top,
		\end{align}
	\end{subequations}
    where $\times$ denotes free parameters in matrices $A_K$, $B$ and $K$. The same linear relation in equation~\eqref{E28} between the null space of the Hankel matrix and the coefficients in matrix $A_K$ also applies to MISO systems. Therefore, the first three steps of Algorithm~1 can be directly used to estimate the coefficients of $A_K$. After vectorization, similar steps to Steps~4 and 5 can then be used to estimate matrices $B$ and $K$.    
\end{remark}

\section{Extension to MIMO Systems} \label{Sct4}

As we mentioned, to apply WNSF, the key step is to establish a linear relation between the HOARX parameters and the parameters of system matrices. Unlike single-output systems, a linear parameterization of the null space of the Hankel matrix \cite{Viberg1997analysis,Hannan2012statistical} is generally unavailable for multi-output systems. Therefore, adapting WNSF\textsubscript{SS} to multi-output systems introduces significant complexity and requires additional considerations. In this section we discuss how WNSF\textsubscript{SS} can be effectively generalized to accommodate multi-output systems.

\subsection{Canonical Parametrizations} \label{Sct4.1}

In an attempt to generalize WNSF\textsubscript{SS} to multi-output systems, we first introduce a canonical parametrization for MIMO systems. In some literature, this parametrization is also called overlapping parametrization or echon state-space realizations. For more details, we refer to \cite{Ljung1999system,Hannan2012statistical,Gevers1984uniquely}, and Appendix \ref{AppF}.

Let $\bar\nu = \left\{\nu_1,\dots,\nu_{n_y}\right\}$ denote the Kronecker index, a set of $n_y$ positive integers satisfying $\sum_{i=1}^{n_y}\nu_i = n_x$. Then, a canonical parametrization for a multi-output state-space model \eqref{E2} is given by \eqref{MIMO_canonical_form},
\begin{figure*}
\begin{align} \label{MIMO_canonical_form}
		A_K &= \left[
		\begin{array}{cccc:c:cccc}
			\multicolumn{4}{c}{\overbrace{\hspace{2cm}}^{\nu_1}}&\multicolumn{1}{c}{\overbrace{}^{\nu_2,\dots,\nu_{n_y-1}}}&\multicolumn{4}{c}{\overbrace{\hspace{2cm}}^{\nu_{n_y}}} \\
			0 & 1 & \cdots & 0 & \cdots & 0 & 0 & \cdots & 0 \\[-5pt]
			\vdots & \vdots & \ddots & \vdots & \ddots & \vdots & \vdots & \ddots & \vdots \\
			0 & 0 & \cdots & 1 & \cdots & 0 & 0 & \cdots & 0\\
			\times & \times & \times & \times & \times & \times & \times & \times & \times \\\cdashline{1-9}	
			\vdots & \vdots & \vdots & \vdots & \ddots & \vdots & \vdots & \vdots & \vdots \\\cdashline{1-9}
			0 & 0 & \cdots & 0 & \cdots & 0 & 1 & \cdots & 0 \\[-5pt]
			\vdots & \vdots & \ddots & \vdots & \ddots & \vdots & \vdots & \ddots & \vdots\\
			0 & 0 & \cdots & 0 & \cdots & 0 & 0 & \cdots & 1 \\
			\times & \times & \times & \times & \times & \times & \times & \times & \times \\
		\end{array}
		\right],
		\
		\begin{aligned} C &= \left[\begin{array}{ccc:c:ccc}
			\multicolumn{3}{c}{\overbrace{\hspace{2cm}}^{\nu_1}}
			&\multicolumn{1}{c}{\overbrace{}^{\nu_2,\dots,\nu_{n_y-1}}}&\multicolumn{3}{c}{\overbrace{\hspace{2cm}}^{\nu_{n_y}}} \\
			1 & \cdots & 0 & \cdots & 0 & \cdots & 0 \\
			0 & \cdots & 0 & \cdots & 0 & \cdots  & 0\\
			\vdots & \ddots & \vdots & \ddots & \vdots & \ddots& \vdots \\
			0 & \cdots & 0 & \cdots & 1 & \cdots  & 0\\
		\end{array}
		\right], \\ 
		B &= \left[\begin{array}{ccc}
			\times & \cdots & \times\\
			\times & \cdots & \times\\
			\vdots & \ddots & \vdots \\
			\times & \cdots & \times\\
		\end{array} 
		\right], K = \left[\begin{array}{ccc}
			\times & \cdots & \times\\
			\times & \cdots & \times\\
			\vdots & \ddots & \vdots \\
			\times & \cdots & \times\\
		\end{array} 
		\right].
	\end{aligned}
	\end{align}
\end{figure*}	
where $\times$ denotes free parameters. Since matrices $B$ and $K$ have no particular structure, the number of free parameters in the canonical parametrization is $(2n_y+n_u)n_x$. Given $n_x$ and $n_y$, there exists $\binom{n_x-1}{n_y-1}$ Kronecker indices $\bar\nu$. The following lemma suggests that for a particular Kronecker index, the state-space representation \eqref{MIMO_canonical_form} is capable of describing almost all $n_x$ dimensional linear systems.

\begin{lemma} [\cite{Gevers1984uniquely,Ljung1999system}] \label{Lem1}
	The state-space model \eqref{MIMO_canonical_form} with a particular Kronecker index $\bar\nu$ can describe almost all $n_x$-dimensional stochastic LTI systems.	
\end{lemma}

According to the above lemma, any $n_x$-dimensional stochastic LTI state-space system can be expressed by means of a state-space model \eqref{MIMO_canonical_form} with a particular Kronecker index $\bar\nu$. To precisely characterize the condition under which this representation holds, we introduce the following Hankel matrix interpretation \cite{Ljung1999system}. Define the following Hankel matrix in analogous to \eqref{E26}:
\begin{equation} \label{Hankel_MIMO}
	{\mathcal{H}}_{{n_x}n} :=  \begin{bmatrix}
		{{g_1}}&{{g_2}}& \cdots &{{g_p}}\\
		{{g_2}}&{{g_3}}& \cdots &{{g_{p + 1}}}\\
		\vdots & \vdots & \ddots & \vdots \\
		{g_{n_x}}&{g_{n_x+1}}& \cdots &{{g_{n}}}
	\end{bmatrix} \in \mathbb{R}^{n_xn_y\times pn_z},
\end{equation}
and similarly define ${\mathcal{H}}_{{(n_x+1)}n}$. Moreover, for a given Kronecker index $\bar\nu$, denote a set of indexes by $\mathbb{I}_{\bar\nu} = \left\{(k-1)n_y+i;\ 1 \leq k \leq \nu_i; \ 1 \leq i \leq n_y\right\}$. Then, we have the following fundamental result.

\begin{lemma} [\cite{Wertz1982determination,Ljung1999system}] \label{Lem2}
	Suppose that the $n_x$ rows $\mathbb{I}_{\bar\nu}$ of ${\mathcal{H}}_{{n_x}n}$ span all the rows of ${\mathcal{H}}_{{(n_x+1)}n}$. Then, the state-space model \eqref{E2} can be represented by the canoncial form \eqref{MIMO_canonical_form} corresponding to the Kronecker index $\bar\nu$. Under this circumstance, the canoncial form \eqref{MIMO_canonical_form} is called ``admissible".
\end{lemma}

Based on the fact that $\text{rank}\left({\mathcal{H}}_{{(n_x+1)}n}\right) = n_x$, the above lemma suggests that all rows of ${\mathcal{H}}_{{n_x}n}$ span an $n_x$-dimensional linear subspace. The generic situation then is that the same space is spanned by any subsets of $n_x$ rows of ${\mathcal{H}}_{{n_x}n}$. In other words, if we randomly pick a Kronecker index $\bar\nu$ from all possible $\binom{n_x-1}{n_y-1}$ indices, the probability is 1 that the $n_x$ rows $\mathbb{I}_{\bar\nu}$ of ${\mathcal{H}}_{{n_x}n}$ span the same space. However, it should be mentioned that there exist non-generic situations, for example, see \cite[Example~C.1]{Katayama2005subspace}.  The structure selection problem lies beyond the scope of this work. From now on, we assume that the given canoncial form \eqref{MIMO_canonical_form} is admissible, which essentially means that the true model \eqref{E2} can be exactly described by the specified canoncial form. Our major interest is to estimate those free parameters in the canoncial form in a statistically optimal way.

\begin{remark} [Overlapping Parametrizations] \label{canonical-form-rmk}
	Let $M_{\bar{\nu}_i}$ denote the state-space model \eqref{MIMO_canonical_form} corresponding to $\bar{\nu}_i$. Moreover, let the sum of $M_{\bar{\nu}_i}$ over possible Kronecker indices be 
	\begin{equation} \label{union-parameterization}
		\overline{M} = \bigcup_{\bar{\nu}_i} \mathcal{R}\!\bigl(M_{\bar{\nu}_i}\bigr),
	\end{equation}
    where $i = 1,2,\dots,\binom{n_x-1}{n_y-1}$. Then, the union $\overline{M}$ covers all linear $n$-dimensional systems. Since a particular parameterization $M_{\bar{\nu}_i}$ is not guaranteed to be equivalent to a given state-space model, these structures \eqref{MIMO_canonical_form} are problem-dependent for multi-output systems, i.e., there is no universal structure that could be used for all linear systems of the same order. Of course, the ranges of $M_{\bar{\nu}_i}$ may overlap considerably, and the question arises as to which structure leads to the most accurate parameter estimates. It was shown in \cite{Wertz1982determination} that, for two admissible parameterizations, the determinants of their corresponding Fisher information matrices are identical. It follows immediately that, in the Gaussian case and with a MLE scheme, any two parameterizations will asymptotically yield the same value for the determinant of the parameter error covariance matrix. The structure selection problem lies beyond the scope of this work. For related discussions, we refer to \cite{Wertz1982determination,Van1982line,Delchamps1982critical}.
\end{remark}

We now derive a linear relation between the HOARX parameters and system matrices on canonical form for multi-output systems. For single-output systems, the coefficients vector $\bm{a}$ is capable to completely parameterize the left null space of ${\mathcal{H}}_{{n_x}n}$ (see \eqref{E28}). For multi-output systems, by contrast, a completely linear parameterization of the left null space of ${\mathcal{H}}_{{n_x}n}$ is generally unavailable \cite{Viberg1997analysis}. Nevertheless, the low-rank propery of ${\mathcal{H}}_{{n_x}n}$ permits a linear relation between the left null space of a suitable chosen submatrix of ${\mathcal{H}}_{{n_x}n}$ and parameters of matrix $A_K$. Such a submatrix is selected according to the specified Kronecker index $\bar\nu$. To illustrate this, let $h_{i,j}$ denote the $j$-th row in the $i$-th block of rows of ${\mathcal{H}}_{{n_x}n}$, thus, $h_{i,j} \in \mathbb{R}^{1\times pn_z}$ is the $(i-1)n_y+j$-th row of ${\mathcal{H}}_{{n_x}n}$. Then, the Hankel matrix ${\mathcal{H}}_{{n_x}n}$ can be denoted by its rows ${\mathcal{H}}_{{n_x}n} = \left\{h_{1,1}^\top,\cdots,h_{1,n_y}^\top,\cdots,h_{n_x,1}^\top,\cdots,h_{n_x,n_y}^\top\right\}^\top$. According to Lemma \ref{Lem2}, the $n_x$ rows $\mathbb{I}_{\bar\nu}$ of ${\mathcal{H}}_{{n_x}n}$ servers as a basis for its entire row space. To be specific, the $n_x$ selected basis rows of ${\mathcal{H}}_{{n_x}n}$ are
\begin{equation*} \label{Hankel-MIMO-Basis}
	\left\{h_{1,1}^\top,\dots,h_{\nu_1,1}^\top,h_{1,2}^\top,\dots,h_{\nu_2,2}^\top,h_{1,n_y}^\top,\dots,h_{\nu_{n_y},n_y}^\top\right\}^\top.
\end{equation*}
We now define two submatrices of ${\mathcal{H}}_{{n_x}n}$, which are (the rows are not in the same order as ${\mathcal{H}}_{{n_x}n}$)
\begin{equation*} \label{Hankel-Sub-Matrix}
	\begin{split}
		{\mathcal{H}}_{{n_x}n}^{+}(\bar\nu) &=\left[h_{1,1}^\top,\cdots,h_{\nu_1,1}^\top,\cdots,h_{1,n_y}^\top,\cdots,h_{\nu_{n_y},n_y}^\top\right]^\top, \\
		{\mathcal{H}}_{{n_x}n}^{-}(\bar\nu) &=
		\left[h_{2,1}^\top,\cdots,h_{\nu_1+1,1}^\top,\cdots,h_{2,n_y}^\top,\cdots,h_{\nu_{n_y}+1,n_y}^\top\right]^\top.
	\end{split}	
\end{equation*}
It can be observed that certain rows of ${\mathcal{H}}_{{n_x}n}^{-}(\bar\nu)$ are already contained in ${\mathcal{H}}_{{n_x}n}^{+}(\bar\nu)$. Meanwhile, since ${\mathcal{H}}_{{n_x}n}^{+}(\bar\nu)$ consists of basis rows of ${\mathcal{H}}_{{n_x}n}$, the remaining rows of ${\mathcal{H}}_{{n_x}n}^{-}(\bar\nu)$-those not included in ${\mathcal{H}}_{{n_x}n}^{+}(\bar\nu)$-can be expressed in terms of a linear combination of the basis rows. This gives rise to the following equation \cite[Th. 2.5.2]{Hannan2012statistical}, where matrix $A_K$ on canonical form \eqref{MIMO_canonical_form} satisfies:
\begin{equation} \label{Hankel-Sub-Matrix-A_K}
	A_K {\mathcal{H}}_{{n_x}n}^{+}(\bar\nu) = {\mathcal{H}}_{{n_x}n}^{-}(\bar\nu).
\end{equation}
In essence, the entries ``1'' and ``0'' in matrix $A_K$ represent rows that are common to both ${\mathcal{H}}_{{n_x}n}^{-}(\bar\nu)$ and ${\mathcal{H}}_{{n_x}n}^{+}(\bar\nu)$, while free parameters ``$\times$'' denote rows expressed as linear combinations. 

Equation \eqref{Hankel-Sub-Matrix-A_K} establishes a linear relation between the HOARX parameters and the system matrices, which can also be interpreted in terms of null-space fitting. To illustrate this, we rewrite \eqref{Hankel-Sub-Matrix-A_K} as
\begin{equation} \label{Hankel-Sub-Matrix-A_K-null}
	\begin{bmatrix}A_K&-I\end{bmatrix} \mathcal{H}_{{n_x}n}(\bar\nu) = 0,
\end{equation}
where $\mathcal{H}_{{n_x}n}(\bar\nu) = \begin{bmatrix}{\mathcal{H}}_{{n_x}n}^{+}(\bar\nu)\\{\mathcal{H}}_{{n_x}n}^{-}(\bar\nu)\end{bmatrix} \in \mathbb{R}^{2n_x\times pn_z}$. 
Since ${\mathcal{H}}_{{n_x}n}^{+}(\bar\nu)$ consists of basis rows of ${\mathcal{H}}_{{n_x}n}$, we have that $\text{rank}(\mathcal{H}_{{n_x}n}(\bar\nu)) = n_x$. The dimension of the left null space of $\mathcal{H}_{{n_x}n}(\bar\nu)$ therefore equals to $n_x$, which is exactly the rank of matrix $\begin{bmatrix}A_K&-I\end{bmatrix}$. This means that the left null space of $\mathcal{H}_{{n_x}n}(\bar\nu)$ is completely parameterized by parameters in $A_K$. Consequently, these parameters can be estimated with the same two-step least-squares used in the SISO system. Then, matrices $B$ and $K$ can be estimated in a similar manner. In the following, we use a case study to detail each step of WNSF\textsubscript{SS} for multi-output systems.

\subsection{A Case Study}  \label{Sct4.2}

Take $n_y = n_u = 2$ and $n_x = 4$, then all possible Kronecker indices are
\begin{equation}
	\left\{\bar\nu_1, \bar\nu_2, \bar\nu_3\right\} = \left\{\left\{1,3\right\},\left\{2,2\right\},\left\{3,1\right\}\right\}.
\end{equation}
For brevity, we define two unknown rows in matrix $A_K$ as
\begin{equation*}
	\bm{a}_1 := \begin{bmatrix}a_{11}&a_{12}&a_{13}&a_{14}\end{bmatrix}, \bm{a}_2 := \begin{bmatrix}a_{21}&a_{22}&a_{23}&a_{24}\end{bmatrix}.
\end{equation*}
 Corresponding to Kronecker indices, three possible forms of matrix $A_K$ are
\begin{equation*}
	\left[\begin{array}{cccc}
		\multicolumn{4}{c}{\bm{a}_1}\\\cdashline{1-4}
		0&0&1&0 \\
		0&0&0&1 \\
		\multicolumn{4}{c}{\bm{a}_2}
	\end{array}
	\right]
	\left[\begin{array}{cccc}
		0&1&0&0\\
		\multicolumn{4}{c}{\bm{a}_1} \\\cdashline{1-4}
		0&0&0&1 \\
		\multicolumn{4}{c}{\bm{a}_2}
	\end{array}
	\right]
	\left[\begin{array}{cccc}
		0&1&0&0 \\
		0&0&1&0 \\
		\multicolumn{4}{c}{\bm{a}_1}\\\cdashline{1-4}
		\multicolumn{4}{c}{\bm{a}_2}
	\end{array}
	\right].
\end{equation*}
Meanwhile, three possible forms for matrix $C$ are
\begin{equation*}
	\left[\begin{array}{c:ccc}
		1&0&0&0 \\
		0&1&0&0
	\end{array}
	\right]
	\left[\begin{array}{cc:cc}
		1&0&0&0 \\
		0&0&1&0
	\end{array}
	\right]
	\left[\begin{array}{ccc:c}
		1&0&0&0 \\
		0&0&0&1
	\end{array}
	\right].
\end{equation*}
We now show how to estimate the parameters of $A_K$ in the first form, and the other two forms can be similarly derived.

\textbf{Step~1 (HOARX Modeling):} This is identical to the single-output case. For a given order $n$, the predictor Markov parameters $\bm{g}_n$ in HOARX are estimated using OLS. Meanwhile, the asymptotic covariance of the estimation error $\Vc{\tilde{\bm{g}}_n}$ is obtained, denoted by $\bm{R}_n^{-1}$.

\textbf{Step~2 (OLS for $A_K$):} After constructing the Hankel matrix $\mathcal{H}_{{n_x}n}$ in \eqref{Hankel_MIMO} using Markov parameters $\bm{g}_n$, we select the basis rows of $\mathcal{H}_{{n_x}n}$ for the specified index $\bar\nu_1 = \left\{1,3\right\}$, giving 
\begin{subequations} \label{Hankel-Sub-Matrix-Case}
	\begin{align}
		{\mathcal{H}}_{{n_x}n}^{+}(\bar\nu) =\begin{bmatrix}h_{1,1}^\top&h_{1,2}^\top&h_{2,2}^\top&h_{3,2}^\top\end{bmatrix}^\top, \\
		{\mathcal{H}}_{{n_x}n}^{-}(\bar\nu) =\begin{bmatrix}h_{2,1}^\top&h_{2,2}^\top&h_{3,2}^\top&h_{4,2}^\top\end{bmatrix}^\top.
	\end{align}	
\end{subequations}
Then, according to \eqref{Hankel-Sub-Matrix-A_K-null}, we have that
\begin{subequations} \label{Hankel-Sub-Matrix-OLS}
	\begin{align}
		\bm{a}_1{\mathcal{H}}_{{n_x}n}^{+}(\bar\nu) &= h_{2,1}, \label{Hankel-Sub-Matrix-OLS-1}\\
		\bm{a}_2{\mathcal{H}}_{{n_x}n}^{+}(\bar\nu) &= h_{4,2}. \label{Hankel-Sub-Matrix-OLS-2}
	\end{align}	
\end{subequations}
After replacing true Markov parameters $\bm{g}_n$ with their estimates $\hat{\bm{g}}_n$ in ${\mathcal{H}}_{{n_x}n}^{+}(\bar\nu)$, $h_{2,1}$ and $h_{4,2}$, two parallel OLS can be used to estimate parameters $\bm{a}_1$ and $\bm{a}_2$, respectively.

\textbf{Step~3 (WLS for $A_K$):} Similar to \eqref{E31}, the residuals in \eqref{Hankel-Sub-Matrix-OLS-1} and \eqref{Hankel-Sub-Matrix-OLS-2} can be cast into
\begin{subequations} \label{Hankel-Sub-Matrix-Residual}
	\begin{align*}
		\hat h_{2,1} - h_{2,1} - \bm{a}_1\left(\hat {\mathcal{H}}_{{n_x}n}^{+}(\bar\nu)  - {\mathcal{H}}_{{n_x}n}^{+}(\bar\nu)\right) &= \Vc{\tilde{\bm{g}}_n}{\mathcal{K}_n}(\bm{a}_1),\\
		\hat h_{4,2} - h_{4,2} - \bm{a}_2\left(\hat {\mathcal{H}}_{{n_x}n}^{+}(\bar\nu)  - {\mathcal{H}}_{{n_x}n}^{+}(\bar\nu)\right) &= \Vc{\tilde{\bm{g}}_n}{\mathcal{K}_n}(\bm{a}_2),
	\end{align*}	
\end{subequations}
where ${\mathcal{K}_n}(\bm{a}_1)$ and ${\mathcal{K}_n}(\bm{a}_2)$ are corresponding block Toeplitz matrices. Then, two optimal weighting matrices $\bar\Lambda_n(\bm{a}_1)  = {\mathcal{K}_n^\top}(\bm{a}_1){{\bar{\bm{R}}_n}^{-1}}{\mathcal{K}_n}(\bm{a}_1)$ and $\bar\Lambda_n(\bm{a}_2)  = {\mathcal{K}_n^\top}(\bm{a}_2){{\bar {\bm{R}}_n}^{-1}}{\mathcal{K}_n}(\bm{a}_2)$ can be constructed to refine the estimates of $\bm{a}_1$ and $\bm{a}_2$ in Step~2 using WLS.

With an available estimate of $A_K$, we now briefly summarize how to estimate matrices $B$ and $K$, which is same as the SISO case.

\textbf{Step~4 (OLS for $B_K$):} Since matrix $C$ is known, with an estimate of $A_K$, an estimate for the extended obervability matrix $\mathcal{O}_{n-1}$ can be constructed. Then, after vectorization by row for the following equation,
\begin{equation*}
	\mathcal{O}_{n-1} B_K = \begin{bmatrix}g_0^\top&g_1^\top&\cdots&g_{n-1}^\top\end{bmatrix}^\top,
\end{equation*}
we have that
\begin{equation} \label{Hankel-Sub-MatrixB-OLS}
	\bm{\eta}\Phi_n = \Vc{\bm{g}_n},
\end{equation}
where $\Phi_n = \mathcal{O}_{n-1}^\top\otimes I_4 \in \mathbb{R}^{4n_x\times4n}, \bm{\eta}= \Vc{B_K}$. After replacing $\bm{g}_n$ and $\mathcal{O}_{n-1}$ with their estimates, an OLS estimate of $B_K$ can be obtained. 

\textbf{Step~5 (WLS for $B_K$):} Similar to SISO case, it can be shown that the residual in \eqref{Hankel-Sub-MatrixB-OLS} can be cast into
\begin{equation*}
	\Vc{\hat{\bm{g}}_n - \bm{g}_n} - \bm{\eta}\left(\hat{\Phi}_n - \Phi_n \right) \simeq \Vc{\tilde{\bm{g}}_n}{\mathcal{K}_n}(\bm{a}_1,\bm{a}_2,\bm{\eta}),
\end{equation*}
where ${\mathcal{K}_n}(\bm{a}_1,\bm{a}_2,\bm{\eta})$ is a associated transformation matrix. In this way, after constructing an optimal weighting matrix $\bar\Lambda_n(\bm{a}_1,\bm{a}_2,\bm{\eta})  = {\mathcal{K}_n^\top}(\bm{a}_1,\bm{a}_2,\bm{\eta}){{\bar {\bm{R}}_n}^{-1}}{\mathcal{K}_n}(\bm{a}_1,\bm{a}_2,\bm{\eta})$, WLS can used to refine the estimate of $B_K$ in Step~4.

In summary, when applying WNSF\textsubscript{SS} to MIMO systems, we first need to specify a Kronecker index $\bar\nu$, and then parametrize system matrices on canonical form \eqref{MIMO_canonical_form}. Meanwhile, a submatrix of the Hankel matrix ${\mathcal{H}}_{{n_x}n}$ should be selected according to $\bar\nu$, which essentially consists of basis rows of ${\mathcal{H}}_{{n_x}n}$. Combining with matrix vectorization, the remaining steps are essentially same as those in SISO systems. In practice, if there is no prior information about the Kronecker index $\bar\nu$, one approach is to enumerate all possible parameterizations and apply WNSF\textsubscript{SS} to obtain a collection of state-space models. Among these, the model that yields the smallest prediction error can then be selected. This procedure is feasible when the state dimension $n_x$ is small, but it can become computationally expensive as $n_x$ grows. It is worth noting that, according to Lemma \ref{Lem1}, the state-space model \eqref{MIMO_canonical_form} with a given Kronecker index $\bar\nu$ is capable of representing almost all $n_x$-dimensional stochastic LTI systems. In our simulations, we observed that a particular choice of Kronecker indices already achieves competitive performance compared to state-of-the-art SIMs.

\section{Asymptotic Properties} \label{Sct5}

In this section we present asymptotic properties of WNSF\textsubscript{SS}. First, we have the following assumption regarding the order of HOARX model \eqref{E16}, which ensures that the truncation error is sufficiently small, so that asymptotically no information is lost in Step~1, loosely speaking meaning that the estimated HOARX model forms an approximate sufficient statistic.

\begin{assumption} (Order of HOARX\cite{Galrinho2018parametric}) \label{Asp4}
	We let the order $n$ of the HOARX \eqref{E16} depend on the sample size $N$ according to the following conditions \footnote{In this assumption, $n$ is denoted by $n(N)$ to highlight the dependency of $n$ on $N$, whereas for simplicity, such a dependence is concealed in other parts of the paper.}:
	\begin{enumerate} [(1)]
		\item $n(N) \to \infty$ as $N \to \infty$.
		\item $n^{4+\delta}(N)/N \to 0$ for some $\delta>0$, as $N \to \infty$.
		\item $\sqrt{N}d(N) \to 0$ as $N \to \infty$, where $d(N):= \sum_{k=n(N)+1}^{\infty} \norm{C{A_K^{k- 1}}B_K}$.	
	\end{enumerate}
\end{assumption}

\begin{remark} \label{Rmk2}
	The above assumption ensures that the order of HOARX model $n(N)$ grows at a suitable rate with $N$. To be specific, the first condition ensures that the growth of $n(N)$ is not too slow, while the second condition ensures that the growth of $n(N)$ is not too fast. In priciple, one can take $n=\beta\text{log}N$, where $\beta>0$, to satisfy these two conditions for sufficiently large $N$. Moreover, for the third condition, since $\rho(A_K)<1$, we have 
	\begin{equation*}
		\norm{A_K^{n(N)}} = \mathcal{O}(\rho^{n(N)}) = \mathcal{O}(N^{-\beta/{\rm{log}(1/\bar\rho)}}),
	\end{equation*}
	where $\bar\rho(A_K) < \rho <1$. In this way, the third condition will be satisfied for a large enough $\beta$. In practice though, $n(N)$ can be determined by minimizing the prediction errors of the estimated state-space model as proposed in \cite{Galrinho2018parametric} for other models estimated with WNSF.
\end{remark}

As shown in \eqref{E20} and \eqref {E21}, the asymptotic properties of the HOARX model \eqref{E16} in Step 1 were well understood. We now provide asymptotic properties of our estimates $\hat{\bm{a}}_{\text{ols}}$,  $\hat{\bm{a}}_{\text{wls}}$, $\hat{\bm{\eta}}_{\text{ols}}$ and $\hat{\bm{\eta}}_{\text{wls}}$ in Steps~2, 3, 4 and 5, respectively. It is noted that the following Theorems~\ref{Thm1}--\ref{Thm3} are stated for single-output systems. Due to the parameterization issue, the results for multi-output systems are presented separately in Theorem~\ref{Thm6}.
\begin{theorem} \label{Thm1}
	The estimates $\hat{\bm{a}}_{\text{ols}}$ and $\hat{\bm\eta}_{\text{ols}}$ in Steps~2 and 4 are consistent:
    \begin{subequations} \label{E37}
		\begin{align}
			\hat{\bm{a}}_{\text{ols}} &\to \bm{a}, \ {\rm{as}} \ N \to \infty \ {\rm{w.p.1}}, \\
			{\hat{\bm\eta}}_{\text{ols}} &\to {\bm\eta}, \ {\rm{as}} \ N \to \infty \ {\rm{w.p.1}},
		\end{align}		
	\end{subequations}
\end{theorem}
\begin{proof}
	See Appendix \ref{AppA}.
\end{proof}
\begin{theorem} \label{Thm2}
	The estimates $\hat{\bm{a}}_{\text{wls}}$ and $\hat{\bm\eta}_{\text{wls}}$ in Steps~3 and 5 are consistent:
	\begin{subequations} \label{E38}
		\begin{align}
			\hat{\bm{a}}_{\text{wls}} &\to \bm{a}, \ {\rm{as}} \ N \to \infty \ {\rm{w.p.1}}, \\
			{\hat{\bm\eta}}_{\text{wls}} &\to {\bm\eta}, \ {\rm{as}} \ N \to \infty \ {\rm{w.p.1}},
		\end{align}		
	\end{subequations}
\end{theorem}
\begin{proof}
	See Appendix \ref{AppB}.
\end{proof}
\begin{theorem} \label{Thm3}
	The estimates $\hat{\bm{a}}_{\text{wls}}$ and $\hat{\bm\eta}_{\text{wls}}$ in Steps~3 and 5 are asymptotically efficient:
	\begin{subequations} \label{E40}
		\begin{align}
			\sqrt{N}\left(\hat{\bm{a}}_{\text{wls}} - \bm{a}\right) &\sim \AsN{0}{\sigma_e^2M_{CR,\bm{a}}^{-1}}, \\
			\sqrt{N}\left({\hat{\bm\eta}}_{\text{wls}} - {\bm\eta}\right) &\sim \AsN{0}{\sigma_e^2M_{CR,\bm{\eta}}^{-1}},
		\end{align}		
	\end{subequations}
	where $M_{CR,\bm{a}}$ and $M_{CR,{\bm\eta}}$ are the CRLBs of $\bm{a}$ and $\bm{\eta}$, respectively, specified in Appendix~\ref{AppC}.
\end{theorem}
\begin{proof}
	See Appendix \ref{AppC}.
\end{proof}

\begin{remark}
	According to the above theorems, the estimates of matrices $A_K$ in Step~3, $B$ and $K$ in Step~5 are consistent and asymptotically efficient. Then, using the invariance principle \cite{Zacks1971theory}, we conclude that the estimate of system matrix $\hat A = \hat A_K + C\hat K$ is also consistent and asymptotically efficient.
\end{remark}

For multi-output systems, unlike the unique canonical form in the single-output case, a Kronecker index $\bar\nu$ is required to specify a canonical form $M_{\bar{\nu}_i}$. Howoever, a specific parameterization $M_{\bar{\nu}_i}$ may not correspond to the true state-space model, leading to potential inconsistency in the presence of model mismatch. Nevertheless, when the parameterization $M_{\bar{\nu}_i}$ \eqref{MIMO_canonical_form} is admissible, which occurs with high probability, the consistency and asymptotic variance can be derived similarly to the SISO case. The results are stated in the following theorem.

\begin{theorem} \label{Thm6}
    For a given multi-output system \eqref{E2}, if the parameterization $M_{\bar{\nu}_i}$ \eqref{MIMO_canonical_form} is admissible, then the WNSF\textsubscript{SS} estimates from Steps 2 and 4 are consistent, and those from Steps 3 and 5 are both consistent and asymptotically efficient. 
\end{theorem}
\begin{proof}
	See Appendix \ref{AppF}.
\end{proof}

\begin{remark}
    Under the admissible parameterization $M_{\bar{\nu}_i}$, the estimates obtained in Steps~3 and 5 are asymptotically efficient in the sense that, their asymptotic error covariance matrices coincide with those of the PEM applied to the same parameterization, where PEM employs a quadratic cost with optimal weighting, which is known to be asymptotically efficient \cite{Ljung1999system}.
\end{remark}

Based on Theorem \ref{Thm6} and \cite[Th. 3.1]{Wertz1982determination}, it is straightforward to have the following corollary:
\begin{corollary}
	Given two admissible parameterizations for a multi-output system \eqref{E2}, then the determinants of the asymptotic error covariance matrices obtained using WNSF\textsubscript{SS} are identical.
\end{corollary}

\section{Simulations} \label{Sct6}

In this section, we perform simulation studies and discuss practical issues. First, we demonstrate the asymptotic properties of WNSF\textsubscript{SS}. Next, we compare WNSF\textsubscript{SS} with the-state-of-art methods on two numerical examples, one is a SISO system, and the other is a MIMO system. Finally, we evaluate the performance of WNSF\textsubscript{SS} on random systems and practical data sets from DaISy \cite{De1997daisy}. 

We perform open- and closed-loop simulations, where the data are generated by
\begin{equation*}
	\begin{split}
		u_k &= \frac{1}{1 + F_y(q)\,G_\circ(q)} r_k 
		- \frac{F_y(q)H_\circ(q)}{1 + F_y(q)G_\circ(q)} e_k,\\
		y_k &= \frac{G_\circ(q)}{1 + F_y(q)\,G_\circ(q)} r_k + \frac{H_\circ(q)}{1 + F_y(q)G_\circ(q)} e_k.
	\end{split}	
\end{equation*}
For open-loop, we mean $F_y(q)=0$. Details for $G_\circ(q)$, $H_\circ(q)$, $F_y(q)$, $r_k$ and $e_k$ are specified in each example. The following methods are included for comparison:
\begin{enumerate}[(1)]
	\item N4SID \cite{Van2012subspace}: N4SID with the CVA weighting. This corresponds to the classical CCA method introduced in \cite{Larimore1990canonical}, which is known to be asymptotically efficient for time series identification (= no inputs) \cite{Bauer2005comparing} and optimal for white inputs \cite{Bauer2002some} among classical SIMs.
	\item SSARX \cite{Jansson2003subspace}: SSARX shares the same pre-estimation step as WNSF\textsubscript{SS} and is effective for both open-loop and closed-loop cases.
	\item PBSID\textsubscript{o} \cite{Chiuso2007role}: An ``optimally weighted" PBSID. Its asymptotic variance is less or equal
	than that of the classical CCA method. 
	\item WNSF\textsubscript{ar} \cite{Galrinho2018system}: A variant of WNSF that applies to ARMAX models, proven to be asymptotically efficient.
	\item PEM from the MATLAB 2021a System Identification Toolbox \cite{Ljung1995system}, with two initialization strategies:
	\begin{enumerate}
		\item PEM\textsubscript{d}: PEM initialized with default settings.
		\item PEM\textsubscript{t}: PEM initialized using the true system.
	\end{enumerate}
\end{enumerate}

\subsection{Illustration of Asymptotic Properties} \label{Sct6.1}

In this subsection, we use a single-output system and multi-output system to illustrate that WNSF\textsubscript{SS} is asymptotically efficient. 
\subsubsection{A SISO System} Consider the following ARMAX model: 
\begin{equation*}
	G_\circ(q) =\frac{b_1q^{-1}+b_2q^{-1}}{1+f_1q^{-1}+f_2q^{-1}},
	H_\circ(q) =\frac{1+a_1q^{-1}+a_2q^{-1}}{1+f_1q^{-1}+f_2q^{-1}}.
\end{equation*}
As is well known, there is an equivalent state-space model on canonical form \eqref{E13} to this ARMAX model. We show that WNSF\textsubscript{SS} is asymptotically efficient for estimating coefficients 
\begin{equation*}
	\begin{split}
		{\bm{\theta}}_\circ &= \begin{bmatrix}
			f_1&f_2&b_1&b_2&a_1&a_2 \end{bmatrix}^\top \\
		&= \begin{bmatrix}
			-1.5&0.7&1&0.5&-0.8&0.2 \end{bmatrix}^\top.
	\end{split}	
\end{equation*}
The innovations $\left\{e_k\right\}$ and references $\left\{r_k\right\}$ are independent Gaussian white sequences with unit variance. For the closed-loop case, we take the controller $u_{k}=5r_{k}-F_y(q)y_{k}$, where
\begin{equation*}
	F_y(q) =\frac{0.63-2.08q^{-1}+2.82q^{-2}-1.86q^{-3}+0.5q^{-4}}{1-2.65q^{-1}+3.11q^{-2}-1.75q^{-3}+0.39q^{-4}}.
\end{equation*}

\begin{figure}
	\centering
	\includegraphics[scale=0.63]{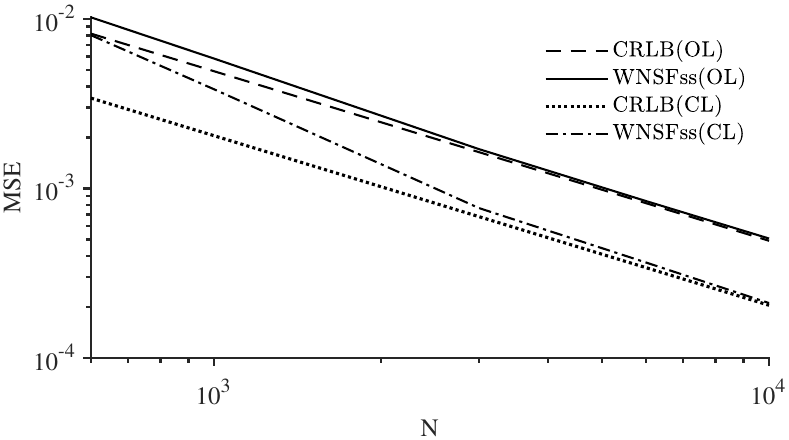}
	\caption{Average MSE of $\hat{\bm{\theta}}$ from 1000 Monte Carlo trials (SISO system): Open-loop (OL) and Closed-loop (CL) Cases.}
	\label{F1S}
\end{figure}

We perform 1000 Monte Corlo trails, with the sample size $N\in \left\{600,1000,3000,6000,10000\right\}$ and the order of HOARX $n\in \left\{40,50,60,70,80\right\}$, respectively. The results shown in Figure~\ref{F1S} are the average mean-squared error (MSE) of estimates of $\bm{\theta}_\circ$ using WNSF\textsubscript{SS} and theoretical CRLBs for both open-loop and closed-loop cases. As shown, the respective CRLBs are attained as the sample size increases.

\subsubsection{A SIMO System} 

Consider the following three order state-space model: 
\begin{equation*}
	\begin{split}
		{A}_K &= \begin{bmatrix}
			0.4&0.1&0 \\ 
			0&0&1 \\ 
			0.5&0.2&0.6\end{bmatrix},
		{B} = \begin{bmatrix}
			1\\ 0.2\\ 0.5\end{bmatrix}, \\
		{K} &= \begin{bmatrix}
			0.5&0.1 \\ 
			0&0.6\\ 
			-0.5 &-0.56 \end{bmatrix},
		{C} = \begin{bmatrix}
			1&0&0\\
			0&1&0\end{bmatrix}.
	\end{split}
	\nonumber
\end{equation*}
The innovations $\left\{e_k\right\}$ consist of independent Gaussian white sequences with unit variance. For the open-loop case, we take $u_{k}=r_{k}$, where $\left\{r_k\right\}$ consist of independent Gaussian white sequences with unit variance, and independent with $\left\{e_k\right\}$. For the closed-loop case, we take the controller $u_{k}=r_{k}-F_yy_{k}$, where $F_y=\text{diag}(0.5,0.5)$. Since the above model is already in a canonical form, we show that WNSF\textsubscript{SS} is asymptotically efficient for estimating free parameters contained in matrices $A_K$, $B$ and $K$, i.e.,  
\begin{equation*}
	{\bm{\theta}}_\circ = \Vc{\begin{matrix}
		0.4&0.1&0&0.5&0.2&0.6&1&0.2 \\
		0.5&0.5&0.1&0&0.6&-0.5&-0.56 \end{matrix}}.
\end{equation*}

\begin{figure}
	\centering
	\includegraphics[scale=0.63]{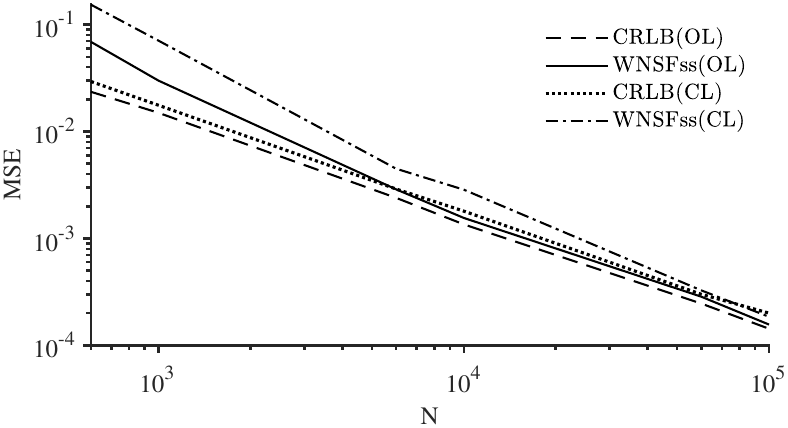}
	\caption{Average MSE of $\hat{\bm{\theta}}$ from 100 Monte Carlo trials (SIMO system): Open-loop (OL) and Closed-loop (CL) Cases.}
	\label{F1M}
\end{figure}

We perform 200 Monte Corlo trails, with the sample size $N\in \left\{600,1000,6000,10000,60000,100000\right\}$ and the order of HOARX $n\in \left\{60,80,100,120,140,160\right\}$, respectively. The results shown in Figure~\ref{F1M} are the average mean-squared error (MSE) of estimates of $\bm{\theta}_\circ$ using WNSF\textsubscript{SS} and theoretical CRLBs for both open-loop and closed-loop cases. As shown, when the parameterization is consistent with the true model, the respective CRLBs are attained as the sample size increases. Regarding the method we used for deriving the CRLB for parameterized state-space models, it is mainly based on \cite{Soderstrom2006computing}. For more details, we refer to Appendix.

\subsection{Comparison with Other Methods} \label{Sct6.2}

In this subsection, we compare the performance of WNSF\textsubscript{SS} against PEM and SIMs using two numerical examples. The first example is a fourth-order SISO system characterized by two resonance peaks in the transfer functions $G_\circ(q)$ and $H_\circ(q)$. The second is a fourth-order MIMO system under a poor excitation condition. Such challenging scenarios often cause PEM to converge to a non-global minimum, and some SIMs typically exhibit poor performance.

\subsubsection{A SISO System}

\begin{figure}
	\centering
	\includegraphics[scale=0.68]{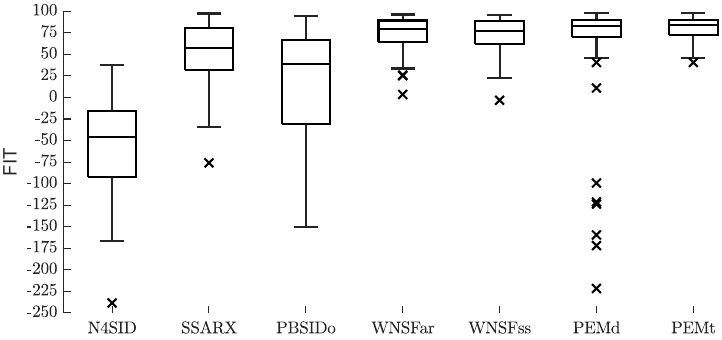}
	\caption{FITs from 100 Monte Carlo trials: Open-loop.}
	\label{F2}
\end{figure}

\begin{figure}
	\centering
	\includegraphics[scale=0.68]{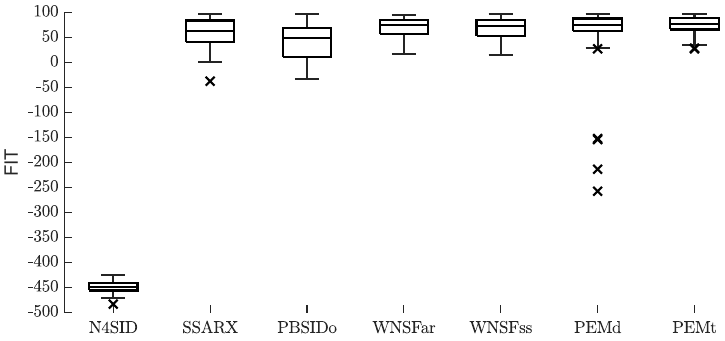}
	\caption{FITs from 100 Monte Carlo trials: Closed-loop.}
	\label{F3}
\end{figure}

Consider the following ARMAX model:
\begin{equation*}
	\begin{split}
		G_\circ(q) & =\frac{0.1q^{-1}+0.05q^{-2}+0.02q^{-3}+0.01q^{-4}}{1+0.2401q^{-4}}, \\
		H_\circ(q) &= \frac{1-2.48q^{-1}+3.08q^{-2}-2.24q^{-3}+0.81q^{-4}}{1+0.2401q^{-4}},
    \end{split}
\end{equation*} 
where both $G_\circ(q)$ and $H_\circ(q)$ have two resonance peaks. We show the comparison between WNSF\textsubscript{SS} and other algorithms in terms of realization of system matrices. The innovations $\left\{e_k\right\}$ and references $\left\{r_k\right\}$ are independent Gaussian white sequences with unit variance. For the closed-loop case, we take the controller $F_y(q) = -0.5$. The performance is evaluated by 
\begin{equation*}
	{\rm{FIT}} = 100\left(1-\frac{\|g_{\circ} - \hat g\|}{\|g_{\circ} - {\rm{mean}}[g_{\circ}]\|}\right),
\end{equation*}
where $g_{\circ}$ is impulse response parameters of the true system, and $\hat g$ is impulse response parameters of the estimated systems using different methods. The number of samples is fixed at $N=6000$, and 100 Monte Carlo simulations are performed. For a fair comparison, for the past and future horizons used in SIMs, we take $f=p \in \left\{5:5:50\right\}$, and for the order of HOARX used in WNSF methods, we take $n\in \left\{50:10:150\right\}$. Corresponding to the sets of parameters $f$ and $n$, a set of state-space models are identified using each method in every Monte Carlo simulation. Then, the model that gives the smallest prediction error is selected to compute the FIT. The FITs for several methods under open-loop and closed-loop data are presented in Figures~\ref{F2} and \ref{F3}, respectively. Among SIMs, N4SID performs poorly on this example. Although SSARX and PBSID\textsubscript{o} perform better than N4SID, they provide models that give median FITs of no more than 50\% for both open-loop and closed-loop cases. Meanwhile, PEM with the default initialization (PEM\textsubscript{d}) has a considerable amount of low-accuracy outliers where the algorithm fails to find the global minima. In contrast, WNSF\textsubscript{ar} and our method WNSF\textsubscript{ss} have similar performance, which provide models that give FITs comparable with PEM with initialized by the true system (PEM\textsubscript{t}).

%

\subsubsection{A MIMO System}

The following MIMO system is frequently used for the evaluation of SIMs \cite{Veen2013closed}:
\begin{equation*}
	\begin{split}
		{A} &= \begin{bmatrix}
			0.67&0.67&0&0 \\ 
			-0.67&0.67&0&0 \\ 
			0&0&-0.67&-0.67 \\
			0&0&0.67&-0.67\end{bmatrix}, \\
		{B} &= \begin{bmatrix}
			0.6598& -0.5256 \\ 
			1.9698 &0.4845 \\ 
			4.3171& -0.4879 \\
			-2.6436& -0.3416\end{bmatrix},
		{K} = \begin{bmatrix}
			-0.6968 &-0.1474 \\ 
			0.1722 &0.5646 \\ 
			0.6484 &-0.4660 \\
			-0.9400 &0.1032 \end{bmatrix}, \\
		{C} &= \begin{bmatrix}
			-0.3749& 0.0751 &-0.5225 &0.5830 \\
			-0.8977 &0.7543& 0.1159 &0.0982\end{bmatrix}.
	\end{split}
	\nonumber
\end{equation*}
We consider the closed-loop setting, i.e., the input $u_k = -F_yy_k + r_k$, where $F_y=\text{diag}(-0.1,-0.1)$. Similar to \cite{Veen2013closed}, the performance of several methods under a poor excitation condition is evaluated. The innovation $e_{k} \sim \mathcal{N}(0,\sigma_e^2I)$, where $\sigma_e^2 = 10^{-4}$, and the excitation signal is given by
\begin{equation*}
	r_{k} = \begin{bmatrix}
		\sin\left(\frac{4\pi k}{10}\right) + \sin\left(\frac{11\pi k}{20}\right) \\ 
		\sin\left(\frac{9\pi k}{20}\right) + \sin\left(\frac{6\pi k}{10}\right)
	\end{bmatrix}
	+ v_{k},
\end{equation*}
where $v_{k} \sim \mathcal{N}(0,\sigma_v^2I)$, and $\sigma_v^2 = 8\times10^{-8}$. The number of samples is fixed at $N=4000$, and the order of HOARX $n=50$. For the past and future horizon used in SIMs, we take $f=p=7$. We perform 50 independent Monte Corlo trails. For this MIMO system, all possible canonical parameterizations have been enumerated in Section~\ref{Sct4.2}. In the simulation, we choose the parameterization associated with the Kronecker index $\bar\nu_1 = \left\{1,3\right\}$ for the WNSF\textsubscript{SS} method. It can be verified that the above MIMO system is equivalent the canonical parameterization in \eqref{MIMO_canonical_form} for the given Kronecker index $\bar\nu_1$. For MIMO systems, since the transformation from a ARMAX model to a state-space model is not straightforward, the multivariable WNSF\textsubscript{ar} method is not included for comparison in this example. For PEM, we use the function $\text{ssest}(\dots,``\text{Form}",``\text{canonical}")$ to identify canonical state-space models.

\begin{figure}
	\centering	
	\includegraphics[scale=0.55]{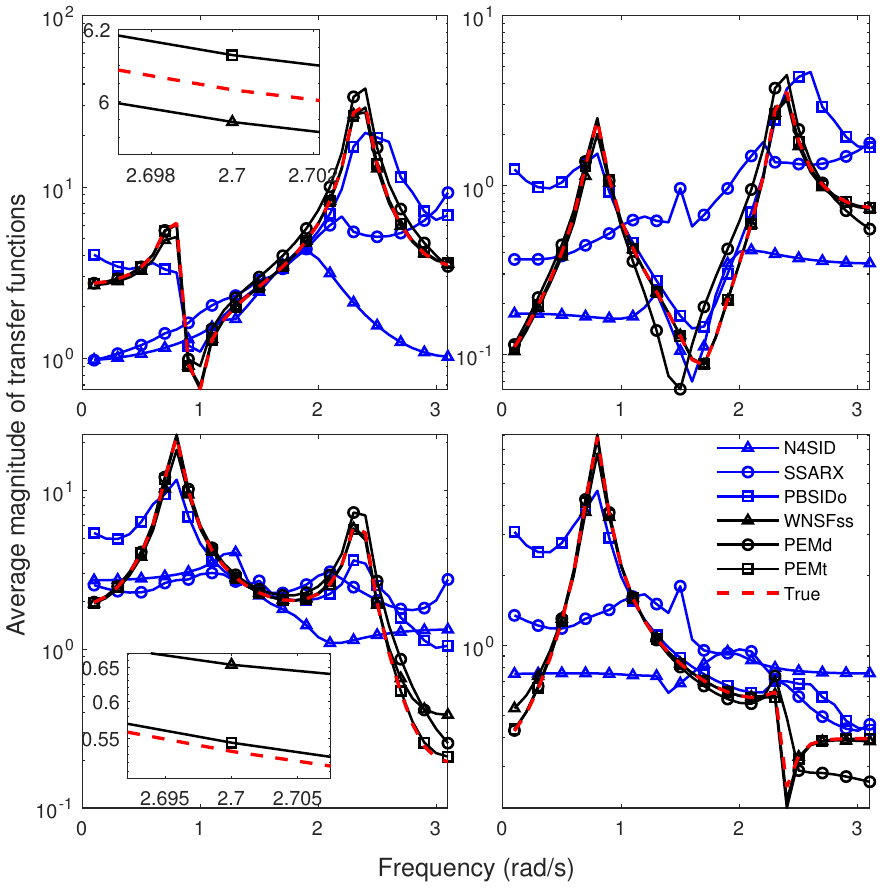}
	\caption{Average transfer functions of the identified MIMO systems from 50 Monte Carlo trials.}
	\label{F4}
\end{figure}

The average transfer functions of identified models using different methods are shown in Figure \ref{F4}. It can be observed that among SIMs, PBSID\textsubscript{o} performs better than N4SID and SSARX, but it is not as accurate as WNSF\textsubscript{SS}, PEM\textsubscript{d} and PEM\textsubscript{t}. Moreover, WNSF\textsubscript{SS} performs slightly better than PEM\textsubscript{d} in identifying resonance peaks, but slightly worse than PEM\textsubscript{t}. This verifies that WNSF\textsubscript{SS} can be effectively applied to identifying MIMO state-space models.
 
These simulation results illustrate that WNSF\textsubscript{SS} shows robustness against algorithmic failures and maintains a median performance that is competitive with other methods. 

%
%

\subsection{Benchmark Problems from DaISy} \label{Sct6.4}

In order to evaluate the performance of our method on practical systems, we testify the performance of WNSF\textsubscript{SS} and other methods on eight benchmark problems from the DaISy collection \cite{De1997daisy}. An introduction to these benchmark problems is summarized in Table~\ref{Table1}. The first five systems are SISO systems, the sixth system is a MISO system, and the last two are MIMO systems. 

\begin{table}[H]
	\caption {Description of benchmark problems from DaISy} \label{Table1}
	\begin{center}
		\begin{tabular}{cccccc}
			\toprule
			Data sets & Description & $n_u$  & $n_y$ & $n_x$ & $N$ \\
			\midrule
			96-006 &Hair dryer & 1  & 1 & 4 & 1000 \\
			96-004 &Ball $\&$ Beam & 1  & 1 & 2 & 1000 \\
			99-001  &Steam heating system & 1  & 1 & 4 & 801 \\
			96-008  &Wing flutter & 1  & 1 & 4 & 1024 \\
			96-009  &Robot arm & 1  & 1 & 4 & 1024 \\
			96-011  &Heat flow density & 2  & 1 & 8 & 1680 \\
			97-003  &Industrial winding process & 5  & 2 & 3 & 2500 \\
			96-007  &CD player arm & 2  & 2 & 3 & 2048 \\
			\bottomrule
		\end{tabular}
	\end{center}
\end{table}

Each dataset is split into 70\% for identification and 30\% for validation. Moreover, the performance is evaluated by the identification error and validation error, defined as \cite{Liu2010interior}
\begin{equation*}
	\begin{split}
		e_I &= \left(\frac{\sum_{t=0}^{N_I-1}\|y_I(t) - \hat y(t)\|^2}{\sum_{t=0}^{N_I-1}\|y_I(t) - \bar y_I\|^2}\right)^{1/2},\\
		e_V &= \left(\frac{\sum_{t=0}^{N_V-1}\|y_V(t) - \hat y(t)\|^2}{\sum_{t=0}^{N_V-1}\|y_V(t) - \bar y_V\|^2}\right)^{1/2},\\
	\end{split}
\end{equation*}
where $N_I = 0.7N$ and $N_I = 0.3N$. Moreover, $y_I(t)$ and $y_V(t)$ are the given output from the identification set and the validation set, $\bar y_I = \frac{1}{N_I}\sum_{t=0}^{N_I-1}y_I(t)$ and $\bar y_V = \frac{1}{N_V}\sum_{t=0}^{N_V-1}y_V(t)$, and $\hat y(t)$ is the output of the identified model from various methods. For a fair comparison, for the past and future horizons in SIMs, we create a candidate set for $f = p \in \left\{n_x+1:1:40\right\}$, and for the order of HOARX used in WNSF\textsubscript{SS}, we take a set $n \in \left\{10:1:150\right\}$. We then choose $f$ and $n$ that give the minimal identification error to be the future horizon of SIMs and order of HOARX for each data set. For PEM, since the true system is unknown, only PEM\textsubscript{d} which is initialized by default in MTALAB is included for comparison. For the canonical parameterization used in WNSF\textsubscript{SS} for two MIMO systems, we take $\bar\nu_1 = \left\{1,2\right\}$ for realization. The identification errors and validation errors of these methods are summarized in Tables \ref{Table2} and \ref{Table3}, respectively, with the lowest error for each dataset highlighted in bold.

\begin{table}[H]
	\caption{Errors of Different Methods}
	\label{tab:combined_errors}
	\centering
	\begin{subtable}[t]{0.48\textwidth}
		\centering
		\caption{Identification Errors $e_I$}
		\label{Table2}
		\begin{tabular}{cccccc}
			\toprule
			{Dataset} & {N4SID} & {SSARX} & {PBSID\textsubscript{o}} & {WNSF\textsubscript{SS}} & PEM\textsubscript{d} \\
			\midrule
			96-006 & 0.5148 & 0.5150& 0.5148&\textbf{0.5138} &0.5927\\
			96-004 &1.0702&848.0910&1.0865&\textbf{0.8823}&7504.1\\
			99-001&\textbf{0.6082} &0.6141&0.6131&0.6201&0.6240\\
			96-008 &0.2562&0.2564&0.2429&\textbf{0.2232}&0.4184\\
			96-009 &\textbf{0.1541}&0.6468&0.6374 &0.7118&0.5365\\
			96-011  &0.4895 &\textbf{0.3709}&0.3979&0.3750 &0.4282\\
			97-003 &0.8081&0.7989&0.8012&\textbf{0.7839}&0.7947\\
			96-007 &1.0003&\textbf{0.4937}&0.4955&0.5068&3.2686\\
			\bottomrule
		\end{tabular}
	\end{subtable}%
    \vspace{3mm}
	\hfill
	\begin{subtable}[t]{0.48\textwidth}
		\centering
		\caption{Validation Errors $e_V$}
		\label{Table3}
		\begin{tabular}{cccccc}
			\toprule
			{Dataset} & {N4SID} & {SSARX} & {PBSID\textsubscript{o}} & {WNSF\textsubscript{SS}} & PEM\textsubscript{d} \\
			\midrule
			96-006 & 0.9808 & 0.9824&0.9817&\textbf{0.9794}&1.0792\\
			96-004&9.0412&31.8028&\textbf{3.1320}&5.0331&729.66\\
			99-001 &\textbf{1.3406}&1.3504&1.3482&1.3556&1.3501\\
			96-008 &3.3466&0.7561&0.7200&\textbf{0.5936}&0.8790\\
			96-009 &0.9277&0.7956&0.8058&\textbf{0.7792}&0.9208\\
			96-011  &0.9534 &0.6107&0.6799&\textbf{0.6082}&0.7329\\
			97-003 &0.8012&0.7991&0.8046&\textbf{0.7841}&0.7917\\
			96-007 &0.9992 &0.5770&\textbf{0.5144}&0.5191&3.3743\\
			\bottomrule
		\end{tabular}
	\end{subtable}
\end{table}

As shown in Table \ref{Table2}, WNSF\textsubscript{SS} generally provides moderate identification accuracy across datasets. Its identification errors are consistently better than PEM\textsubscript{d} in nearly all cases, especially for problematic Datasets such as 96-004 and 96-007. However, in some cases, it is outperformed by N4SID and SSARX. For instance, in Dataset 96-009, the identification error of N4SID is noticeably lower than that of WNSF\textsubscript{SS}.

As shown in Table~\ref{Table3}, WNSF\textsubscript{SS} demonstrates clear advantages in terms of validation errors. In Datasets 96-006, 96-008, 96-009, 96-011 and 97-003, WNSF\textsubscript{SS} achieves the lowest validation error, and in Datasets 96-004 and 96-007, it yields the near-lowest validation error. In contrast, although N4SID and SSARX achieve the lowest identification error in four Datasets, it often trails behind WNSF\textsubscript{SS} in terms of validation accuracy. Moreover, the validation errors of WNSF\textsubscript{SS} are consistently better than PEM\textsubscript{d} in nearly all cases.

In summary, WNSF\textsubscript{SS} is effective in producing models that generalize well across datasets coming from practical problems. Together with comparison on previous numerical examples, these results highlight the robustness of WNSF\textsubscript{SS}, suggesting that it can be considered as an appealing alternative for identifying state-space models.

\subsection{Random Systems} \label{Sct6.3}
\begin{figure}
	\centering
	\includegraphics[scale=0.65]{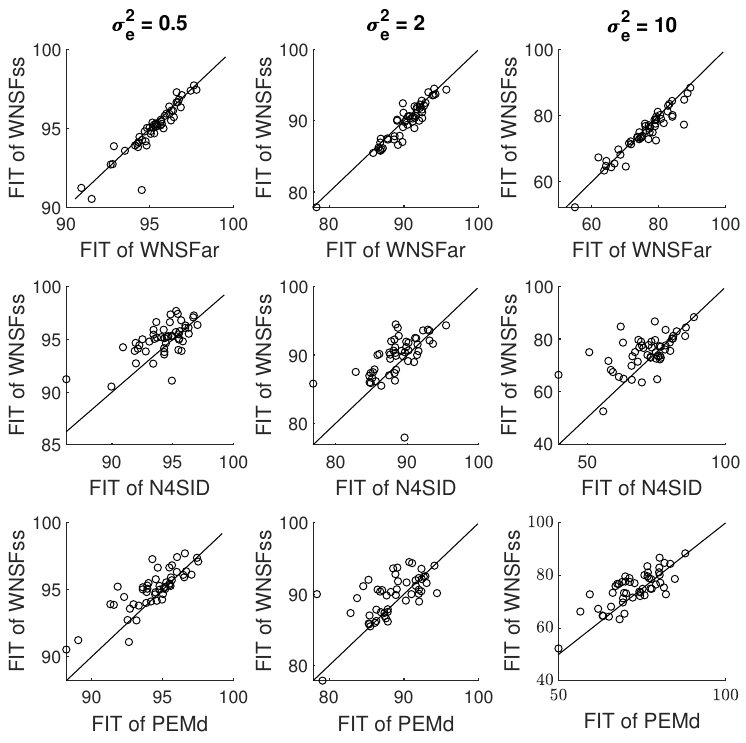}
	\caption{Joint FIT distribution from 50 Monte Carlo trials (10-order SISO systems): A random system ($\circ$), and the solid line is a bisector line.}
	\label{F5}
\end{figure}

\begin{figure}
	\centering
	\includegraphics[scale=0.65]{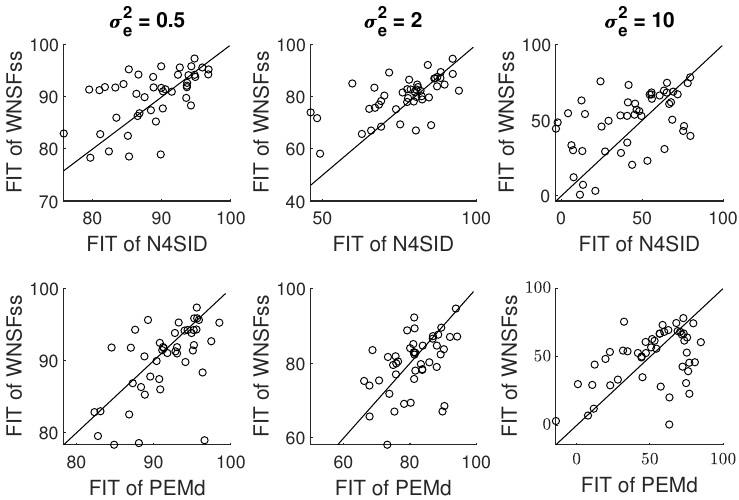}
	\caption{Joint FIT distribution from 50 Monte Carlo trials (3-order MIMO systems): A random system ($\circ$), and the solid line is a bisector line.}
	\label{F6}
\end{figure}

In order to test the robustness of WNSF\textsubscript{SS}, we now perform simulations on two sets of random systems generated by MATLAB function $\text{drss}(\dots)$. One set consists of 10-order random SISO systems, and the other set consists of three-order random MIMO systems with $n_y = 2, n_u = 5$. Below is the script we use as a reference for generating SISO systems:
\begin{equation*}
	\begin{split}
		&{\rm{m = idss(drss(n_x,1,1));}}\\
		&{\rm{m.d = zeros(1,1);}}\\
		&{\rm{m.b = 5*randn(n_x,1);}}\\
		&{\rm{y = sim(m,u) + \sigma_e*randn(N,1);}}	
	\end{split}
\end{equation*}
In order to avoid extremely slow systems, we limit the system in both sets to have poles with a maximum magnitude of 0.97. Moreover, to guarantee that all systems have similar gains, we restrict them to have $2 < \norm{G(q)}_{\mathcal{H}_2} <4$. The number of samples is fixed at $N=1000$. For a fair comparison, for the past and future horizons in SIMs, we create a candidate set for $f = p \in \left\{n_x+1:2:40\right\}$, and for the order of HOARX used in WNSF methods, we take a set $n \in \left\{10:2:100\right\}$. We then choose $f$ and $n$ that give the minimal prediction error for each random system to be the future horizon of SIMs and order of HOARX for WNSF methods. For SISO systems, the inputs are given by $u_k = \frac{0.8}{1-0.9q^{-1}}r_k$, where $\left\{r_k\right\}$ consists of i.i.d. Guassian sequences with zero mean and unit variance.
and for MIMO systems, the inputs are given by $\text{idinput}([N,n_u],\text{'rbs'},[0 \ 0.1])$. Moreover, three different levels of noises are used, i.e., $\sigma_e^2 \in \left\{0.5,2,10\right\}$. For the canonical parameterization used in WNSF\textsubscript{SS} for MIMO systems, we take both $\bar\nu_1 = \left\{1,2\right\}$ and $\bar\nu_2 = \left\{2,1\right\}$ for realization. Then, the model that gives the smallest prediction error are chosen for comparison. We mainly compare the performance of WNSF\textsubscript{SS} against N4SID with CVA weighting and PEM\textsubscript{d}, as well as WNSF\textsubscript{ar} for SISO systems. The performance is evaluated by FIT. Since PEM initialized by default in MATLAB gives poor performace for these random systems, especially for MIMO systems, for a meaningful comparison, PEM initialzed by the estimate of N4SID is used for comparison. The results of SISO and MIMO systems are shown in Figures~\ref{F5} and \ref{F6}, respectively. 

As shown in Figure~\ref{F5}, except for few outliers, WNSF\textsubscript{SS} demonstrates nearly identical performance on most systems to WNSF\textsubscript{ar}, confirming that the two approaches are asymptotically equivalent for SISO systems. Furthermore, since WNSF\textsubscript{ar} is proven to be asymptotically efficient, these results also supprt that WNSF\textsubscript{SS} is asymptotically efficient. Moreover, WNSF\textsubscript{SS} generally outperforms both N4SID and PEM\textsubscript{d}, which gives higher FIT on more random systems than N4SID and PEM\textsubscript{d} do. This comparison shows the robustness of WNSF\textsubscript{SS} for identifying high order systems.

As shown in Figure~\ref{F6}, WNSF\textsubscript{SS} is competitive with N4SID and PEM\textsubscript{d} for identifying MIMO systems, giving higher FIT on slightly more random systems than N4SID does, but less than PEM\textsubscript{d} does. This is not surprsing, since PEM\textsubscript{d} is initialzed by the estimate of N4SID, and is asymptotically efficient. Even so, the comparison shows the robustness of WNSF\textsubscript{SS} for MIMO systems identification.

\section{Relations to Other Methods} \label{Sct7}

Our method is closely related to SIMs, PEM, and existing WNSF approaches. In the following, we briefly review these methods and clarify how WNSF\textsubscript{SS} relates to them.

\subsection{Subspace Identification} \label{Sct7.1}

Same as WNSF\textsubscript{SS}, the following Hankel matrix plays a key role in SIMs:
\begin{equation*}
	\mathcal{H}_{fp} = \mathcal{O}_f\mathcal{C}_p = \begin{bmatrix}
		CB_K & CA_KB_K & \cdots & CA_K^{p}B_K \\
		CA_KB_K & CA_K^2B_K & \cdots & CA_K^{p+1}B_K \\
		\vdots & \vdots & \ddots & \vdots \\
		CA_K^{f}B_K & CA_K^{f+1}B_K & \cdots & CA_K^{n-1}B_K \\
	\end{bmatrix},
\end{equation*}
where $n=f+p-1$ is the number of Markov parameters stacked in $\mathcal{H}_{fp}$. Under the Assumption \ref{Asp1}, we have that ${\text{rank}}(\mathcal{H}_{fp}) = n_x$. Obtaining the above Hankel matrix is a starting point for most SIMs. Earlier versions of SIMs mainly start with estimating a series of Markov parameters using least-squares \cite{Ho1966effective,Kung1978new}, and then construct $\mathcal{H}_{fp}$ from those Markov parameters, whereas modern SIMs directly estimate this Hankel (Hankel-like) matrix using projection or regressions. Having an estimate $\hat{\mathcal{H}}_{fp}$, the key step to obtain system matrices is to take SVD on $\hat{\mathcal{H}}_{fp}$, i.e.,
\begin{equation} \label{8E5} 
	\hat{\mathcal{H}}_{{f}p} = \hat U \hat S {\hat V}^{\top} \approx \hat U_1\hat S_1 {\hat V}_1^{\top},
\end{equation}
where $\hat S  = \text{diag}\left(\hat \sigma_1,\hat \sigma_2,\cdots,\hat \sigma_{n_x},\cdots,\hat \sigma_{f+1}\right)$, and $\hat S_1$ contains the first $n_x$ singular values of $\hat S$. Moreover, $\hat U$ and $\hat V$ contain left and right singular vectors, respectively. In this way, a balanced realization of $\mathcal{O}_{n_x}$ and $\mathcal{C}_p $ are
\begin{subequations} \label{8E6}
	\begin{align}
		\hat{\mathcal{O}}_{f} &= \hat U_1 \hat S_1^{1/2}, \\
		\hat{\mathcal{C}}_p &= \hat S_1^{1/2} \hat V_1^{\top}.
	\end{align}
\end{subequations}
Having estimates $\hat{\mathcal{O}}_{f}$ and $\hat{\mathcal{C}}_p$, the system matrices $A_K$, $B_K$ and $C$ can be estimated via least-squares by using the shift-property of ${\mathcal{O}}_{f}$ and ${\mathcal{C}}_p$. Furthermore, statistical properties can be improved by pre- and post-multiplying the Hankel matrix with some weighting matrices before SVD. Alternatively to \eqref{8E5}, taking SVD on 
\begin{equation} \label{8E7} 
	W_1\hat{\mathcal{H}}_{fp}W_2 = \hat U \hat S{\hat V}^{\top} \approx \hat U_1\hat S_1{\hat V}_1^{\top},
\end{equation}
the estimates $\hat{\mathcal{O}}_{f}$ and $\hat{\mathcal{C}}_p$ are then given by
\begin{subequations} \label{8E8}
	\begin{align}
		\hat{\mathcal{O}}_{f} = W_1^{-1}\hat U_1 \hat S_1^{1/2}, \\
		\hat{\mathcal{C}}_p = \hat S_1^{1/2} \hat V_1^{\top}W_2^{-1}.
	\end{align}
\end{subequations}
The difference between variants of SIMs is essentially in the estimates of $\mathcal{H}_{{f}p}$ and the choices of weighting matrices $W_1$ and $W_2$. However, determining optimal weighting matrices that achieve asymptotic efficiency remains an open question.

Compared to SIMs, WNSF\textsubscript{SS} has the following features:

(1) Same pre-estimation step as SSARX \cite{Jansson2003subspace} but different purposes behind: In order to decouple the correlation between future inputs $U_f$ and future noises $E_f$ in the closed-loop setting, SSARX uses the predictor form \eqref{E2} and pre-estimates the HOARX model \eqref{E16} to get consistent estimates of Markov parameters, which corresponds to Step 1 in WNSF\textsubscript{SS}. However, after this pre-estimation, SSARX reverts to the traditional SIM framework, estimating the range space of the extended observability matrix. In contrast, WNSF\textsubscript{SS} focuses on the null space and leverages the asymptotic distribution of estimation errors in Markov parameters, which SSARX overlooks, making the two approaches fundamentally different. 

(2) Estimation of the null-space of the extended observability matrix rather than the range space of this matrix as used in most SIMs: In \cite{Viberg1997analysis} a null-space fitting method is proposed which uses a matrix fraction description of a state-space model and optimally estimates the null space of the extended observability matrix with a two-step least-squares procedured. The major difference of this method and WNSF\textsubscript{SS} is that the former requires an explicit estimate of the extended observability matrix, necessitating the use of SVD to obtain such an estimate. In contrast, WNSF\textsubscript{SS} bypasses the SVD and directly estimates the null space using least-squares without the extended observability matrix being available, making the approach statistically solid and more straightforward to analyze.

\begin{remark}
  One point worth highlighting is that, in this work, we assume the system order $n_x$ is known in advance. In contrast, SIMs typically estimate the system order at an intermediate step through SVD. A common, albeit somewhat crude, strategy for order selection involves examining gaps between singular values of the Hankel matrix to determine its rank. While this method is practical in many scenarios, it depends heavily on a problem-specific threshold for classifying singular values as sufficiently small.
  
  Since our approach avoids the SVD step, additional steps are required to determine the system order. However, alternative strategies exist to address this. Since order selection lies beyond the scope of this study, a more detailed discussion of this issue will be presented in future work.
\end{remark}

\subsection{Prediction Error Method} \label{Sct7.2}

We now proceed with an short introduction of PEM. The model \eqref{E1} can be represented in transfer functions as
\begin{equation} \label{8E8}
	y_k = G\left(q,\theta\right)u_k + H\left(q,\theta\right)e_k,
\end{equation}
where $\theta$ denotes free parameters in the canonical parameterization of system matrices $\left\{A,B,C,K\right\}$, and $G\left(q,\theta\right)$ and $H\left(q,\theta\right)$ are transfer functions given by
\begin{equation*}
	\begin{split}
		G\left(q,\theta\right) &= C\left(qI-A\right)^{-1}B, \\ H\left(q,\theta\right) &= C\left(qI-A\right)^{-1}K + I.
	\end{split}
\end{equation*}
To estimate $\theta$, we first derive an one-step-ahead predictor
\begin{equation} \label{8E9}
	\hat{y}_k(\theta) = \left(I-H^{-1}\left(q,\theta\right)\right)y_k + H^{-1}\left(q,\theta\right)G\left(q,\theta\right)u_k,
\end{equation}
and then the prediction error is
\begin{equation} \label{8E10}
	\varepsilon_k(\theta) = y_k-\hat{y}_k(\theta) = H^{-1}\left(q,\theta\right)\left(y_k - G\left(q,\theta\right)u_k\right).
\end{equation}
The idea of PEM is to minimize a cost function
\begin{equation} \label{8E11}
	J(\theta) = \frac{1}{\bar N} \sum_{t=1}^{\bar N} l\left(\varepsilon_t(\theta)\right),
\end{equation}
where $l\left(\cdot \right)$ is a scalar-valued function of prediction errors. The estimate of $\theta$ is then obtained by minimizing $J(\theta)$. Moreover, when the error sequence is Gaussian, PEM with a quadratic cost function is equivalent to the MLE. In this case, the consistency is guaranteed, and the asymptotic covariance is $M_{\text{CR},\theta_\circ}^{-1}$  \cite{Ljung1999system}, corresponding to the CRLB given by
\begin{equation} \label{8E12}
	M_{\text{CR},\theta_\circ} := \bar {\mathbb{E}} \left[\frac{\zeta_k(\theta_\circ)\zeta_k^\top(\theta_\circ)}{\sigma_e^2}\right],
\end{equation}
where $\zeta_k(\theta_\circ) = -\frac{\text{d}}{\text{d}\theta}\varepsilon_k(\theta)\big|_{\theta = \theta_\circ}$, where $\theta_\circ$ is true value of system parameters.

For PEM, solving this optimization problem requires local nonlinear optimization algorithms and good initial estimates. This problem is excacerbated for multi-input multi-output (MIMO) models, which typically require extensive parametrizations, leading to many false local minima.

Compared to PEM, WNSF\textsubscript{SS} has the following features:

(1) Same canonical parameterization, but easier implementation: Although both PEM and WNSF\textsubscript{SS} use the same canonical parameterization of state-space models, their implementation differs significantly. PEM relies on local nonlinear optimization and requires careful initialization, while WNSF\textsubscript{SS} uses only multi-step least-squares, where each step consists of the solution of a quadratic optimization problem. This makes WNSF\textsubscript{SS} much simpler to implement.

(2) Comparable performance with PEM: As demonstrated in Section \ref{Sct5}, for single-output systems, WNSF\textsubscript{SS} is asymptotically efficient. Moreover, as shown in the simulation, WNSF\textsubscript{SS} is competitive with PEM in terms of finite sample estimation accuracy. 

\subsection{WNSF for ARMAX Models} \label{Sct7.3}

The WNSF method, originally proposed in \cite{Galrinho2014weighted,Galrinho2018parametric}, has been applied to various model structures, such as OE, ARMA, ARMAX, and BJ models \cite{Galrinho2018parametric,Galrinho2018system}, but not  to state-space models. It is well-known that for a single-output state-space model \eqref{E1}, there is an equivalent ARMAX model. In this case, one can first apply the WNSF method to get an ARMAX model, and then cast it into a state-space model \eqref{E1}, which gives asymptotic efficient estimates of system matrices in their canonical forms. For convenience, we refer to the WNSF method for ARMAX models as WNSF\textsubscript{ARMAX} throughout this section. However, for a multiple-output system, the equivalent transformation between an ARMAX model and a state-space model is significantly more complex \cite{Ljung1999system,Hannan2012statistical}. Therefore, although the WNSF method can be extended to multivariate ARMAX models \cite{Galrinho2018system}, a WNSF approach that directly applies for state-space models is typically preferred. 

Compared to WNSF\textsubscript{ARMAX}, WNSF\textsubscript{SS} has the following features:

(1) Equivalence in the SISO case: As detailed in Section \ref{Sct3}, the main steps of WNSF\textsubscript{ARMAX} and WNSF\textsubscript{SS} are substantially similar when applied to SISO systems. A major difference is that WNSF\textsubscript{ARMAX} estimates all parameters of the ARMAX model simultaneously, while WNSF\textsubscript{SS} first estimates free parameters of matrix $A_K$. Then, a similar procedure is used to estimate matrices $B$ and $K$. Both methods yield asymptotically efficient estimates for the parameters of interest. 

(2) Direct applicability to the multiple-output case: In contrast to WNSF\textsubscript{ARMAX}, which faces challenges in extending to multiple-output systems due to the complexity of converting an ARMAX model to a state-space model, WNSF\textsubscript{SS} can be directly applied to such cases. This direct applicability makes WNSF\textsubscript{SS} a more straightforward method for applications where a state-space model is preferred.

\section{Conclusion} \label{Sct8}

The WNSF method is known to be applicable to many common SISO and MIMO models, including OE, ARMA, ARMAX, and BJ models, both with rational elements and matrix fraction descriptions. In this work we have extended the portfolio of model structures to the important class of black-box state-space models. The method begins by estimating a HOARX model using OLS, which functions as a sufficient statistic and captures the true system's dynamics with sufficient accuracy. The HOARX model is subsequently reduced to a state-space model in observer canonical form through multi-step least-squares, where WLS plays a crucial role in providing an asymptotically efficient estimate. Since the optimal weighting matrix in WLS depends on the true system parameters, we substitute these with consistent estimates obtained from the prior OLS step, which does not impact the asymptotic optimality. We assess WNSF\textsubscript{SS}'s performance on both numerical and practical systems, highlighting its asymptotic efficiency and balanced accuracy in identification and validation, which suggest that WNSF\textsubscript{SS} is an appealing alternative for building state-space models.

WNSF\textsubscript{SS} lies conceptually between PEM and SIM. Like PEM, it uses the cononical parameterization of state-space models, and is proven to be consistent and asymptotically efficient. As with SIM, it estimates the null space of the Hankel matrix, and exhibits robust numerical properties.

Finally, we note that the asymptotic efficiency of existing SIMs remains an open question. In contrast, the proposed method has been shown to be asymptotically efficient and has demonstrated competitive performance in the examples presented. As such, WNSF\textsubscript{SS} may serve as a useful reference point for evaluating the asymptotic efficiency of other SIMs.


\bibliographystyle{plain}

\bibliography{autosam}

\appendix
\numberwithin{equation}{section}

\section{Consistency of Steps~2 and 3} \label{AppA}

\subsection{Auxiliary Results}
To prove Theorem \ref{Thm1}, we introduce some auxiliary results.

(1) $\norm{\tilde{\bm{g}}_n} \to 0$, ${\rm{as}} \ N \to \infty$ $\text{w.p.1}$: For the first $n$ true Markov parameters $\bm{g}_n$ and their estimates $\hat{\bm{g}}_n$ in Step 1, using the triangular inequality, we have
\begin{equation} \label{AE1}
	\norm{\tilde{\bm{g}}_n} \leq \norm{\hat{\bm{g}}_n - \bar{\bm{g}}_n} + \norm{\bar{\bm{g}}_n - \bm{g}_n},
\end{equation}
where $\bar{\bm{g}}_n$ is defined in \eqref{E20}. According to \cite[Lemma5.1]{Ljung1992asymptotic}, we have $\norm{\bar{\bm{g}}_n - \bm{g}_n} \to 0$, as $n \to \infty$. Moreover, according to \cite[Th.5.1]{Ljung1992asymptotic}, we have $\norm{\hat{\bm{g}}_n - \bar{\bm{g}}_n} \to 0$, ${\rm{as}} \ N \to \infty$ $\text{w.p.1}$. As a result, we have
\begin{equation} \label{AE2}
	\norm{\tilde{\bm{g}}_n} \to 0, {\rm{as}} \ N \to \infty \ {\text{w.p.1}}.
\end{equation}

(2) $\norm{{\tilde{\mathcal{H}}}_{n_xn}}  \to 0$, ${\rm{as}} \ N \to \infty$ $\text{w.p.1}$: Using the norm inequality of a block matrix in Lemma \ref{LemG1}, we have
\begin{equation} \label{AE3}
   	\norm{{\tilde{\mathcal{H}}}_{n_xn}} \leq \sqrt{n_x+1} \norm{\tilde{\bm{g}}_n}.
\end{equation}
According to \eqref{AE2}, $\norm{\tilde{\bm{g}}_n} \to 0$, as $N \to \infty \ {\text{w.p.1}}$, we therefore conclude that $\norm{{\tilde{\mathcal{H}}}_{n_xn}}  \to 0$, ${\rm{as}} \ N \to \infty$ $\text{w.p.1}$. Moreover, since ${\mathcal{H}}_{n_xn}^{-}$ and ${\mathcal{H}}_{n_xn}^{+}$ are sub matrices of ${\mathcal{H}}_{n_xn}$, we  have
\begin{subequations} \label{AE4}
   	\begin{align}
   		&\norm{ {\tilde{\mathcal{H}}}_{n_xn}^{-}} \to 0, {\rm{as}} \ N \to \infty \ {\text{w.p.1}}, \label{AE4a}\\
   		&\norm{ {\tilde{\mathcal{H}}}_{n_xn}^{+}}\to 0, {\rm{as}} \ N \to \infty \ {\text{w.p.1}} \label{AE4b}.
   	\end{align}
\end{subequations}

(3) $\norm{{\mathcal{H}}_{n_xn}}$ is bounded for $\forall n$: Similarly, using the norm inequality of a block matrix in Lemma \ref{LemG1}, we have
\begin{equation} \label{AE5}
   	\norm{{\mathcal{H}}_{n_xn}} \leq \sqrt{n_x+1} \norm{\bm{g}_n}, \forall n.
\end{equation}
Under the Assumption \ref{Asp1}, the system is asymptotically stable, thus, the Markov parameters $\left\{g_i = C{A_K^{i - 1}}B_K\right\}$ are exponentially decaying with $i$, which ensures that $\norm{\bm{g}_n}$ is bounded for $\forall n$. Therefore, $\norm{{\mathcal{H}}_{n_xn}}$ is bounded for $\forall n$.

(4) ${\hat{\mathcal{H}}}_{n_xn}$ is bounded as $N \to \infty$ $\text{w.p.1}$: Using the triangular inequality, we have
\begin{equation} \label{AE6}
   	\norm{{\hat{\mathcal{H}}}_{n_xn}} \leq \norm{{\tilde{\mathcal{H}}}_{n_xn}}  + \norm{{\mathcal{H}}_{n_xn}}.
\end{equation}
According to auxiliary results (2) and (3) in this section, we have that $\norm{{\hat{\mathcal{H}}}_{n_xn}}$ is bounded as $N \to \infty$ $\text{w.p.1}$.

(5)  ${\mathcal{T}}_{n,p}(\bm{a})$ is bounded for $\forall n$: We first define a characteristic polynomial $A(q,\bm{a}) :=1 + a_1 q^{-1} + \cdots + a_{{n_x}}q^{-n_x}$. According to \cite[Th. 3]{Rojas2012analyzing}, we then have 
\begin{equation} \label{AE7}
   	\norm{{\mathcal{T}}_{n,p}(\bm{a})} \leq \norm{A(q,\bm{a})}_{\mathcal{H}_\infty}.
\end{equation}
Due to asymptotic stability of $A(q,\bm{a})$, we conclude that $\norm{A(q,\bm{a})}_{\mathcal{H}_\infty} < c$, thus,  ${\mathcal{T}}_{n,p}(\bm{a})$ is bounded $\forall n$.

(6)  Define $M(\bm{g}_n) := \lim_{n \to \infty}{\mathcal{H}}_{n_xn}^{+}({\mathcal{H}}_{n_xn}^{+})^\top$, where ${\mathcal{H}}_{n_xn}^{+}$ is defined in \eqref{E26}. Then, $M(\bm{g}_n)$ is invertible: According to \eqref{E27}, ${\mathcal{H}}_{n_xn}^{+}$ can be rewritten as ${\mathcal{H}}_{n_xn}^{+} = {\Gamma}_{n_x-1}L_p$. Under Assumption \ref{Asp1}, for $\forall n \geq n_x$, $L_p$ is full-row rank, we then have $\text{rank}({\mathcal{H}}_{n_xn}^{+}) = \text{rank}({\Gamma}_{n_x-1})= n_x$, and $\text{rank}(M(\bm{g}_n)) = \text{rank}({\mathcal{H}}_{n_xn}^{+}) = n_x$.

\subsection{Proof of Theorem \ref{Thm1}}
\begin{proof}
	The estimation error in \eqref{E29} can be written as
	\begin{equation} \label{AE9}
		\begin{split}
			\tilde{\bm{a}}_{\text{ols}} &= -{\hat {\mathcal{H}}}_{n_xn}^{-}({\hat {\mathcal{H}}}_{n_xn}^{+})^{\top}\left({\hat {\mathcal{H}}}_{n_xn}^{+}({\hat {\mathcal{H}}}_{n_xn}^{+})^{\top}\right)^{-1} - \bm{a} \\
			&= -\left({\hat{\mathcal{H}}}_{n_xn}^{-} + \bm{a}{\hat{\mathcal{H}}}_{n_xn}^{+}\right)({\hat {\mathcal{H}}}_{n_xn}^{+})^{\top}\left({\hat{\mathcal{H}}}_{n_xn}^{+}({\hat {\mathcal{H}}}_{n_xn}^{+})^{\top}\right)^{-1} \\
			&= -\begin{bmatrix}\bm{a}&1\end{bmatrix} \tilde{\mathcal{H}}_{n_xn}({\hat {\mathcal{H}}}_{n_xn}^{+})^{\top}\left({\hat{\mathcal{H}}}_{n_xn}^{+}({\hat {\mathcal{H}}}_{n_xn}^{+})^{\top}\right)^{-1} \\
			&= -\tilde{{\bm{g}}}_n{\mathcal{K}_n}(\bm{a})({\hat {\mathcal{H}}}_{n_xn}^{+})^{\top}\left({\hat{\mathcal{H}}}_{n_xn}^{+}({\hat {\mathcal{H}}}_{n_xn}^{+})^{\top}\right)^{-1},
		\end{split}
	\end{equation}
	where the last two equalities follow from \eqref{E30} and \eqref{E31}. Similar to $M(\bm{g}_n)$, define $\hat M(\hat{\bm{g}}_n) := \lim_{n \to \infty} {\hat{\mathcal{H}}}_{n_xn}^{+} ({\hat {\mathcal{H}}}_{n_xn}^{+})^\top$. In this way, using the triangular inequality to \eqref{AE9}, we have
	\begin{equation} \label{AE10}
		\norm{\tilde{\bm{a}}_{\text{ols}}}\leq \norm{\tilde{{\bm{g}}}_n}\norm{{\mathcal{K}_n}(\bm{a})}\norm{{\hat{\mathcal{H}}}_{n_xn}^{+}}\norm{\hat M^{-1}(\hat{\bm{g}}_n)}.
	\end{equation}
	According to auxiliary results in this section, we have $\norm{\tilde{\bm{g}}_n} \to 0$ and $\norm{{\hat{\mathcal{H}}}_{n_xn}^{+}}$ is bounded, ${\rm{as}} \ N \to \infty$ $\text{w.p.1}$. Moreover, $\norm{{\mathcal{K}_n}(\bm{a})}$ is bounded since ${\mathcal{T}}_{n,p}(\bm{a})$ is bounded for all $n$. Thus, consistency is ensured if $\hat M(\hat{\bm{g}}_n)$ is invertible ${\rm{as}} \ N \to \infty$ $\text{w.p.1}$. To show this, based on auxiliary results (2), (3) and (4) in this section and Lemma \ref{LemG2}, we have
	\begin{equation} \label{AE12}
		\norm{\hat M(\hat{\bm{g}}_n) - M({\bm{g}}_n)} \to 0, {\rm{as}} \ N \to \infty \ {\text{w.p.1}}.
	\end{equation}
	According to the auxiliary result (6), we have that $M({\bm{g}}_n)$ is invertible. Since the mapping from the entries of a matrix to its eigenvalues is continuous, we therefore conclude that $\hat M(\hat{\bm{g}}_n)$ is invertible ${\rm{as}} \ N \to \infty$ $\text{w.p.1}$.
	
	Returning \eqref{AE10}, we now have that 
	\begin{equation} \label{AE13}
		\norm{\tilde{\bm{a}}_{\text{ols}}} \leq c_1\norm{\tilde{{\bm{g}}}_n} \to 0, {\rm{as}} \ N \to \infty \ {\text{w.p.1}}.
	\end{equation}
	Moreover, using \eqref{AE1}, we have
	\begin{equation} \label{AE14}
		\norm{\tilde{\bm{a}}_{\text{ols}}} \leq c_1 \left(\norm{\hat{\bm{g}}_n - \bar{\bm{g}}_n} + \norm{\bar{\bm{g}}_n - \bm{g}_n}\right).
	\end{equation}
	According to \cite[Lemma5.1]{Ljung1992asymptotic} and \cite[Th.5.1]{Ljung1992asymptotic}, we have $\norm{\bar{\bm{g}}_n - \bm{g}_n} \leq cd(N)$, where $d(N)$ is defined in Assumption \ref{Asp4} and it decays faster than $\norm{\hat{\bm{g}}_n - \bar{\bm{g}}_n}$. Since $\norm{\hat{\bm{g}}_n - \bar{\bm{g}}_n}$ decays as $\mathcal{O}\left(\sqrt{\frac{n{\text{log}}N}{N}}\left(1+d(N)\right)\right)$, we have that
	\begin{equation} \label{AE15}
		\norm{\tilde{\bm{a}}_{\text{ols}}} = \mathcal{O}\left(\sqrt{\frac{n{\text{log}}N}{N}}\left(1+d(N)\right)\right).
	\end{equation}

	Regarding the estimation error for Step~4, it  equals to
	\begin{equation} \label{DE-estimation_B_K_error_ols}
		\tilde{\bm{\eta}}_{\text{ols}} \simeq \tilde{\bm{g}}_n {\mathcal{K}_n}(\bm{a},\bm{\eta}) {\hat\Phi}_n^\top\left({\hat\Phi}_n{\hat\Phi}_n^\top\right)^{-1}.
	\end{equation}
    It is easy to see that matrices ${\mathcal{K}_n}(\bm{a},\bm{\eta})$ and ${\hat\Phi}_n{\hat\Phi}_n^\top$ are of fixed dimension and bounded. Then, similar to the proof for $\tilde{\bm{a}}_{\text{ols}}$, we have $\norm{\tilde{\bm{\eta}}_{\text{ols}}} \leq c_2\norm{\tilde{\bm{g}}_n} \to 0, {\rm{as}} \ N \to \infty \ {\text{w.p.1}}$.
\end{proof}

\section{Consistency of Steps~3 and 5} \label{AppB}

\subsection{Auxiliary Results}
To prove Theorem \ref{Thm2}, we introduce some auxiliary results.

(1) $\norm{\bar R_n}$ and $\norm{\bar R_n^{-1}}$ are bounded for $\forall n$ \cite{Hannan2012statistical}.

(2) $\norm{R_n}$ and $\norm{R_n^{-1}}$ are bounded for $\forall n$, as $N \to \infty \ {\text{w.p.1}}$ \cite[Lemma4.2]{Ljung1992asymptotic}.

(3) $\bar\Lambda_n(\bm{a})  = \sigma _e^2 {{\mathcal{K}_n^\top}(\bm{a})}{{\bar R_n}^{-1}}{\mathcal{K}_n}(\bm{a})$ is invertible and bounded for $\forall n$: Since ${\mathcal{K}_n}(\bm{a}) = {{\mathcal{T}_{n,p}}(\bm{a})} \otimes I$, where ${{\mathcal{T}_{n,p}}(\bm{a})}$ is a full-column rank Toeplitz matrix, we have that ${\mathcal{K}_n}(\bm{a})$ is full-column rank, which further implies that $\bar\Lambda_n(\bm{a})$ is invertible. Moreover, using the triangular inequality, we have
\begin{equation} \label{BE1}
   	\norm{\bar\Lambda_n(\bm{a})} \leq \sigma _e^2 \norm{{\mathcal{K}_n}(\bm{a})}^2\norm{{\bar R_n}^{-1}}.
\end{equation}
According to the auxiliary result (5) in Appendix \ref{AppA} and auxiliary result (1) in this section, both ${{\mathcal{T}_{n,p}}(\bm{a})}$ and ${\bar R_n}^{-1}$ are bounded for $\forall n$, thus, $\norm{\bar\Lambda_n(\bm{a})}$ is bounded.  
    
(4) $ M(\bm{g}_n,\bm{a}) := \lim_{n \to \infty} {\mathcal{H}}_{n_xn}^{+}{\bar\Lambda}_n^{-1}(\bm{a})({{\mathcal{H}}}_{n_xn}^{+})^\top$ is invertible: Under Assumption \ref{Asp1}, we have that ${\mathcal{H}}_{n_xn}^{+}$ is full-row rank for $\forall n \geq n_x$. Moreover, since the covariance matrix ${\bar\Lambda}_n(\bm{a})$ is invertible, we conclude that
$\text{rank}(M(\bm{g}_n,\bm{a})) = n_x$.
    
(5) $\hat {M}(\hat{\bm{g}}_n,\hat{\bm{a}}_{\text{ols}}) := \lim_{n \to \infty} {\hat{\mathcal{H}}_{n_xn}^{+}} {\hat\Lambda_n^{-1}(\hat{\bm{a}}_{\text{ols}})}({\hat {\mathcal{H}}_{n_xn}^{+}})^{\top}$, we have that $\hat {M}(\hat{\bm{g}}_n,\hat{\bm{a}}_{\text{ols}})$ is invertible as  $N \to \infty \ {\text{w.p.1}}$: To show this, we first use Lemma \ref{LemG2} to prove that
\begin{equation} \label{BE2}
   	\norm{\hat {M}(\hat{\bm{g}}_n,\hat{\bm{a}}_{\text{ols}}) - M(\bm{g}_n,\bm{a})} \to 0, {\rm{as}} \ N \to \infty \ {\text{w.p.1}}.
\end{equation}
According to auxiliary results (2) and (3) in Appendix \ref{AppA}, we have that $\norm{{\mathcal{H}}_{n_xn}^{+}}$ is bounded and $\norm{{\tilde{\mathcal{H}}}_{n_xn}^{+}}\to 0, {\rm{as}} \ N \to \infty \ {\text{w.p.1}}$. Moreover, since $\bar\Lambda_n(\bm{a})$ is invertible and bounded, we have that $\norm{{\bar\Lambda}_n^{-1}(\bm{a})}$ is bounded. Additionally, we need to ensure that $\norm{{\hat\Lambda_n^{-1}(\hat{\bm{a}}_{\text{ols}})} - {\bar\Lambda}_n^{-1}(\bm{a})} \to 0$, as  $N \to \infty \ {\text{w.p.1}}$. Using the triangular inequality, we have
\begin{equation} \label{BE3}
   	\begin{split}
   		\norm{{\hat\Lambda_n(\hat{\bm{a}}_{\text{ols}})} - {\bar\Lambda}_n(\bm{a})} \leq &\sigma_e^2\norm{{\tilde{\mathcal{K}}_n}(\hat{\bm{a}}_{\text{ols}})}\norm{{{R_n^{-1}}}}
   		\norm{{\hat{\mathcal{K}}_n}(\hat{\bm{a}}_{\text{ols}})}\\
   		& + \sigma_e^2\norm{{\tilde{\mathcal{K}}_n}(\hat{\bm{a}}_{\text{ols}})}\norm{{{R_n^{-1}}}}
   		\norm{{\mathcal{K}_n}(\bm{a})}\\
   		& + \sigma_e^2\norm{{\mathcal{K}_n}(\bm{a})}^2\norm{{\bar R_n}^{-1}-{{R_n^{-1}}}}.
   	\end{split}    	
\end{equation}
Since $\norm{{{R_n^{-1}}}}$ is bounded, as $N \to \infty \ {\text{w.p.1}}$, we have 
\begin{equation} \label{BE4}
	\norm{{\bar R_n}^{-1}-{{R_n^{-1}}}} \leq \norm{{\bar R_n}^{-1}}\norm{{R_n^{-1}}}\norm{{\bar R_n}-{{R_n}}} \to 0.
\end{equation}
Moreover, using Theorem \ref{Thm1}, we have that
\begin{equation} \label{BE5}
	\norm{{\tilde{\mathcal{K}}_n}(\hat{\bm{a}}_{\text{ols}})}  \to 0, {\rm{as}} \ N \to \infty \ {\text{w.p.1}}.  	
\end{equation}
According to \eqref{BE3}, \eqref{BE4} and \eqref{BE5}, we conclude that 
\begin{equation} \label{BE6}
   	\begin{split}
   		\norm{{\hat\Lambda_n(\hat{\bm{a}}_{\text{ols}})} - {\bar\Lambda}_n(\bm{a})}  \to 0, {\rm{as}} \ N \to \infty \ {\text{w.p.1}}.
   	\end{split}    	
\end{equation}
Since $\bar\Lambda_n(\bm{a})$ is invertible and bounded, using continuity of eigenvalues, we conclude that ${\hat\Lambda_n(\hat{\bm{a}}_{\text{ols}})}$ is invertible and bounded as $N \to \infty \ {\text{w.p.1}}$.
Furthermore, according to Lemma \ref{LemG3}, we have
\begin{equation} \label{BE7}
   	\norm{{\hat\Lambda_n^{-1}(\hat{\bm{a}}_{\text{ols}})} - {\bar\Lambda}_n^{-1}(\bm{a})} \to 0, {\rm{as}} \ N \to \infty \ {\text{w.p.1}}.    	
\end{equation}
Returning to \eqref{BE2}, we now have that
 \begin{equation} \label{BE8}
   	\norm{\hat{M}(\hat{\bm{g}}_n,\hat{\bm{a}}_{\text{ols}}) - M(\bm{g}_n,\bm{a})} \to 0, {\rm{as}} \ N \to \infty \ {\text{w.p.1}}.
\end{equation}
Furthermore, since $M(\bm{g}_n,\bm{a})$ is invertible, according to Lemma \ref{LemG2}, we have that $\hat{M}(\hat{\bm{g}}_n,\hat{\bm{a}}_{\text{ols}})$ is invertible, as $N \to \infty \ {\text{w.p.1}}$.

\subsection{Proof of Theorem \ref{Thm2}}
\begin{proof}
	The estimation error in \eqref{E34} can be written as
	\begin{equation} \label{BE9}
		\begin{split}
			\tilde{\bm{a}}_{\text{wls}} &= -{\hat {\mathcal{H}}}_{n_xn}^{-}{\hat\Lambda_n^{-1}(\hat{\bm{a}}_{\text{ols}})}({\hat {\mathcal{H}}_{n_xn}^{+}})^{\top} {\hat{M}}^{-1}(\hat{\bm{g}}_n,\hat{\bm{a}}_{\text{ols}}) - \bm{a} \\
			& = -\tilde{{\bm{g}}}_n{\mathcal{K}_n}(\bm{a}){\hat\Lambda_n^{-1}(\hat{\bm{a}}_{\text{ols}})}({\hat{\mathcal{H}}_{n_xn}^{+}})^{\top} {{\hat{M}}^{-1}(\hat{\bm{g}}_n,\hat{\bm{a}}_{\text{ols}})},
		\end{split}
	\end{equation}
	where the last equality follows from \eqref{E30} and \eqref{E31}. Using the triangular inequality, we have
	\begin{equation} \label{BE10}
		\begin{split}
			\norm{\tilde{\bm{a}}_{\text{wls}}} \leq& \norm{\tilde{{\bm{g}}}_n} \norm{{\mathcal{K}_n}(\bm{a})}\norm{{\hat\Lambda_n^{-1}(\hat{\bm{a}}_{\text{ols}})}}\\
			& \times \norm{{\hat{\mathcal{H}}_{n_xn}^{+}}} \norm{{{\hat{ M}}^{-1}(\hat{\bm{g}}_n,\hat{\bm{a}}_{\text{ols}})}}.
		\end{split}
	\end{equation}
	According to auxiliary results summarized in this section, we have that $\norm{{\mathcal{K}_n}(\bm{a})}$, $\norm{{\hat\Lambda_n^{-1}(\hat{\bm{a}}_{\text{ols}})}}$ and $\norm{{\hat{\mathcal{H}}_{n_xp}^{+}}}$ are bounded, and ${{\hat{M}}(\hat{\bm{g}}_n,\hat{\bm{a}}_{\text{ols}})}$ is invertible as $N \to \infty \ {\text{w.p.1}}$, we therefore conclude that 
	\begin{equation*}
		\norm{\tilde{\bm{a}}_{\text{wls}}} \leq c_2\norm{\tilde{{\bm{g}}}_n} \to 0, {\rm{as}} \ N \to \infty \ {\text{w.p.1}}.
	\end{equation*}

	Regarding the estimation error Step~5, it equals to
	\begin{equation} \label{DE-estimation_B_K_error_wls}
		\begin{split}
			\tilde{\bm{\eta}}_{\text{wls}} \simeq &\tilde{\bm{g}}_n {\mathcal{K}_n}(\bm{a},\bm{\eta}) \hat\Lambda_n^{-1}(\hat{\bm{a}}_{\text{wls}},\hat{\bm{\eta}}_{\text{ols}}) {\hat\Phi}_n^\top\\
			& \times \left({\hat\Phi}_n\hat\Lambda_n^{-1}(\hat{\bm{a}}_{\text{wls}},\hat{\bm{\eta}}_{\text{ols}}){\hat\Phi}_n^\top\right)^{-1}.
		\end{split}		
	\end{equation}
    Then, similar to the proof for Theorem~\ref{Thm1} in Appendices~\ref{AppA}, we have $\norm{\tilde{\bm{\eta}}_{\text{wls}}} \leq c_4\norm{\tilde{\bm{g}}_n} \to 0, {\rm{as}} \ N \to \infty \ {\text{w.p.1}}$.

\end{proof}

\section{Asymptotic Efficiency of Steps~3 and 5} \label{AppC}
\subsection{Auxiliary Results (the Cram\'er-Rao Lower Bound)}

For convenience, we use the following ARMAX model to derive the CRLB of ${\bm{a}}$ and $\bm{\eta}$ in the state-space model:
\begin{equation} \label{CE1}
	\bm{F}(q,\bm{\theta})y_k = \bm{L}(q,\bm{\theta})u_k + \bm{A}(q,\bm{\theta})e_k,
\end{equation}
where 
\begin{equation*}
	\begin{split}
		\bm{F}(q,\bm{\theta}) &= 1 + f_1 q^{-1} + \cdots + f_{{n_x}}q^{-n_x}, \\
		\bm{L}(q,\bm{\theta}) &= l_1 q^{-1} + \cdots + l_{{n_x}}q^{-n_x}, \\
		\bm{A}(q,\bm{\theta}) &= 1 + a_1 q^{-1} + \cdots + a_{{n_x}}q^{-n_x},\\
		\bm{\theta} &= \begin{bmatrix}
			f_1&\cdots&f_{{n_x}}&l_1&\cdots&l_{{n_x}}&a_1&\cdots&a_{{n_x}}
		\end{bmatrix}.
	\end{split}	
\end{equation*}
According to \cite[Sec. 4.3]{Ljung1999system}, the above ARMAX model can be cast into the state-space model \eqref{E2}, where the relations between their parameters are as follows:
\begin{equation*}
	f_i = a_i - k_i, l_i = b_i, i = 1,2,\cdots,n_x.
\end{equation*}

Furthermore, the ARMAX model \eqref{CE1} has the following transfer function form:
\begin{equation} \label{CE2}
	y_k = \bm{G}(q,\bm{\theta})u_k + \bm{H}(q,\bm{\theta})e_k,
\end{equation}
where 
\begin{equation*}
	\bm{G}(q,\bm{\theta}) = \bm{F}^{-1}(q,\bm{\theta})\bm{L}(q,\bm{\theta}),  \bm{H}(q,\bm{\theta}) = \bm{F}^{-1}(q,\bm{\theta})\bm{A}(q,\bm{\theta}).
\end{equation*}
Define $\bm{T}(q,\bm{\theta}) := \begin{bmatrix}\bm{G}(q,\bm{\theta})&\bm{H}(q,\bm{\theta}) \end{bmatrix}$ and let $\bm{T}^{'}(q,\bm{\theta})$ as the gradient of $\bm{T}(q,\bm{\theta}) $ with respect to $\bm{\theta}$, i.e.,
\begin{equation} \label{CE2}
   	\bm{T}^{'}(q,\bm{\theta}) = \begin{bmatrix}
   		-\frac{\bm{L}(q,\bm{\theta})}{\bm{F}^{2}(q,\bm{\theta})}\mathcal{V}_{n_x}(q)& -\frac{\bm{A}(q,\bm{\theta})}{\bm{F}^{2}(q,\bm{\theta})}\mathcal{V}_{n_x}(q) \\
   		\frac{1}{\bm{F}(q,\bm{\theta})}\mathcal{V}_{n_x}(q)&0\\
   		0&\frac{1}{\bm{F}(q,\bm{\theta})}\mathcal{V}_{n_x}(q)
   	\end{bmatrix},
\end{equation}
where $\mathcal{V}_{n_x}(q) := \begin{bmatrix}q^{-1}&q^{-2}&\cdots&q^{-n_x}\end{bmatrix}^\top$. For simplicity, we omit $q$ in transfer functions, such as $\bm{F}(q,\bm{\theta})$, $\bm{G}(q,\bm{\theta})$  and $\mathcal{V}_{n_x}(q)$, we therefore obtain
\begin{equation} \label{CE3}
	\begin{split}
		\zeta_k(\bm{\theta}) :=& \bm{H}^{-1}(\bm{\theta})\bm{T}^{'}(\bm{\theta})\begin{bmatrix}u_k\\e_k\end{bmatrix} \\
		=&\begin{bmatrix}
			-\frac{\bm{L}(\bm{\theta})}{\bm{F}(\bm{\theta})\bm{A}(\bm{\theta})}\mathcal{V}_{n_x}& -\frac{1}{\bm{F}(\bm{\theta})}\mathcal{V}_{n_x} \\
			\frac{1}{\bm{A}(\bm{\theta})}\mathcal{V}_{n_x}&0\\
			0&\frac{1}{\bm{A}(\bm{\theta})}\mathcal{V}_{n_x}
		\end{bmatrix}\begin{bmatrix}u_k\\e_k\end{bmatrix}.
	\end{split}		
\end{equation}
Furthermore, we have that
\begin{equation} \label{CE4}
	\begin{bmatrix}u_k\\e_k\end{bmatrix} = \bm{X}(q)\begin{bmatrix}r_k\\e_k\end{bmatrix},
\end{equation}
where 
\begin{equation*}
	\begin{split}
		\bm{X}(\bm{\theta}) &= \begin{bmatrix}
			\bm{S}(\bm{\theta})&-F_y(q)\bm{S}(\bm{\theta})\bm{H}(\bm{\theta})\\
			0&1\end{bmatrix}, \\
		\bm{S}(\bm{\theta}) &= \left(1+F_y(q)\bm{G}(\bm{\theta})\right)^{-1}.
	\end{split}
\end{equation*}
After replacing \eqref{CE4} into \eqref{CE3}, we have that 
\begin{equation*}
	\zeta_k(\bm{\theta}) = \bm{\Phi}(\bm{\theta}) \begin{bmatrix}r_k\\e_k\end{bmatrix},
\end{equation*}
where $\bm{\Phi}(q,\bm{\theta})=\bm{H}^{-1}(\bm{\theta})\bm{T}^{'}(\bm{\theta})\bm{X}(\bm{\theta})$. Using Parseval's relation, we can express the CRLB of $\bm{\theta}$ as
\begin{equation} \label{CE5}
	\begin{split}
		M_{CR,\bm{\theta}} &=\bar {\mathbb{E}}\left[ \zeta_k(\bm{\theta})\zeta_k^\top(\bm{\theta})\right] \\
		&= \frac{1}{2\pi}\int_{-\pi}^{\pi} 
		\bm{\Phi}(e^{iw},\bm{\theta})\text{diag}\left(\Psi_r(w),\sigma_e^2\right)\bm{\Phi}^{*}(e^{iw},\bm{\theta})\,dw.
	\end{split}	
\end{equation}
In particular, we recognize that the CRLB of $\bm{a}$ is
\begin{equation} \label{CE5}
	M_{CR,\bm{a}} = \frac{\sigma_e^2}{2\pi}\int_{-\pi}^{\pi} \frac{\mathcal{V}_{n_x}}{\bm{A}(e^{iw},\bm{\theta})}\frac{\mathcal{V}_{n_x}^{*}}{\bm{A}^{*}(e^{iw},\bm{\theta})} dw.
\end{equation}
We now show that
\begin{equation} \label{CRLB-a-true}
	M_{CR,\bm{a}} = M(\bm{g}_n,\bm{a}) = \lim_{n \to \infty} {\mathcal{H}}_{n_xn}^{+}{\bar\Lambda}_n^{-1}(\bm{a})({{\mathcal{H}}}_{n_xn}^{+})^\top.
\end{equation}
First, we express ${\bar R_n}$, ${\mathcal{K}_n}(\bm{a})$ and ${\mathcal{H}}_{n_xn}$ involved in $M(\bm{g}_n,\bm{a})$ in the frequency domain. First, notice that the regressor \eqref{E16} can be rewritten as
\begin{equation} \label{CE7}
	\bm{z}_n(k) = \bm{P}_1\begin{bmatrix}
			\mathcal{V}_{n}&0\\0&\mathcal{V}_{n}
		\end{bmatrix}\begin{bmatrix}y_k\\u_k\end{bmatrix}= \bm{P}_1\begin{bmatrix}\mathcal{V}_{n}&0\\0&\mathcal{V}_{n}\end{bmatrix}\bm{Z}(q,\bm{\theta})\begin{bmatrix}
			r_k\\e_k
		\end{bmatrix},	
\end{equation}
where $\bm{P}_1$ is a permutation matrix, and  
\begin{equation*}
	\bm{Z}(q,\bm{\theta}) = \begin{bmatrix}
		-\bm{G}(q)\bm{S}(q)&-\bm{H}(q)\bm{S}(q)\\\bm{S}(q)&-F_y(q)\bm{H}(q)\bm{S}(q)
	\end{bmatrix}.
\end{equation*}
Based on \eqref{CE7}, ${\bar R_n}$ can be rewritten as
\begin{equation} \label{CE8}
	\begin{split}
	    {\bar R_n} =& \bar {\mathbb{E}}\left[ \bm{z}_n(k) \bm{z}_n^\top(k)\right] \\
		=&\frac{1}{2\pi}\int_{-\pi}^{\pi} 
		\bm{P}_1\begin{bmatrix}\mathcal{V}_{n}&0\\0&\mathcal{V}_{n}\end{bmatrix}
		\bm{Z}(e^{iw},\bm{\theta})\text{diag}\left(\Psi_r(w),\sigma_e^2\right)\\
		&\times \bm{Z}^{*}(e^{iw},\bm{\theta})
		\begin{bmatrix}\mathcal{V}_{n}^{*}&0\\0&\mathcal{V}_{n}^{*}
		\end{bmatrix}\bm{P}_1^\top \,dw.
	\end{split}		
\end{equation}
Second, notice that $\mathcal{K}_n(\bm{a})  = \mathcal{T}_{n,p}(\bm{a}) \otimes I$, we then write $\mathcal{K}_n(\bm{a})$ as 
\begin{equation} \label{CE9}
	\mathcal{K}_n(\bm{a}) = \bm{P}_1 \text{diag}\left(\mathcal{T}_{n,p}(\bm{a}),\mathcal{T}_{n,p}(\bm{a})\right)\bm{P}_2,
\end{equation}
where $\bm{P}_2$ is a permutation matrix. Moreover, the Teoplitz matrix $\mathcal{T}_{n,p}(\bm{a})$ can be expressed in the frequency domain as
\begin{equation} \label{CE10}
	\mathcal{T}_{n,p}(\bm{a}) = \frac{1}{2\pi}\int_{-\pi}^{\pi} \mathcal{V}_{n}{\bm{A}(e^{iw},\bm{\theta})}\mathcal{V}_{p}^{*}\,dw.
\end{equation}
Third, notice that $\left\{g_i = \begin{bmatrix}
	C{A_K^{i - 1}}B&C{A_K^{i - 1}}K\end{bmatrix}\right\}_{i=1}^n$ are the truncated impulse responses of $\begin{bmatrix}\frac{\bm{F}(q)}{\bm{A}(q,)}&\frac{\bm{L}(q)}{\bm{A}(q)}
\end{bmatrix}$, thus, the Hankel matrix ${\mathcal{H}}_{n_xn}^{+}$ can be expressed by a product of Toeplitz matrices and permutation matrices, i.e.,
\begin{equation} \label{CE11}
	\begin{split}
		({\mathcal{H}}_{n_xn}^{+})^\top &=  \bm{P}_2\begin{bmatrix}
			\mathcal{T}_{p,n_x}\left(\frac{\bm{F}(q)}{\bm{A}(q)}\right)\bm{P}_3\\
			\mathcal{T}_{p,n_x}\left(\frac{\bm{L}(q)}{\bm{A}(q)}\right)\bm{P}_3
		\end{bmatrix} \\
		&=  \bm{P}_2\begin{bmatrix}
			\frac{1}{2\pi}\int_{-\pi}^{\pi} \mathcal{V}_{p}\frac{\bm{F}(e^{iw})}{\bm{\mathrm{A}}(e^{iw})}\mathcal{V}_{n_x}^{*}\,dw\bm{P}_3\\
			\frac{1}{2\pi}\int_{-\pi}^{\pi} \mathcal{V}_{p}\frac{\bm{L}(e^{iw})}{\bm{\mathrm{A}}(e^{iw})}\mathcal{V}_{n_x}^{*}\,dw\bm{P}_3
		\end{bmatrix},
	\end{split}	
\end{equation}
where the permutation matrix $\bm{P}_3$ converts a Toeplitz matrix into a Hankel matrix, and the permutation matrix $\bm{P}_2$ reorders the rows of the Hankel matrix to align with $({\mathcal{H}}_{n_xn}^{+})^\top$. 

Using expressions \eqref{CE8}, \eqref{CE9} and \eqref{CE10}, we rewrite $M(\bm{g}_n,\bm{a})$ as
\begin{equation} \label{CE12}
	\begin{split}
		&M(\bm{g}_n,\bm{a}) = \lim_{n \to \infty} {\mathcal{H}}_{n_xn}^{+}\left( {{\mathcal{K}_n^\top}(\bm{a})}{{\bar R_n}^{-1}}{\mathcal{K}_n}(\bm{a})\right)^{-1}({\mathcal{H}}_{n_xn}^{+})^\top \\
		&=\lim_{n \to \infty} \innerproduct{\gamma}{\Sigma_n}\left(\innerproduct{\Sigma_n}{\Omega_n}{\innerproduct{\Omega_n}{\Omega_n}}^{-1}\innerproduct{\Omega_n}{\Sigma_n}\right)^{-1}\innerproduct{\Sigma_n}{\gamma},
	\end{split}	
\end{equation}
where
\begin{equation*}
	\begin{split}
		\Omega_n &= \bm{P}_1\begin{bmatrix}
			-\mathcal{V}_{n}\bm{G}(q)\bm{S}(q)\psi_r(q)&-\mathcal{V}_{n}\bm{H}(q)\bm{S}(q)\sigma_e\\
			\mathcal{V}_{n}\bm{S}(q)\psi_r(q)&-\mathcal{V}_{n}F_y(q)\bm{H}(q)\bm{S}(q)\sigma_e
		\end{bmatrix}, \\
		\Sigma_n &= \bm{P}_2\begin{bmatrix}		\mathcal{V}_{p}&0\\0&\mathcal{V}_{p}\end{bmatrix}
		\begin{bmatrix}-\frac{F_y^{*}(q)\bm{H}^{*}(q)\bm{F}^{*}(q)}{\psi_r^{*}(q)}&\frac{\bm{F}^{*}(q)}{\sigma_e}\\
			-\frac{{\bm{A}^{*}(e^{iw})}}{{\psi_r^{*}(q)}}&-\frac{\bm{L}^{*}(q)}{\sigma_e}
		\end{bmatrix}, \\
		\gamma &= \begin{bmatrix}
			0&-\frac{\sigma_e\bm{P}_3^\top\mathcal{V}_{n_x}}{{\bm{A}(e^{iw})}}
		\end{bmatrix}.
	\end{split}	
\end{equation*}
It can be verified that $\innerproduct{\Omega_n}{\Omega_n} = {\bar R_n}$, $\innerproduct{\Omega_n}{\Sigma_n} = \mathcal{K}_n(\bm{a})$ and $\innerproduct{\Sigma_n}{\gamma} = ({\mathcal{H}}_{n_xn}^{+})^\top$. In a similar way to \cite[Th. 2]{Galrinho2019estimating}, using the geometric approach originally proposed in \cite{Hjalmarsson2010geometric}, we have
\begin{equation*}
	\begin{split}
		&\lim_{n \to \infty} \innerproduct{\gamma}{\Sigma_n}\left(\innerproduct{\Sigma_n}{\Omega_n}{\innerproduct{\Omega_n}{\Omega_n}}^{-1}\innerproduct{\Omega_n}{\Sigma_n}\right)^{-1}\innerproduct{\Sigma_n}{\gamma} \\
		& =\innerproduct{\gamma}{\gamma} =\frac{\sigma_e^2}{2\pi}\int_{-\pi}^{\pi} 
		\frac{\mathcal{V}_{n_x}}{\bm{A}(e^{iw},\bm{a})} \frac{\mathcal{V}_{n_x}^{*}}{\bm{A}^{*}(e^{iw},\bm{a})} \,dw.
	\end{split}	
\end{equation*}
Therefore, we verify that $M_{CR,\bm{a}} =  M(\bm{g}_n,\bm{a})$.

\subsection{Proof of Theorem \ref{Thm3}}
\begin{proof}
	Now we show the asymptotic distribution of our estimates $\hat{\bm{a}}_{\text{wls}}$. Specifically, we show that its asymptotic variance corresponds to the CRLB $M_{CR,\bm{a}}$ in \eqref{CRLB-a-true}. According to \eqref{BE9}, we rewrite the estimation error as  
	\begin{equation} \label{CE13}
		\sqrt{N}\tilde{\bm{a}}_{\text{wls}}
		=\hat\kappa\left(\hat{\bm{g}}_n,\hat{\bm{a}}_{\text{ols}}\right){\hat{ M}^{-1}(\hat{\bm{g}}_n,\hat{\bm{a}}_{\text{ols}})},
	\end{equation}
	where $\hat\kappa\left(\hat{\bm{g}}_n,\hat{\bm{a}}_{\text{ols}}\right) = -\sqrt{N}\tilde{{\bm{g}}}_n{\mathcal{K}_n}(\bm{a}){\hat\Lambda_n^{-1}(\hat{\bm{a}}_{\text{ols}})}({\hat {\mathcal{H}}_{n_xn}^{+}})^{\top}$.
	Note that both $\hat\kappa\left(\hat{\bm{g}}_n,\hat{\bm{a}}_{\text{ols}}\right)$ and $\hat{M} (\hat{\bm{g}}_n,\hat{\bm{a}}_{\text{ols}})$ are of fixed dimension. Moreover, according to \eqref{BE2} we have that 
	\begin{equation*} \label{CE14}
		\norm{\hat{ M}^{-1}(\hat{\bm{g}}_n,\hat{\bm{a}}_{\text{ols}}) - {M}^{-1}(\bm{g}_n,\bm{a})} \to 0, \text{as} \ N \to \infty \ {\text{w.p.1}}.
	\end{equation*}
	If we further assume that 
	\begin{equation} \label{CE15}
		\hat\kappa\left(\hat{\bm{g}}_n,\hat{\bm{a}}_{\text{ols}}\right) \sim \AsN{0}{P_\kappa},
	\end{equation}
	according to \cite[Lemma B.4]{Soderstrom2007system}, we then have
	\begin{equation} \label{CE16}
		\sqrt{N} \tilde{\bm{a}}_{\text{wls}} \sim \AsN{0}{{M}^{-1}(\bm{g}_n,\bm{a})P_\kappa {M}^{-1}(\bm{g}_n,\bm{a})}.
	\end{equation}
	We now use Lemma \ref{LemG4} repeatedly to show that \eqref{CE15} holds, and further 
	\begin{equation} \label{CE17}
		P_\kappa = \sigma _e^2 M(\bm{g}_n,\bm{a}).
	\end{equation}
	Define $\kappa\left({\bm{g}}_n,\bm{a}\right) := -\sqrt{N}\tilde{{\bm{g}}}_n{\mathcal{K}}(\bm{a}){\bar\Lambda_n^{-1}(\bm{a})}({{\mathcal{H}}_{n_xp}^{+}})^{\top}$. Since $\sqrt N \tilde{{\bm{g}}}_n \sim \AsN{0}{\sigma _e^2{{\bar R_n}^{-1}}}$, we have
	\begin{equation} \label{CE18}
		\kappa\left({\bm{g}}_n,\bm{a}\right) \sim \AsN{0}{\sigma _e^2 M(\bm{g}_n,\bm{a})}.
	\end{equation}
	Based on \eqref{AE4b} and \eqref{BE7}, we have $\norm{{\tilde{\mathcal{H}}}_{n_xp}^{+}}\to 0$ and  $\norm{{\hat\Lambda_n^{-1}(\hat{\bm{a}}_{\text{ols}})} - {\bar\Lambda}_n^{-1}(\bm{a})} \to 0$, as $N \to \infty \ {\text{w.p.1}}$. 
	Use Lemma~\ref{LemG4} repeatedly, we conclude that $\hat\kappa\left(\hat{\bm{g}}_n,\hat{\bm{a}}_{\text{ols}}\right)$ and $\kappa\left({\bm{g}}_n,\bm{a}\right)$ have the same asymptotic distribution and covariance. Therefore, $P_\kappa = \sigma _e^2 M(\bm{g}_n,\bm{a})$. Returning to \eqref{CE16}, we have that
	\begin{equation} \label{CE20}
		\sqrt{N}\tilde{\bm{a}}_{\text{wls}} \sim \AsN{0}{\sigma _e^2{M}^{-1}(\bm{g}_n,\bm{a})}.
	\end{equation}
	According to \eqref{CRLB-a-true}, the CRLB of $\bm{a}$, $M_{CR,\bm{a}} = M(\bm{g}_n,\bm{a})$, we thereby complete the proof.

	Regarding the the estimation error $\tilde{\bm{\eta}}_{\text{wls}}$ in \eqref{DE-estimation_B_K_error_wls}, we rewrite it as  
    \begin{equation} \label{CE13}
    	\sqrt{N}\tilde{\bm{\eta}}_{\text{wls}}
    	=\hat\kappa\left(\hat{\bm{g}}_n,\hat{\bm{a}}_{\text{wls}},\hat{\bm{\eta}}_{\text{ols}}\right)\hat{M}^{-1}\left(\hat{\bm{g}}_n,\hat{\bm{a}}_{\text{wls}},\hat{\bm{\eta}}_{\text{ols}}\right),
    \end{equation}
    where 
    \begin{equation*}
    	\begin{split}
    		\hat\kappa\left(\hat{\bm{g}}_n,\hat{\bm{a}}_{\text{wls}},\hat{\bm{\eta}}_{\text{ols}}\right) &:=\tilde{\bm{g}}_n{\mathcal{K}_n}(\bm{a},\bm{\eta}) \hat\Lambda_n^{-1}(\hat{\bm{a}}_{\text{wls}},\hat{\bm{\eta}}_{\text{ols}}) {\hat\Phi}_n^\top, \\
    		\hat{M}\left(\hat{\bm{g}}_n,\hat{\bm{a}}_{\text{wls}},\hat{\bm{\eta}}_{\text{ols}}\right) &:= {\hat\Phi}_n\hat\Lambda_n^{-1}(\hat{\bm{a}}_{\text{wls}},\hat{\bm{\eta}}_{\text{ols}}){\hat\Phi}_n^\top,
    	\end{split}
    \end{equation*}
    are of fixed dimension. Same as $\hat{\bm{a}}_{\text{wls}}$, $\hat{\bm{\eta}}_{\text{wls}}$ is obtained using the asymptotic maximum likelihood scheme defined in \cite{Wahlberg1989model}, which leads to an asymptotically (when both the number of samples $N$ and the order of HOARX $n$ tend to infinity) efficient estimator. Specifically, we have that
    \begin{equation} \label{CE20}
    	\sqrt{N}\tilde{\bm{\eta}}_{\text{wls}} \sim \AsN{0}{\sigma_e^2M_{CR,\bm{\eta}}^{-1}},
    \end{equation}
    where $M_{CR,\bm{\eta}} = {M}(\bm{g}_n,\bm{a},\bm{\eta}) := \lim_{n \to \infty} {\Phi}_n\bar\Lambda_n^{-1}({\bm{a}},{\bm{\eta}}){\Phi}_n^\top$ coincides with the CRLB of $\bm{\eta}$.
\end{proof}

\section{Asymptotic Properties of WNSF for Multi-output Systems} \label{AppF}

\subsection{Auxiliary Results (Overlapping Parametrization)}
In this part, we illustrate how a canonical parametrization is derived for multi-output systems. The key property of a canonical parametrization is that the corresponding state vector $x_k$ can be interpreted in a pure input-output context. This is be seen as follows. Based on \eqref{E1}, the one-step-ahead predictor is given by
\begin{subequations} \label{FE1}
	\begin{align}
		\hat x_{k + 1 | k} = &A(\bm{\theta}) \hat x_{k|k-1}  + B(\bm{\theta})u_{k} + \nonumber \\
		&K(\bm{\theta})(y_{k} - \hat y_{k|k-1}), \\
		\hat y_{k|k-1} = &C\hat x_{k|k-1},
	\end{align}
\end{subequations}
where $\bm{\theta}$ denotes the free parameters in the canonical parametrization \eqref{MIMO_canonical_form}. For convenience, the $i$-th component of $\hat y_{k|k-1}$ is denoted by $\hat y_{k|k-1}^{[i]}$, where $i=1, 2, \dots, n_y$. Let $\bar\nu = \left\{\nu_1,\dots,\nu_{n_y}\right\}$ denote the Kronecker index, a set of $n_y$ positive integers satisfying $\sum_{i=1}^{n_y}\nu_i = n_x$. Corresponding to $\bar\nu$, we pick the following $n$ vectors:
\begin{equation*}
	\begin{Bmatrix}
		\hat y_{k|k-1}^{[1]}, & \hat y_{k+1|k-1}^{[1]}, & \cdots, & \hat y_{k+\nu_1-1|k-1}^{[1]} \\
		\hat y_{k|k-1}^{[2]}, & \hat y_{k+1|k-1}^{[2]}, & \cdots, & \hat y_{k+\nu_2-1|k-1}^{[2]} \\
		\vdots & \vdots & \ddots & \vdots \\
		\hat y_{k|k-1}^{[n_y]}, & \hat y_{k+1|k-1}^{[n_y]}, & \cdots, & \hat y_{k+\nu_{n_y}-1|k-1}^{[n_y]}
	\end{Bmatrix}.
\end{equation*}
If these $n$ vectors are linely independent, then this selection is generic situation. Based on the above linearly independent components, we define a state vector of the system by
\begin{equation*}
	\hat x_{k|k-1} := 
	\begin{bmatrix}
		\hat y_{k|k-1}^{[1]} \\ 
		\vdots \\ 
		\hat y_{k+\nu_1-1|k-1}^{[1]} \\
		\vdots \\
		\hat y_{k|k-1}^{[n_y]} \\
		\vdots \\
		\hat y_{k+\nu_{n_y}-1|k-1}^{[n_y]}
	\end{bmatrix},
	\hat x_{k+1|k} := 
	\begin{bmatrix}
		\hat y_{k+1|k}^{[1]} \\ 
		\vdots \\ 
		\hat y_{k+\nu_1|k}^{[1]} \\
		\vdots \\
		\hat y_{k+1|k}^{[n_y]} \\
		\vdots \\
		\hat y_{k+\nu_{n_y}|k}^{[n_y]}
	\end{bmatrix}.
\end{equation*}
Then, according to \cite[Eq. 4A.39]{Ljung1999system}, we have
\begin{equation} \label{FE2}
	\hat y_{k+t|k} = \hat y_{k+t|k-1} + {M}_tu_k + {N}_te_k,
\end{equation}
where ${M}_t = CA^{t-1}B \in \mathbb{R}^{n_y\times n_u}$ and ${N}_t = CA^{t-1}K \in \mathbb{R}^{n_y\times n_y}$. In terms of components, this can be written as
\begin{equation} \label{FE3}
	\hat y_{k+t|k}^{[i]} = \hat y_{k+t|k-1}^{[i]} + M_t^{[i]}u_k + N_t^{[i]}e_k ,
\end{equation}
where 
\begin{equation*}
	\begin{split}
		M_t^{[i]} &= \begin{bmatrix}	M_{t,1}^{[i]}&\cdots&M_{t,n_y}^{[i]}\end{bmatrix}, \\
		N_t^{[i]} &= \begin{bmatrix}	N_{t,1}^{[i]}&\cdots&N_{t,n_y}^{[i]}\end{bmatrix},
	\end{split}
\end{equation*} are the $i$-th rows of $M_t$ and $N_t$. Thus from \eqref{FE3}, we can verify that
\begin{equation*}
	\begin{split}
		\hat x_{k+1|k} = 
	&\begin{bmatrix}
		\hat y_{k+1|k-1}^{[1]} \\ 
		\vdots \\ 
		\hat y_{k+\nu_1|k-1}^{[1]} \\
		\vdots \\
		\hat y_{k+1|k-1}^{[n_y]} \\
		\vdots \\
		\hat y_{k+\nu_{n_y}|k-1}^{[n_y]}
	\end{bmatrix} + 
	\begin{bmatrix}
		M_{1,1}^{[1]}&\cdots&M_{1,n_y}^{[1]} \\ 
		\vdots &\ddots &\vdots \\ 
		M_{\nu_1,1}^{[1]}&\cdots&M_{\nu_1,n_y}^{[1]} \\
		\vdots & \ddots & \vdots \\
		M_{1,1}^{[n_y]}&\cdots&M_{1,n_y}^{[n_y]} \\
		\vdots &\ddots &\vdots  \\
		M_{\nu_{n_y},1}^{[n_y]}&\cdots&M_{\nu_{n_y},n_y}^{[n_y]} 
	\end{bmatrix} u_k + \\
	&\begin{bmatrix}
		N_{1,1}^{[1]}&\cdots&N_{1,n_y}^{[1]} \\ 
		\vdots &\ddots &\vdots \\ 
		N_{\nu_1,1}^{[1]}&\cdots&N_{\nu_1,n_y}^{[1]} \\
		\vdots & \ddots & \vdots \\
		N_{1,1}^{[n_y]}&\cdots&N_{1,n_y}^{[n_y]} \\
		\vdots &\ddots &\vdots  \\
		N_{\nu_{n_y},1}^{[n_y]}&\cdots&N_{\nu_{n_y},n_y}^{[n_y]}
	\end{bmatrix} e_k.
	\end{split}	
\end{equation*}
For brevity, the above equation is denoted by
\begin{equation} \label{FE4}
	\hat x_{k+1|k} = \xi_{k+1} + Bu_k + Ke_k.
\end{equation}
Now, putting $t=0$ in \eqref{FE3} and noting that $\hat y_{k|k} = y_k$ and $M_0 = 0$ and $N_0 = I$ yield
\begin{equation} \label{FE5}
	y_{k}^{[i]} = \hat y_{k|k-1}^{[i]} + e_k^{[i]},
\end{equation} \label{FE5}
which further gives
\begin{equation} \label{FE6}
	y_{k} = \hat y_{k|k-1} + e_k = C \hat x_{k|k-1} + e_k,
\end{equation}
where $C$ is described in the cononical parameterization \eqref{MIMO_canonical_form}. After replacing $e_k$ in \eqref{FE4}, we have that
\begin{equation} \label{FE7}
	\hat x_{k+1|k} = \left(\xi_{k+1} - KC \hat x_{k|k-1}\right) + Bu_k + Ky_k.
\end{equation}
Since the $n_x$ vectors contained in $\hat x_{k|k-1}$ are linearly independent, the components in $\xi_{k+1}$ can be expressed in terms of a linear combination of the components of the basis vector $\hat x_{k|k-1}$, which gives
\begin{equation} \label{FE8}
	\xi_{k+1} - KC \hat x_{k|k-1} = A_K \hat x_{k|k-1}.
\end{equation}
Moreover, several components of $\xi_{k+1}$ are already contained in the vector $\hat x_{k|k-1}$ as its elements, so that they are expressed in terms of shift operations described in the cononical parameterization \eqref{MIMO_canonical_form}.

With a similar reasoning (replacing $e_k$ with $y_k$), we conclude that the following predictor form has the cononical parameterization \eqref{MIMO_canonical_form}:
\begin{subequations} \label{FE9}
	\begin{align}
		\hat x_{k + 1 | k} = &A_K(\bm{\theta}) \hat x_{k|k-1}  + B(\bm{\theta})u_{k} + K(\bm{\theta})y_{k},\\
		\hat y_{k|k-1} = &C\hat x_{k|k-1},
	\end{align}
\end{subequations}
where $\bm{\theta}$ denotes the free parameters in the canonical parametrization as shown in \eqref{MIMO_canonical_form}.

\subsection{Auxiliary Results (the Cram\'er-Rao Lower Bound)}
In this part, we derive the CRLB for free parameters in a canoncial parameterization. In what follows we will let an index $i$ denote the derivative with respect to $\bm{\theta}_i$ (rather than the $i$:th component). Differentiating the predictor \eqref{FE9} gives:
\begin{subequations} \label{FE10}
	\begin{align}
		\hat x_i(k+1|k) = &A_i\hat x(k|k-1) + A_K\hat x_i(k|k-1) \nonumber \\
		&+B_iu_k + K_iy_k, \\		
		\psi_i^\top(k) = &\epsilon_i(k) = - C\hat x_i(k|k-1).
	\end{align}
\end{subequations}
For brevity, we only derive the CRLB for free parameters in each row of $A_K$, denoted by $\bm{a}_i$. Since we are only interested in $\bm{a}_i$, the derivative respective to $\bm{a}_i$ can be written as
\begin{subequations} \label{FE11}
	\begin{align}
		\hat x_i(k+1|k) = &A_{K_i}\hat x(k|k-1) + A_K\hat x_i(k|k-1),\\	
		\psi_i^\top(k) = &\epsilon_i(k) = - C\hat x_i(k|k-1).
	\end{align}
\end{subequations}
In this way, we have that
\begin{equation} \label{FE12}
	\psi_i^\top(k) = -C(qI-A_K)^{-1}A_{K_i}\hat x(k|k-1).
\end{equation}
Furthermore, based on the predictor \eqref{FE9}, we further have that
\begin{equation} \label{FE13}
	\hat x(k|k-1)= (qI-A_K)^{-1}\begin{bmatrix}
	B&K\end{bmatrix}z_k.
\end{equation}
Substituting the above equation into \eqref{FE12}, we have
\begin{equation} \label{FE17}
	\psi_i^\top(k) = -C(qI-A_K)^{-1}A_{K_i}(qI-A_K)^{-1}\begin{bmatrix}B&K\end{bmatrix}z_k.
\end{equation}
Define $\zeta_k(\bm{a}_i) = \begin{bmatrix}
	\psi_1(k) \\ \psi_2(k) \\ \vdots \\\psi_{n_x}(k)
\end{bmatrix} \in \mathbb{R}^{n_x\times n_y}$. Then, we can express the CRLB of $\bm{a}_i$ as
\begin{equation} \label{FE18}
	M_{CR,\bm{a}_i} =\bar {\mathbb{E}}\left[ \zeta_k(\bm{a}_i)\zeta_k^\top(\bm{a}_i)\right].
\end{equation}
For SISO systems, $M_{CR,\bm{a}_i}$ is equivalent to the expression \eqref{CRLB-a-true} we obtained based on the ARMAX model \eqref{CE1}. To be specific, we have that 
\begin{equation} \label{CE3}
	\zeta_k(\bm{\alpha}) =-\frac{1}{\bm{A}(\bm{\alpha})}\begin{bmatrix}
	\frac{\bm{L}(\bm{\theta})}{\bm{A}(\bm{\alpha})}\mathcal{V}_{n_x}& \frac{\bm{F}(\bm{\theta})}{\bm{A}(\bm{\alpha})}\mathcal{V}_{n_x}
	\end{bmatrix}\begin{bmatrix}u_k\\y_k\end{bmatrix}.	
\end{equation}
It is straightforward to see that $-C(qI-A_K)^{-1}$ in $\psi_i^\top(k)$ corresponds to $-\frac{1}{\bm{A}(\bm{\alpha})}$ in $\zeta_k(\bm{\alpha})$, and $A_{K_i}(qI-A_K)^{-1}\begin{bmatrix}B&K\end{bmatrix}$ in $\psi_i^\top(k)$ corresponds to $\begin{bmatrix}	\frac{\bm{L}(\bm{\theta})}{\bm{A}(\bm{\alpha})}\mathcal{V}_{n_x}& \frac{\bm{F}(\bm{\theta})}{\bm{A}(\bm{\alpha})}\mathcal{V}_{n_x}
\end{bmatrix}$ in $\zeta_k(\bm{\alpha})$, respectively. Therefore, the CRLB shown in \eqref{FE18} is equivalent to the asymptotic error variance \eqref{CRLB-a-true}, which also coincides with the asymptotic error covariance matrix of WNSF\textsubscript{SS}, as shown in Appendix \ref{AppC}. 

The key point is that based on the predictor's sensitivity, it is convenient to derive the CRLB for state-space models, particularly for multi-output systems. In the case of single-output systems, this approach is equivalent to the ARMAX model method discussed earlier. In practice, for a given state-space model, one can construct an augmented state-space model by stacking $\hat x(k|k-1)$ and $\epsilon_i(k)$ into the state vector, and compute the CRLB by solving a Lyapunov equation; see \cite{Soderstrom2006computing} and Appendix \ref{AppH} for details.

\subsection{Proof of Theorem \ref{Thm6}}
\begin{proof}
    Regarding the consistency and asymptotic normality of WNSF\textsubscript{SS} for multi-output systems, when the canoncial parameterization is admissible, the analysis is similar to the single-output case. This is due to that matrices with fixed dimensions therein are also fixed here, and dimensions that increased with a rate that is function of $N$ in Assumption~\ref{Asp4} still do so with the same rate here. 
	
	What remains is to show that the asymptotic variance matches that of the PEM applied to the same admissible parameterization $M_{\bar{\nu}_i}$, where PEM is used with a quadratic cost function and optimal weighting. First, it can be shown that the asymptotic variance of WNSF\textsubscript{SS} for $\bm{a}_i$ is given by $\sigma_e^2M^{-1}(\bm{g}_n,\bm{a}_i)$, where 
	\begin{equation} \label{FE19}
		\begin{split}
			M(\bm{g}_n,\bm{a}_i) = &\lim_{n \to \infty} {\mathcal{H}}_{n_xn}^{+}(\bar\nu)\left( {{\mathcal{K}_n^\top}(\bm{a}_i)}{{\bar{\bm{R}}_n}^{-1}}{\mathcal{K}_n}(\bm{a}_i)\right)^{-1}\\
		    &({\mathcal{H}}_{n_xn}^{+}(\bar\nu))^\top.
		\end{split}		
	\end{equation}
	From \eqref{FE9}, it is easy to see that an $n_y$ output state-space models can be equivalently rewritten as an $n_y$ output ARMAX model ($n_y$ parallel but not independent single-output ARMAX models). For more details about equivalent parameterizations for cononical ARMAX models and state-space models, we refer to \cite{Hannan2012statistical,Van1982line}. For each ARMAX model and state-space model, the parameters $\bm{a}_i$ are identical. Therefore, a similar proof as in Appendix~\ref{AppC} can be derived to show that $M(\bm{g}_n,\bm{a}_i)$ coincides with the CRLB in \eqref{FE19}. 

	From another perspective, it can be shown that each WLS in Steps~3 and 5 of WNSF\textsubscript{SS} consists of a solution of the quadratic optimization problem which minimizes the approximated likelihood function $\hat{L}_N(\theta)$, we conclude that they yield  asymptotically efficient estimates. For more details about this perspective, we refer to Appendix \ref{AppI}.
\end{proof}

\section{Technical Lemmas}   \label{AppG}

\begin{lemma} [Lemma A.1 in \cite{Tsiamis2019finite}] \label{LemG1}
	Norm of a block matrix: Let $M$ be a block-column matrix defined as $M = \begin{bmatrix}
		M_1^\top&M_2^\top&\cdots &M_f^\top
	\end{bmatrix}^\top$, where all the $M_i$'s have the same dimension. Then, the block matrix $M$ satisfies
	\begin{equation*}
		\norm{M} \leq \sqrt{f} \max\limits_{1\leq i\leq f} \norm{M_i}.
	\end{equation*}
\end{lemma}

\begin{lemma} [Proposition 1 in \cite{Galrinho2018parametric}] \label{LemG2}
	Consider the product $\prod_{i=1}^p \hat{M}_N^{(i)}$, where $p$ is finite and $\hat{M}_N^{(i)}$ are stochastic matrices of appropriate dimensions (possibly a function of $N$) such that
	\begin{equation*}
		\norm{\hat{M}_N^{(i)} - {M}_N^{(i)}} \to 0, {\rm{as}} \ N \to \infty \ {\text{w.p.1}}.
	\end{equation*}
	where ${M}_N^{(i)}$ is a deterministic matrix for each $N$ satisfying $\norm{{M}_N^{(i)}} < c_i$, which may influence its dimensions according to the dimensions of $\hat{M}_N^{(i)}$. Then, we have that
	\begin{equation*}
		\norm{\prod_{i=1}^p \hat{M}_N^{(i)} - \prod_{i=1}^p {M}_N^{(i)}} \to 0, {\rm{as}} \ N \to \infty \ {\text{w.p.1}}.
	\end{equation*}
\end{lemma}

\begin{lemma} [Theorem 4.1 in  \cite{Wedin1973perturbation}] \label{LemG3}
	Consider rank $m$ matrices $M_1\in \mathbb{R}^{m\times n}$ and $M_2\in \mathbb{R}^{m\times n}$, where $m\leq n$. Then, we have 
	\begin{equation}
		\nonumber
		\norm{M_1^\dagger-M_2^\dagger} \leq \sqrt{2}\norm{M_1^\dagger}\norm{M_2^\dagger}\norm{M_1-M_2}.
	\end{equation}
\end{lemma}

\begin{lemma} [Proposition 2 in \cite{Galrinho2018parametric}] \label{LemG4}
	Consider a finite dimensional vector $\hat x_N = \sqrt{N}\hat P_N \hat Q_N \hat \delta_N$, where $\hat P_N$ and $\hat Q_N$ are random matrices, and $\hat \delta_N$ is random vector of compatible dimensions. Except for the constraint that the number of rows of $\hat P_N$ is fixed, other dimensions are allowed to grow to infinity at a suitable rate with $N$. Furthermore, we assume that $\hat P_N$ is bounded, and there is $\bar Q$ such that $\norm{\hat Q_N -\bar Q} \to 0$ as $N \to \infty \ {\text{w.p.1}}$, and $\norm{\hat \delta_N} \to 0$ as $N \to \infty \ {\text{w.p.1}}$. Then, if $\sqrt{N}\norm{\hat Q_N -\bar Q}\norm{\hat \delta_N} \to 0$, as $N \to \infty \ {\text{w.p.1}}$, $\hat x_N$ and $\sqrt{N}\hat P_N \bar Q_N \hat \delta_N$ have the same asymptotic distribution and covariance.
\end{lemma}

\section{On Computing the CRLB in State-Space Models} \label{AppH}

This algorithm is mainly based on \cite{Soderstrom2006computing}. Consider the following discrete-time LTI system on the innovations form:
\begin{subequations} \label{HE1}
	\begin{align}
		x_{k + 1} &= A(\bm{\theta})x_{k}  + B(\bm{\theta})u_{k} + K(\bm{\theta})e_{k}, \label{HE1a}\\
		y_{k} &= Cx_{k} + e_{k}, \label{HE1b}		
	\end{align}
\end{subequations}
where $\bm{\theta} = \begin{bmatrix}
		\theta_1&\theta_2&\cdots&\theta_{n_\theta}
	\end{bmatrix}$ denotes free parameters in a canonical form, $n_\theta = (2n_y + n_u)n_x$, and
\begin{equation*}
	\begin{split}
		\mathbb{E}\left\{\begin{bmatrix}
	e_{k}\\e_{k}
	\end{bmatrix}\begin{bmatrix}
	e_{l}\\e_{l}
	\end{bmatrix}^\top\right\} =&  \begin{bmatrix}
	\sigma_e^2K(\bm{\theta})K^\top(\bm{\theta}) & \sigma_e^2K(\bm{\theta}) \\
	\sigma_e^2K^\top(\bm{\theta}) & \sigma_e^2I
	\end{bmatrix}\delta_{k,l} \\
	:=& \begin{bmatrix}
	R_{1}(\bm{\theta})&R_{12}(\bm{\theta}) \\
	R_{21}(\bm{\theta})&R_{2} \\
	\end{bmatrix}\delta_{k,l}.
	\end{split}	
\end{equation*}
Now, the prediction error is given by
\begin{subequations} \label{HE2}
	\begin{align}
		\hat x({k + 1|k}) &= A_K\hat x({k|k-1}) + B u_{k} + Ky_{k} , \label{HE2a}\\
		\epsilon(k,\bm{\theta}) &= y_{k} - C\hat x({k|k-1}) . \label{HE2b}		
	\end{align}
\end{subequations}
Moreover, we have that
\begin{subequations} \label{HERiccati}
	\begin{align}
		P &= APA^\top + R_{1} - K(CPA^\top + R_{12}^\top), \\
		Q &= \mathbb{E}\left\{\epsilon(k,\bm{\theta}) \epsilon^\top(k,\bm{\theta}) \right\} = CPC^\top + R_{2},
	\end{align}
\end{subequations}
where $K$ satisfies $K = (APC^\top + R_{12})(CPC^\top + R_{2})^{-1}$. To find the expression for CRLB, we introduce the sensitivity
\begin{equation} \label{HE3}
	\psi(k,\bm{\theta}) = -\left(\frac{\partial \epsilon(k,\bm{\theta})}{\partial \bm{\theta}} \right) ^\top \in \mathbb{R}^{n_\theta \times n_y}.
\end{equation}
Then, the CRLB is given by
\begin{equation}  \label{HE4}
	M_{\text{CR},\bm{\theta}} = \mathbb{E}\left\{\psi(k,\bm{\theta})Q^{-1}\psi^\top(k,\bm{\theta}) \right\}.
\end{equation}
To find expressions for the covariance matrix of the parameter estimates, apparently, we need
$\mathbb{E}\left[\psi_i(k)\psi_j^\top(k)\right]$, where $i,j=1,\dots,n_\theta$. Set
\begin{equation}  \label{HE5}
    \Vc{\psi(k)}=\begin{pmatrix}
	\psi_1^{\top}(k)\\
	\vdots\\
	\psi_{n_\theta}^{\top}(k)
	\end{pmatrix}\in \mathbb{R}^{n_yn_\theta},
\end{equation}
where $\psi_i^\top(k) = \epsilon_i(k) = \frac{\partial \epsilon(k,\bm{\theta})}{\theta_i}\in \mathbb{R}^{n_y}$. In what follows we will let an index $i$ denote the derivative with respect to $\theta_i$ (rather than the $i$th component). These quantities can be derived from sensitivity derivatives of the optimal predictor \eqref{HE2}, and the Riccati equation \eqref{HERiccati}. We start by deriving $P_i$. Differentiating the Riccati equation \eqref{HERiccati} gives
\begin{equation} \label{HE6}
	\begin{split}
		P_i	= &A_i P A^{\top} + A P_i F^{\top} + A P A_i^{\top} \\
		&+ R_{1i} + K\bigl( C P_i C^{\top} \bigr)K^{\top} \\	
		&-\bigl(A_i P C^{\top} + A P_i C^{\top} + R_{12i}\bigr)K^{\top} \\	
		&-K\bigl(C P_i A^{\top} + C P A_i^{\top} + R_{12i}^{\top}\bigr)\\	
		= &A_KP_iA_K^{\top} +A_iPA_K^{\top} +A_KPA_i^{\top} \\
		&+\bigl(R_{1i}-K R_{12i}^{\top}-R_{12i}K^{\top}\bigr).
	\end{split}
\end{equation}
This is a Lyapunov equation in $P_i$ that is easy to solve numerically. The sensitivity of $Q$ is easily related to $P_i$:
\begin{equation} \label{HE7}
	Q_i =  C P_i C^{\top}.
\end{equation}

Next we have to differentiate the optimal predictor \eqref{HE2}:
\begin{subequations} \label{HE8}
	\begin{align}		
		\hat x_i(k+1|k) = &(A_i - K_i C)\hat x(k|k-1) \nonumber \\
		&+ A_K\hat x_i(k|k-1) + B_iu_k + K_iy_k,\\
		\psi_i^\top(k) = &\epsilon_i(k)  = - C\hat x_i(k|k-1).
	\end{align}
\end{subequations}
We are now in a position to form an augmented state space model for computing $\psi_i(t)$. 

For the open-loop case, introduce the notations
\begin{equation*}
	\bm{A}_{o} =\begin{bmatrix}A_1\\\vdots\\A_{n_\theta}\end{bmatrix},
	\bm{B}_{o} =\begin{bmatrix} B_1\\\vdots\\B_{n_\theta}\end{bmatrix},
	\bm{K}_{o} =\begin{bmatrix}	K_1\\\vdots\\	K_{n_\theta}\end{bmatrix},
\end{equation*}
and define
\begin{equation*}
	\begin{split}
		\bar{\bm{A}}_o &=\begin{bmatrix}	A & 0\\	\bm{A}_{o} & I_{n_\theta}\otimes A_K\end{bmatrix}, \bar{\bm{B}}_o = \begin{bmatrix}B & K\\	\bm{B}_{o} & \bm{K}_{o}\end{bmatrix},  \\
		\bar{\bm{C}}_o &= \begin{bmatrix}0 & -I_{n_\theta}\otimes C\end{bmatrix}.
	\end{split}
\end{equation*}

Then, an augmented state-space model is defined by
\begin{subequations} \label{HE10}
	\begin{align}
		\begin{bmatrix}
			\hat{x}_{k+1}\\
			\hat{x}_{1}(k+1)\\
			\hat{x}_{2}(k+1)\\
			\vdots \\
			\hat{x}_{n_\theta}(k+1)\\
		\end{bmatrix} &= \bar{\bm{A}}_o \begin{bmatrix}
			\hat{x}_{k}\\
			\hat{x}_{1}(k)\\
			\hat{x}_{2}(k)\\
			\vdots \\
			\hat{x}_{n_\theta}(k)\\
		\end{bmatrix}  + \bar{\bm{B}}_o \begin{bmatrix}
			u_{k}\\e_k
		\end{bmatrix}, \\
		\begin{bmatrix}	\psi_1^{\top}(k)\\	\vdots\\	\psi_{n_\theta}^{\top}(k)	\end{bmatrix} &= \bar{\bm{C}}_o \begin{bmatrix}
			\hat{x}_{k}\\
			\hat{x}_{1}(k)\\
			\hat{x}_{2}(k)\\
			\vdots \\
			\hat{x}_{n_\theta}(k)\\
		\end{bmatrix}.
	\end{align}
\end{subequations}	
The covariance matrix of the augmented state vector can easily be found by solving the following Lyapunov equation:
\begin{equation} \label{HE11}
    \bar{\bm{P}}_o = \bar{\bm{A}}_o \bar{\bm{P}}_o \bar{\bm{A}}_o^{\top} + \bar{\bm{B}}_o \text{Cov}\left\{\begin{bmatrix}
	u_{k}\\e_k\end{bmatrix}\right\}\bar{\bm{B}}_o^{\top},
\end{equation}
to get $\mathbb{E}\left\{\Vc{\psi(k)}\Vc{\psi(k)}^\top\right\} = \bar{\bm{C}}_o \bar{\bm{P}}_o \bar{\bm{C}}_o^\top$.

For the closed-loop case, assume that $u_k = r_k - F_y y_k$. Then, replacing $u_k$ in the predictor \eqref{HE2} and its derivative \eqref{HE7}, we have
\begin{subequations} \label{HE13}
	\begin{align}
		\hat x({k + 1|k}) = &(A-BF_yC)\hat x({k|k-1}) + B r_{k} \nonumber\\
		&+ (K-BF_y)e_{k}, \\
		\epsilon(k,\bm{\theta}) = &y_{k} - C\hat x({k|k-1}),
	\end{align}
\end{subequations}
and
\begin{subequations} \label{HE14}
	\begin{align}
		\hat x_i(k+1|k) = &(A_i - B_iF_yC)\hat x(k|k-1) + A_K\hat x_i(k|k-1) \nonumber \\
		&+B_ir_k + (K_i-B_iF_y)e_k, \\		
		\psi_i^\top(k) = &\epsilon_i(k) = - C\hat x_i(k|k-1),
	\end{align}
\end{subequations}
respectively. Similarly, introduce the notations
\begin{equation*} \label{HE8}
	\begin{split}
		\bm{A}_{c} &=\begin{pmatrix}A_1-B_1F_yC\\\vdots\\A_{n_\theta}-B_{n_\theta}F_yC\end{pmatrix},	\bm{B}_{c} =\begin{pmatrix} B_1\\\vdots\\B_{n_\theta}\end{pmatrix},	\\
		\bm{K}_{c} &=\begin{pmatrix}	K_1-B_1F_y\\\vdots\\K_{n_\theta}-B_{n_\theta}F_y\end{pmatrix},
	\end{split}	
\end{equation*}
and define
\begin{equation} \label{HE9}
	\begin{split}
		\bar{\bm{A}}_c &=\begin{bmatrix}	A-BF_yC & 0\\	\bm{A}_c & I_{n_\theta}\otimes A_K\end{bmatrix}, 
		\bar{\bm{B}}_c = \begin{bmatrix}B & K-BF_y\\	\bm{B}_{c} & \bm{K}_{c}\end{bmatrix},  \\
		\bar{\bm{C}}_c &= \begin{bmatrix}0 & -I_{n_\theta}\otimes C\end{bmatrix}.
	\end{split}
\end{equation}
Then, a similar augmented state-space model can be defined to obtain the covariance matrix of the augmented state vector.

To summarize, we have the following generic algorithm to compute the CRLB for a parameterized state-space model. The matrices $A, B,C, K$ and $A_i, B_i, K_{i}, R_{1i}$ for $i=1,\ldots,n_\theta$ are given.

\textbf{Step~1}: Solve the Riccati equation \eqref{HERiccati} to get $P, Q$.

\textbf{Step~2}: For $i=1,\ldots,n_\theta$: Solve the Lyapunov equation \eqref{HE6} to get $P_i$. Then, form the corresponding block rows of the augmented state space model.

\textbf{Step~3}: Denoting the augmented state space model in brief as
\begin{subequations} \label{HE10}
	\begin{align}
		\bar{\bm{x}}({k+1}) &= \bar{\bm{A}} \bar{\bm{x}}(k)  + \bar{\bm{B}} \bar{\bm{v}}(k) , \\
		\Vc{\psi(k)} &= \bar{\bm{C}} \bar{\bm{x}}(k).
	\end{align}
\end{subequations}
Solve the Lyapunov equation
\begin{equation}
	\bar{\bm{P}} = \bar{\bm{A}} P \bar{\bm{A}}^{\top} + \bar{\bm{B}}\,\operatorname{cov}(\bar{\bm{v}}(k))\,\bar{\bm{B}}^{\top}
\end{equation}
to get $\mathbb{E}\left\{\Vc{\psi(k)}\Vc{\psi(k)}^\top\right\} = \bar{\bm{C}} \bar{\bm{P}} \bar{\bm{C}}^{\top}$.

\textbf{Step~4}: Since
\begin{equation*}
	\begin{split}
		\mathbb{E}\left\{\Vc{\psi(k)}\Vc{\psi(k)}^\top\right\} & \in \mathbb{R}^{n_yn_\theta \times n_yn_\theta}, \\
        M_{\text{CR},\bm{\theta}} = \mathbb{E}\left\{\psi(k,\bm{\theta})Q^{-1}\psi^\top(k,\bm{\theta}) \right\} &\in \mathbb{R}^{n_\theta \times n_\theta},
	\end{split}	
\end{equation*}
we obtain $\mathbb{E}\left\{\psi(k,\bm{\theta})\psi^\top(k,\bm{\theta}) \right\}$ by simply rearranging the elements of $\mathbb{E}\left\{\Vc{\psi(k)}\Vc{\psi(k)}^\top\right\}$. Using this expression, we derive the CRLB. (in our problem the noise covariance is $Q = \sigma_e^2 I$).

\section{Approximating the likelihood function} \label{AppI}

Sometimes it is possible to approximate the likelihood function without jeopardizing asymptotic efficiency of the estimate. At a high level, consider the data model
\begin{equation} \label{IE1}
    \phi_N = \phi(\theta) + e_N,    
\end{equation}
where 
\begin{equation} \label{IE2}
    \sqrt{N}\,e_N \sim \AsN{0}{P(\theta)},    
\end{equation}
where $\theta \in \mathbb{R}^n$ and $\phi(\theta) \in \mathbb{R}^p$, $p \ge 2$, for which the negative log-likelihood function can be approximated by
\begin{equation} \label{IE3}
	\begin{split}
		L_N(\theta) \approx	&\frac{N}{2}\,(\phi_N - \phi(\theta))^{\!\top} P^{-1}(\theta)(\phi_N - \phi(\theta)) \\
		&+ \frac{1}{2}\log \det P(\theta).
	\end{split}	
\end{equation}
For large $N$, for each fixed $\theta$ with non-singular $P(\theta)$, the first term dominates the second term, suggesting that the $\log\det P(\theta)$ term can be neglected. Hence, for large $N$ we use the approximation
\begin{equation} \label{IE4}
	L_N(\theta) \approx
	\frac{N}{2}(\phi_N - \phi(\theta))^{\top} P^{-1}(\theta)(\phi_N - \phi(\theta)).
\end{equation}
This implies that the per-sample Fisher information matrix is obtained by approximating the score function with%
\footnote{All derivatives are with respect to $\theta$ and $\phi'(\theta) := \dfrac{\partial \phi(\theta)}{\partial \theta}$ denotes the $p \times n$ Jacobian.}
\begin{equation*} \label{IE5}
	\begin{split}
		S_N(\theta) \approx &N \phi'(\theta)^{\top} P^{-1}(\theta)(\phi_N - \phi(\theta))\\
		&+ N(\phi_N - \phi(\theta))^{\top}\left(\frac{\mathrm{d}}{\mathrm{d}\theta} P^{-1}(\theta)\right)(\phi_N - \phi(\theta)).
	\end{split}
\end{equation*}
Assume that $\phi'(\theta)\in\mathbb{R}^{p\times n}$ has full column rank. Then the second term is of order $\norm{\phi_N-\phi(\theta)}^2$ whereas the first term is of order $\norm{\phi_N-\phi(\theta)}$. Thus, as $N\to\infty$ the second term can be neglected, giving
\begin{equation} \label{IE6}
	S_N(\theta) \approx
	N\phi'(\theta)^{\top} P^{-1}(\theta)(\phi_N - \phi(\theta)),
\end{equation}
but the right-hand side is the score function for the model \eqref{IE1} when $P(\theta)$ is known constant matrix. This means that the information regarding $\theta$ in the noise covariance is not useful asymptotically and should $P(\theta_\circ)$ be known, the criterion
\begin{equation} \label{IE7}
	\frac{N}{2}(\phi_N - \phi(\theta_\circ))^{\top} P^{-1}(\theta_\circ)(\phi_N - \phi(\theta_\circ)),
\end{equation}
will result in an asymptotically efficient estimate. This remains true if $P(\theta_\circ)$ and $\phi(\theta_\circ)$ are replaced by $\sqrt{N}$-consistent estimates $P_N$ and $\hat{\phi}_N(\theta)$. For details when $e_N$ is normally distributed, see Complement~C4.4 in \cite{Soderstrom2007system}.

In summary, the approximate negative log-likelihood
\begin{equation} \label{IE7}
	\hat{L}_N(\theta) = \big(\phi_N - \hat{\phi}_N(\theta)\big)^{\top} P_N^{-1}\,\big(\phi_N - \hat{\phi}_N(\theta)\big)
\end{equation}
yields an asymptotically efficient estimate. Since each WLS in Steps~3 and 5 of WNSF\textsubscript{SS} consists of a solution of the quadratic optimization problem which minimizes $\hat{L}_N(\theta)$, we conclude that they yield  asymptotically efficient estimates.

\end{document}